\documentclass[epj]{svjour}

\usepackage{amsmath}
\usepackage{amssymb}
\usepackage{amsfonts}
\usepackage{graphicx}

\numberwithin{equation}{section}

\begin{document}

\title{The High Temperature Superconductivity in Cuprates:  Physics of the Pseudogap  Region}
\author{Paolo Cea~\inst{1,}\inst{2,}\thanks{e-mail: paolo.cea@ba.infn.it}}    

\institute{Dipartimento di Fisica, Universit\`a di Bari, 70126 Bari, Italy  \and 
 INFN - Sezione di Bari, 70125 Bari, Italy}
\date{Received: date / Revised version: date}

\abstract{
We discuss the physics of the high temperature superconductivity in hole doped copper oxide ceramics in the pseudogap
region. Starting from an effective reduced Hamiltonian relevant to the dynamics of  holes injected into the copper oxide layers 
proposed in  a previous paper, we determine the superconductive condensate wavefunction. We show that the low-lying elementary
condensate excitations are analogous to the rotons in superfluid $^4He$. We argue that the rotons-like excitations
account for the specific heat anomaly at the critical temperature. We  discuss and compare with experimental
observations the London penetration length, the  Abrikosov  vortices,  the upper and lower critical magnetic fields,
and the critical current density.  We give arguments to explain the origin of the Fermi arcs and Fermi pockets. 
We investigate the nodal gap in the cuprate superconductors and discuss both the doping and temperature dependence of the nodal gap.
We suggest that the nodal gap is responsible for the doping dependence of the so-called nodal
Fermi velocity detected in angle resolved photoemission spectroscopy studies. We discuss the thermodynamics of the nodal 
quasielectron liquid and their role in the low temperature specific heat. We propose that the ubiquitous presence of charge density
wave in hole doped cuprate superconductors in the pseudogap region originates from instabilities of the nodal quasielectrons
driven by the interaction with the planar $CuO_2$ lattice. We investigate the doping dependence of the charge density wave gap
and the competition between charge order and superconductivity. We discuss the effects of external magnetic fields
on the charge density wave gap and elucidate the interplay between charge density wave and Abrikosov 
vortices. Finally, we examine  the physics underlying quantum oscillations in the pseudogap region.
}

\maketitle
\section{Introduction}
\label{s1}
\begin{figure}
\vspace{0.8cm}
\hspace{-0.6cm}
\resizebox{0.5\textwidth}{!}{%
\includegraphics{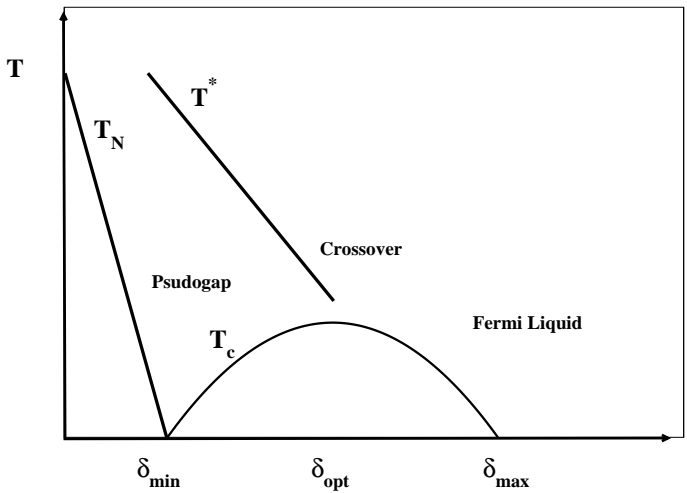}
}
\caption{\label{Fig-1}  Schematic diagram of the temperature versus hole doping level for the copper oxides, indicating where various phases occur. 
$T_N$ is the N\`eel transition temperature. The region below  $T_N$  is where   fully developed antiferromagnetic order set in.  
$T^*$ indicates the temperature where the transition to the pseudogap regime occurs.  $T_c$ is the  superconducting transition
 temperature. The   dome-shaped superconducting region extends from $\delta_{min}$ up to $\delta_{max}$.}
\end{figure}

One of the most exciting development in modern  physics has been the discovery of high  temperature superconductivity 
 in copper oxides (cuprates)  by J. G. Bednorz and K. A. M\"uller~\cite{Bednorz:1986}. 
Indeed, the origin of high temperature superconductivity in cuprates continues to be one of the most debated problem in 
condensed matter physics~\footnote{For recent overviews, see 
Refs.~\cite{Besov:2005,Leggett:2006,Lee:2006,Schrieffer:2007,Plakida:2010,Alloul:2014,Wesche:2015}.}.
Nevertheless, a large amount of informations about  superconductivity in cuprates  has been obtained. In fact,
the phase diagram of the hole doped cuprates is by now quite well established. In  Fig.~\ref{Fig-1}  we illustrate schematically 
the widely accepted phase diagram for hole doped cuprate superconductors~\cite{Taillefer:2010,Keimer:2015,Fradkin:2015}.
The crystal structure of the cuprate high temperature superconductors consists  of $CuO_2$ sheets  
separated by insulating layers. The main driver of superconductivity in the cuprates is the  copper-oxide plane
and, to a good approximation, Cooper pairs form independently on each layer.
It is remarkably that  superconductivity in cuprates arises in the two-dimensional $CuO_2$ planes as a
common behavior to all cuprate families. \\
Parent compounds in these materials are antiferromagnetic Mott insulators.  
As in semiconductors, the carrier concentration in the cuprate superconductors can be  chan-ged by doping, namely
by increasing the number of holes  in the $CuO_2$ planes.  With hole doping to the system
with  small values of the doping $\delta$,  it remains antiferromagnetic but the critical N\`eel temperature
decreases rapidly. Then, for hole doping $\delta \gtrsim \delta_{min} \simeq 0.05$, the long range
antiferromagnetic order is quickly  suppressed and  the system comes into  the superconducting phase.
The superconducting state can be observed as a dome  that  appears roughly between 
$\delta_{min} \lesssim \delta \lesssim \delta_{max} \simeq 0.30$. The superconductive critical temperature attains 
 a maximum at around  $\delta = \delta_{opt} \simeq  0.16$.  This doping level is referred to as the optimally doped region.  
The lower doping region is called underdoped region, while the higher doping region is referred  to  as the overdoped region. 
The overdoped region (especially highly overdoped region) is close to the Fermi liquid state. Indeed, in the overdoped regime the material
 it is believed to have a large Fermi  surface in the normal state and a simple BCS d-wave superconducting gap opens below 
  the superconductive critical temperature $T_c$, even though a clear experimental  evidence is still lacking.
On the other hand,  the underdoped region of the phase diagram is rather unconventional
and it is characterized by the presence of the so-called pseudogap.
The pseudogap was first detected in the temperature dependence of the spin-lattice relaxation and Knight shift in nuclear magnetic
resonance and magnetic susceptibility studies (for up-to-date reviews, see 
Refs.~\cite{Timusk:1999,Tallon:2001,Phillips:2003,Hufner:2008,Sacuto:2013,Kordyuk:2015}).  
In fact, a gradual depletion of the density of states at the Fermi
energy was observed below a crossover temperature $T^*$, revealing the opening of the pseudogap
well above the superconductive critical temperature $T_c$ (see Fig.~\ref{Fig-1}). The  existence of the pseudogap phase
in the underdoped region of the phase diagram is, in fact, one of the most puzzling feature of the high temperature cuprate superconductors.
 It turns out that  the underdoped cuprates  in the  pseudogap region $\delta < \delta^*$   display a host of anomalous properties.
Indeed, in the underdoped regime  the normal state Fermi surface is no longer complete since only Fermi arcs remain.
Moreover, the pseudogap gap increases as the doping gets  smaller, while the superconductive critical
temperature increases almost linearly with the doping. In addition,  along the Fermi arcs there is 
another gap, the nodal gap~\cite{Hufner:2008}, which  does not change much with doping or, if any, deceases in deeply underdoped samples. \\
The above qualitative description together with the persistent evidence of several forms of electronic order, including charge density wave
which  are ubiquitous in this class of superconductors, point to an inextricable complexity of 
high temperature cuprate superconductors~\cite{Fradkin:2012}. Nevertheless, soon after the discover 
of the high temperature superconductors, it was realized~\cite{Anderson:1987,Anderson:1997} that superconductivity were
intimately related to the square planar $CuO_2$ lattice whose physics were well described by the nearly
half-filled Hubbard model with moderately on-site Coulomb repulsion.  Actually, the microscopic model for the description of electrons in the   
 $CuO_2$  layers can be assumed to be the effective single-band Hubbard  model~\cite{Anderson:1987,Anderson:1997}:
\begin{eqnarray}
\label{1.1}
\hat{H}  =  -t  \sum_{<i,j>,\sigma}   \bigg [  \hat{c}^{\dagger}_{i,\sigma}  \, \hat{c}_{j,\sigma}    +    \hat{c}^{\dagger}_{j,\sigma}  \, \hat{c}_{i,\sigma} \bigg ]  \; + 
\\ \nonumber
U  \,  \sum_{i}   \hat{n}_{i,\uparrow}  \, \hat{n}_{i,\downarrow}  \; \;  , \; \; 
  \hat{n}_{i,\sigma} \, = \,  \hat{c}^{\dagger}_{i,\sigma}  \, \hat{c}_{i,\sigma}  \; .
\end{eqnarray}
In Eq.~(\ref{1.1})  $ \hat{c}^{\dagger}_{i,\sigma} $ and   $ \hat{c}_{i,\sigma} $ are creation and annihilation operators for electrons with spin $\sigma$, 
$U$ is the onsite Coulomb repulsion for electrons of opposite spin at the same atomic orbital, and $t$ is the hopping parameter.  
As  is well known \cite{Anderson:1959}  the superexchange mechanism 
yields a Heisenberg antiferromagnetic exchange interaction with $J \, = \, \frac{4 t^2}{U}$ between spins on the $Cu$ atoms.  
Since $ U \, \gg t$, at half-filling the onsite Coulomb repulsion gives an insulator where the electron spins are antiferromagnetically ordered. 
 By doping the system 
an increasing number of holes is created in the $CuO_2$ planes. The dynamics of the injected holes in the  antiferromagnetic background
is still poorly understood. Indeed, the interaction responsible for high temperature superconductivity in the cuprates has remained  elusive. 
Therefore, it is desirable to construct  the simplest model that is able to  capture the basic experimental facts  of the  physics 
of the universal properties of the cuprate superconductors.
To this end, to make things as simple as possible and based on plausible physical assumptions, in a previous paper~\cite{Cea:2013}, 
to be referred to hereafter as I, we  proposed an effective Hamiltonian to account for the low-lying excitations of  the dynamics of
holes immersed in an  antiferromagnetic  background  in the $CuO_2$ planes.
In fact, we found that our approach allowed us to reach at least a qualitative  understanding of the unusual behavior seen in the 
various regions of the phase diagram of the cuprate high temperature superconductors. \\
The aim of the present paper is to better elucidate the physics of the pseudogap region. In fact,  as we said, despite  intensive
studies  the origin of the pseudogap, which dominates the  underdoped region of the phase diagram, as well as 
its relation with the superconductivity is still under debate. We will focus here  specifically on the occurrence of the pseudogap phase
 in the phase diagram of the cuprate superconductors and discuss in detail the physics  of the cuprates in this region. \\
The plan of the paper is as follows. In Sect.~\ref{s2}, for reader convenience, we briefly review the phase diagram
within our phenomenological microscopic theory. The physics of the superconductive condensate in the pseudogap region
is discussed in Sect.~\ref{s3}, where we also discuss the ground state and the condensate wavefunctions.  Sect.~\ref{s3.1} is devoted to
the low-lying excitations of the  superconductive condensate.  In  Sect.~\ref{s3.2} we determine the condensate wavefunction in an
external magnetic field.  In  Sect.~\ref{s3.3} we discuss the roton gas at finite temperature
and  the specific heat anomaly at the critical temperature. The penetration length in the London limit is discussed
 in Sect.~\ref{s3.4}, while the structure of vortices and the critical magnetic fields are presented in  Sect.~\ref{s3.5}. 
 Sections  \ref{s3.6}  and \ref{s3.7} are  dedicated  to the comparison
of the temperature dependence of the critical magnetic fields and critical current with selected experimental observations.
The physics of the nodal quasielectron liquid is introduced in Sect.~\ref{s4}.  In Sect.~\ref{s4.1} we analyze the origin and
the temperature dependence of  the so-called nodal gap. 
Sect.~\ref{s4.2} is reserved to the thermodynamics of the nodal quasielectron liquid.  In Sect.~\ref{s4.3} we critically discuss  
the specific heat at low temperatures. The charge density wave instabilities are presented  in Sect.~\ref{s5}. 
 In Sect.~\ref{s5.1} we estimate the wavenumber vectors responsible for the  instability. In  Sect.~\ref{s5.2} we discuss 
 the charge density wave  critical temperature and energy gap.   In  Sect.~\ref{s5.3} we analyze the competition between 
 charge density wave instability and superconductivity. The effects of an applied magnetic field on the charge density wave
 instability  and the phenomenology of the charge density wave instabilities in the vortex region are discussed in Sect.~\ref{s5.4}.   
Sect.~\ref{s5.5} is devoted  to the physics of quantum oscillation.   Finally, Sect.~\ref{s6}  provides the summary and  
the main conclusions of the paper. 
\\
Several technical details are relegated in the Appendices \ref{AppendixA}, \ref{AppendixB}, 
\ref{AppendixC},  \ref{AppendixD}, and  \ref{AppendixE}.
\section{The High Temperature Superconductivity in  Cuprates}
\label{s2}
In this Section we briefly illustrate the effective Hamiltonian proposed in I  and the resulting phase diagram for hole doped 
cuprate superconductors.  Our approach relies on some gross oversimplifications which, nevertheless,  should capture the relevant
physics of hole doped cuprates. Firstly, we assumed that the physics of the high temperature cuprates is deeply rooted in 
the copper-oxide planes. This allowed us to completely neglect the motion along the direction perpendicular to the $CuO_2$ planes. 
 Moreover, we assumed  that the single-band effective Hubbard model is sufficient to account for all the essential physics of the copper-oxide 
  planes. Within these simplifying approximations, the effective Hamiltonian for the propagation of the holes in the antiferromagnetic background
can be written as~\cite{Huang:1987,Hirsch:1987}:
\begin{equation}
\label{2.1}
\hat{H}_0  \; = \; - \frac{t^2}{U}  \; \sum_{\vec{r}}  \sum_{i,j}   \hat{\psi}^{\dagger}_{h}(\vec{r} + \vec{i} a_0 + \vec{j} a_0) \hat{\psi}_{h}(\vec{r})  \; ,
\end{equation}
where $ \hat{\psi}^{\dagger}_{h}(\vec{r})$, $ \hat{\psi}_{h}(\vec{r})$ are creation and annihilation operators for holes at the 
lattice site $\vec{r}$,  $a_0$ is the copper-oxide planar lattice constant,  and the sum over the direction vectors $ \vec{i}$ and $\vec{j}$ 
is restricted to next-nearest neighbor lattice sites. 
Note that in Eq.~(\ref{2.1})  the antiferromagnetic background forces the  holes to have antiparallel spins.  This ensures that
the motion of a hole does not disturb the  antiferromagnetic background.  With this antiferromagnetic background approximation, 
it turns out that there is an effective attractive two-body potential between nearest-neighbor holes~\cite{Huang:1987}. 
More precisely,  two holes with distance  $r$, $a_0 \ll r  \ll \xi_{AF}$, where $\xi_{AF}$ is the antiferromagnetic local order length scale,
 are subject to an effective attractive two-body potential.  In fact, our proposal is quite similar to the spin-bag theory~\cite{Schrieffer:1988} where 
 the pairing is due to a local reduction of the antiferromagnetic order (bag)  shared by two holes. 
This led us to  consider the following reduced interaction Hamiltonian:
\begin{eqnarray}
\label{2.2}
\hat{H}_{int} \;  =  \; \frac{1}{2}   \int d \vec{r}_1\,  d \vec{r}_2  \; 
\hat{\psi}^{\dagger}_{h,\uparrow}(\vec{r}_1)   \,   \hat{\psi}^{\dagger}_{h,\downarrow}(\vec{r}_2) \; \times
\\ \nonumber
 V(\vec{r}_1 - \vec{r}_2)  \, \hat{\psi}_{h,\downarrow}(\vec{r}_2)  \,  \hat{\psi}_{h,\uparrow}(\vec{r}_1)  \; \;  .
\end{eqnarray}
where  the two-body potential $V(\vec{r_1} - \vec{r_2})$ is given by:
\begin{equation}
\label{2.3} 
V(\vec{r_1} - \vec{r_2}) \;  = \; 
 \left \{ \begin{array}{ll}
 \; \; \infty \; \; \; \; &  \vec{r_1} \, = \, \vec{r_2}
  \\
  - \, V_0  \; \; \; \; &  |\vec{r_1} - \vec{r_2}| \, \leq \, r_0(\delta)
  \\
  \; \; \;  \;  0    \; \; \; \;  & otherwise
\end{array}
    \right.
\end{equation}
The range of the potential $ r_0(\delta)$  is expected to be of  the order of the  observed size of pairs which turns out to be rather small
 $\xi_0  \lesssim 6 a_0$:
\begin{equation}
\label{2.4} 
 r_0(\delta) \; = \;  6 \;  a_0 \; \left ( 1 \; - \; \frac{\delta}{\delta_c} \right )^{\frac{1}{2}} \;  .
 \end{equation}
The dependence of $ r_0(\delta)$  on  the doping fraction $\delta$  in Eq. (\ref{2.4})  takes care of the fact that
 the area of the antiferromagnetic islands decreases  with increasing $\delta$ since the injected holes tend to destruct 
 the local antiferromagnetic order, which  eventually vanishes for   $\delta  \gtrsim \delta_{c}$. Actually  the value
 $\delta_c  \simeq  0.35$ assures that   the superconducting instability disappears at the maximal doping $\delta_{max} \simeq 0.30$. 
Regarding the parameter $V_0$ in Eq.~(\ref{2.3}), in I we  fixed this parameter such that there is at least one real space 
d-wave bound state:
\begin{equation}
\label{2.5} 
 V_0 \; \simeq  \;  2 \;  J \; \simeq \; \frac{8 t^2}{U} \; .
 \end{equation}
Since we are interested in the limit of low-lying excitations,  we observe that, by writing
\begin{equation}
\label{2.6} 
 \hat{\psi}_{h}(\vec{k})  \; = \;  \frac{1}{\sqrt{\mathcal{N}}}  \; \sum_{\vec{r}}  \exp{(- i \vec{k} \cdot \vec{r})}    \; \hat{\psi}_{h}(\vec{r}) \; .
 \end{equation}
where $\mathcal{N}$ is the number of sites of the copper lattice,  we get:
\begin{equation}
\label{2.7} 
\hat{H}_0  \; = \;    \; \sum_{\vec{k}}  \varepsilon_{\vec{k}}   \;  \;  \hat{\psi}^{\dagger}_{h}(\vec{k})  \,   \hat{\psi}_{h}(\vec{k}) \; ,
 \end{equation}
where:
\begin{equation}
\label{2.8} 
\varepsilon_{\vec{k}}  =  -  \; \frac{2 t^2}{U}  \; \bigg  [ \; \cos{(2 k_x a_0})  +    \cos{(2 k_y a_0}) \;  \bigg ] 
\simeq    \frac{\hbar^2 \, \vec{k}^2}{2 \, m^*_h}   \;  ,
 \end{equation}
and
\begin{equation}
\label{2.9} 
m^*_h \; = \;  \frac{\hbar^2}{8  \, \frac{t^2}{U} a_0^2}  \; \; . 
\end{equation}
So that  we may write for the effective Hamiltonian:
\begin{equation}
\label{2.10} 
\hat{H}_0  \; = \;    \; \sum_{\sigma=\uparrow, \downarrow }    \int d \vec{r} \;   \hat{\psi}^{\dagger}_{h,\sigma}(\vec{r}) \, \left ( \frac{- \hbar^2 \nabla^2}{2 \, m^*_h}
\, \right )  \hat{\psi}_{h,\sigma}(\vec{r}) \; .
\end{equation}
Therefore, our effective Hamiltonian for low-lying excitations is given by:
\begin{equation}
\label{2.11} 
\hat{H}   \; = \;   \hat{H}_0 \; \; + \; \; \hat{H}_{int} \; ,
\end{equation}
where $\hat{H}_0$ and $\hat{H}_{int}$ are given by Eqs.~(\ref{2.10}) and (\ref{2.2}) respectively. 
We will consider a highly idealized crystal with unit cell $a \simeq b \simeq a_0$, $c \simeq c_0$, and
 use the following numerical values for the microscopic  parameters:
\begin{eqnarray}
\label{2.12} 
 a_0 \; \simeq \; 4.0 \, 10^{-8} \, cm \; \; ,  \; \;   c_0 \; \simeq \; 13.0 \, 10^{-8} \, cm \; \; , 
 \\ \nonumber  
  t \; \simeq \; 0.11 \; ev  \; \; , \; \;  \frac{U}{t} \; \simeq  \; 10  \hspace{2.7cm}
 \end{eqnarray}
that, indeed,  are appropriate for a typical cuprate~\cite{Jang:2015}. Note that, using these numerical values we get:
\begin{equation}
\label{2.13} 
m^*_h \; \simeq \;  5.41 \; m_e \; , 
\end{equation}
where $m_e$ is the electronic mass, in fair good agreements with several observations in hole doped cuprate superconductors.  \\
It is worthwhile to stress that our previous arguments  cannot be considered as a truly microscopic derivation of the effective Hamiltonian. 
 Our approximations were a drastic simplification of the complete many-body Hamiltonian, nevertheless we  used it to
understand the physics from the simplest point of view  without a large number of parameters. Indeed,
 in I  we showed that the effective Hamiltonian offered  a consistent picture of the complex phase diagram
 of high transition temperature cuprate superconductors. \\
 Due to the reduced dimensionality the two-body attractive potential admits real-space d-wave bound states. 
 The binding energy of these bound states $\Delta_2(\delta)$, which plays the role of the pseudogap, decreases with 
 increasing doping until it vanishes at a certain critical doping  $ \delta^*$. To a very good approximation, we have~\cite{Cea:2013}:
\begin{equation}
\label{2.14} 
 \Delta_{ 2}(\delta)  \; \simeq  \;   \Delta_{ 2}(0)  \; \left [ 1 \; - (\frac{\delta}{\delta^*})^{1.5} \right ]    \; \; , \; \; 
 \delta \; \lesssim \; \delta^* \; \simeq \; 0.207 \; . 
 \end{equation}
Using the values of the parameters in Eq.~(\ref{2.12}), we found $ \Delta_{ 2}(0)  \simeq 41.91 \;  mev$.
The pseudogap temperature is simply related to the binding energy: 
\begin{equation}
\label{2.15}  
k_B \, T^*(\delta) \; = \;  \frac{1}{2} \, \Delta_2(\delta) \;  . 
\end{equation}
In fact, in I we showed that  Eq.~(\ref{2.15}) is consistent with observations (see Fig.~8 in Ref.~\cite{Cea:2013}). 
We see, then, that the actual significance of the pseudogap is  the signature of preformed pairs. If the binding energy of each hole pair 
were so strong that the size of the pair were small compared with the inter-particle spacing, then the ground state would consist of paired 
holes  that,  behaving like  bosons,  would  condense into the same state. In this case  the ground state wavefunction  reduces to the ground state 
of a Bose-Einstein gas of paired holes~\cite{Schafroth:1955,Blatt:1955,Blatt:1964}.
However, although the formation of pairs is essential in forming the superconducting state, its remarkable properties require phase coherence among the pairs. 
In general,  the required phase coherence is established by the condensation of pairs. However, in two spatial dimensions it is well known that there is 
no truly Bose-Einstein condensation since the Mermin-Wagner-Hohenberg theorem~\cite{Mermin:1966,Hohenberg:1967} prevents a broken continuous symmetry 
at finite temperature.  Nevertheless, as  discussed in I, for  $ \delta \gtrsim \delta_{min} \simeq 0.05 $ the pairs
begin to overlap ensuring that the phases of the pairs are locked to a constant value.  The onset of phase coherence gives rise to
the superconductivity of the hole pair condensate~\footnote{Interestingly enough, recently the authors of  
Ref.~\cite{Yildirim:2015} suggested, in a different contest, that the superconductivity in underdoped cuprates can be understood
as a superfluid of real-space hole pairs.}.  
Numerous spectroscopic data support the scenario of preformed pairs gaining coherence at low 
temperatures~\cite{Renner:1998,Kanigel:2008,Yang:2008,Mishra:2014,Kondo:2015}.
However, as is well known~\cite{Berezinskii:1971,Kosterlitz:1973}, the phase coherence of the condensate survives 
for temperatures not exceeding the  Berezinskii-Kosterlitz-Thouless (B-K-T) critical temperature:
\begin{equation}
\label{2.16}  
 k_B  \; T_{B-K-T} \; \simeq \; \frac{\pi}{2} \;  K_s(T_{B-K-T})  \; ,
\end{equation}
where $K_s$ is the so-called phase stiffness:
\begin{equation}
\label{2.17}  
K_s \; \simeq \;  \frac{\hbar^2 }{2 m^*_h}  \; n_s \; ,
\end{equation}
and the superfluid density is given by:
\begin{equation}
\label{2.18}  
 n_s \; \simeq \frac{\delta}{2 a^2_0} \; \;  \; .
\end{equation}
For temperatures above $T_{B-K-T}$ the phase coherence and the superconductivity are lost due to the thermal activation of vortex excitations. 
If we neglect the temperature dependence of the phase stiffness, we obtain:
\begin{equation}
\label{2.19}  
 k_B  \; T_{B-K-T} \; \simeq \; \frac{\pi}{8} \;  \frac{\hbar^2 }{ m^*_h} \;  \frac{\delta}{ a^2_0} \;  \simeq \;  \pi \,   \frac{t^2 }{ U} \; \delta \; \;  \; .
\end{equation}
Using our numerical values for $t$ and $U$ in Eq.~(\ref{2.12}) we get:
\begin{equation}
\label{2.20}  
T_{B-K-T} \; \simeq \; 401 \; K \; \; \delta \; \;  \; .
\end{equation}
%
%
\begin{figure}
\vspace{0.8cm}
\hspace{-0.3cm}
\resizebox{0.5\textwidth}{!}{%
\includegraphics{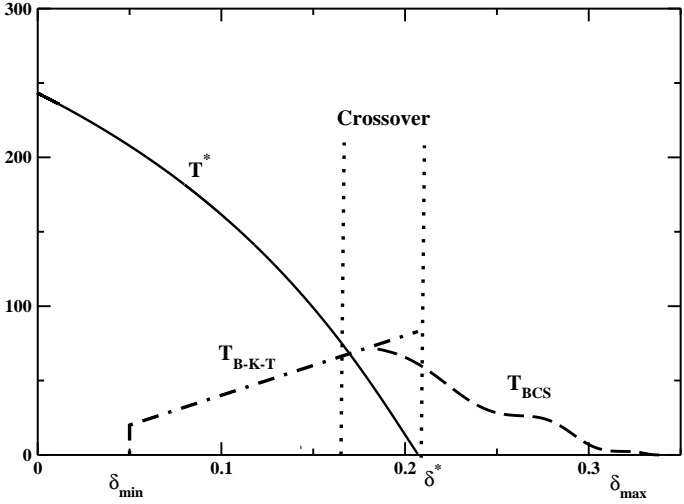}
}
\caption{\label{Fig2}  The pseudogap, Berezinskii-Kosterlitz-Thouless, and  BCS critical temperatures (in Kelvin) versus  the hole doping fraction 
 $\delta$.  The pseudogap temperature $T^*$ vanishes at $\delta = \delta^*$. The region enclosed by the vertical dotted lines  where the three temperatures 
 are comparable corresponds to the crossover region. Note that the crossover region coincides with the optimal doped region.}
\end{figure}
In Fig.~\ref{Fig2}, were  we summarize  the phase diagram of the hole doped cuprate superconductors according to our model, we
display both the pseudogap and  Berezinskii-Kosterlitz-Thouless critical temperatures. Within our approach  the pseudogap region
is the region where   $T_{B-K-T}$ lies below the pesudogap temperature $T^*$. In fact,  there is general agreement on 
the experimental evidence that  in the pseudogap region  the critical temperature is set by the onset of phase coherence
 at temperatures lower than the pseudogap
 temperature~\cite{Renner:1998,Kanigel:2008,Yang:2008,Mishra:2014,Kondo:2015,Emery:1995}.
Moreover,  there is convincing  experimental evidences~\cite{Hetel:2007} that the superconductorg transition in this region
is driven by the order-disorder Berezinskii-Kosterlitz-Thouless (B-K-T)  transition~\cite{Berezinskii:1971,Kosterlitz:1973}. \\
Within our approach  the overdoped region is realized for hole doping in excess of the critical doping $\delta^*$ where the pseudogap vanishes. In this region the conventional superconductivity theory of Bardeen, Cooper, and Schrieffer (BCS)~\cite{Cooper:1956,Bardeen:1957,Schrieffer:1999}
applies since the attractive two-body potential Eq.~(\ref{2.3}) is short-range and can be considered as a small perturbation.
In fact, the superconductive instability is driven by the short-range attractive interaction between the quasiparticles  and the pairing is
in momentum space, so that the relevant superconducting ground state is the BCS variational ground state. For reader convenience,
the relevant calculations are presented in Appendix~\ref{AppendixA}, and the d-wave BCS critical temperature is displayed in 
 Fig.~\ref{Fig2}. It turns out that in the overdoped region  $ \delta > \delta^* $  the d-wave BCS theory is able to
 account  many of the low-energy and low-temperature properties of the copper oxides, in agreement with several observations,
 as discussed in I, Sect.~5.
Moreover,  in this region  the normal state is described by  an ordinary Fermi liquid metals that is characterized by 
hole quasiparticles with effective mass given by Eq.~(\ref{2.13}) and density:
\begin{equation}
\label{2.21}  
n_h \; \simeq \;  \frac{1 + \delta}{a_0^2}  \; \;  .
\end{equation}
Looking at Fig.~\ref{Fig2} we see that in our approach the superconducting dome-shaped region of the hole doped oxide cuprates
is determined by the region enclosed by the critical temperatures  $T_{B-K-T}$ and $T_{BCS}$. Note that   $T_{B-K-T}$ 
increases with  doping, while the BCS critical temperature decreases with increasing hole doping becoming 
 comparable for $\delta \sim \delta_{opt} \simeq 0.17$.  We infer, thus, that   the superconducting critical temperature reaches its maximum in the optimal doped region where there is  a crossover  from a  Bose-Einstein condensate of tightly bound hole pairs  to Cooper pairing
 of weakly attracting  holes~\footnote{For a through discussion of the crossover from  Bose-Einstein condensation
   to  BCS pairing see Ref.~\cite{Leggett:2012}.}.
 Finally,  it is important to point out that in this region the competition between the pseudogap and 
 the d-wave BCS gap together  with  the enhanced role of the phase fluctuations makes  questionable the usually 
 adopted mean-field approximation. 
\section{Physics of the  Superconductive Condensate}
\label{s3}
To reach a reasonable description of the physics of superconductivity in hole doped copper oxide ceramics  it is
necessary, preliminarily, to construct at least a good approximation of the ground state wavefunction.
To do this, it is convenient to adopt the Schr\"odinger picture. In this representation our reduced Hamiltonian Eq.~(\ref{2.11})
 for $N$ holes in the $CuO_2$ plane becomes:
\begin{equation}
\label{3.1} 
\hat{H}   \; = \;   \hat{H}_0 \; \; + \; \; \hat{H}_{int} \; ,
\end{equation}
where 
\begin{equation}
\label{3.2} 
 \hat{H}_0  \; = \;      \sum_{i=1}^{N}   \;  - \, \frac{ \hbar^2}{2 m^*_h}    \nabla^2_{ \vec{r}_i}    \; ,
\end{equation}
and
\begin{equation}
\label{3.3} 
 \hat{H}_{int}  \; = \;   \frac{1}{2} \;   \sum_{i \neq j }^{N}     V(\vec{r}_i - \vec{r}_j )  \; .
\end{equation}
Note that in  Eq.~(\ref{3.3}) the potential   $ V(\vec{r}_i - \vec{r}_j )$, given by    Eq.~(\ref{2.3}), is different from zero only for holes
with antiparallel spins. The ground state wavefunction is the solution of the Schr\"odinger eigenvalue  equation:
\begin{equation}
\label{3.4} 
\hat{H} \;  \Omega(\vec{r}_1, ..., \vec{r}_N)   \; = \;  E_0 \;   \Omega(\vec{r}_1, ..., \vec{r}_N)   \; ,
\end{equation}
with the smallest eigenvalue $E_0$. In the pseudogap region we said the two isolated holes in interaction
with the attractive potential  Eq.~(\ref{2.3}) admit a bound state solution. In fact, let  consider the Schr\"odinger equation
for two holes (omitting the spin indices):
\begin{equation}
\label{3.5} 
 \left [  - \, \frac{ \hbar^2}{2 m^*_h}  \left (   \nabla^2_{ \vec{r}} \; + \; \nabla^2_{\vec{r}'}   \right ) + V(\vec{r} - \vec{r}\, ') \right ] \; 
 \Phi( \vec{r}, \vec{r}\, ') = \; \varepsilon \; 
 \Phi (\vec{r}, \vec{r}\, ')  \;  .  
\end{equation}
Writing:
\begin{equation}
\label{3.6} 
\vec{\rho} \; = \; \vec{r}  - \vec{r}\,'  \; \; \; \; , \; \; \; \; \vec{R} \; = \frac{\vec{r}  + \vec{r}\,'}{2} \; ,
\end{equation}
we obtain:
\begin{equation}
\label{3.7} 
 \left [  - \, \frac{ \hbar^2}{2 m^*_h}  \left (  2 \, \nabla^2_{ \vec{\rho}} \; + \; \frac{1}{2} \, \nabla^2_{\vec{R}}   \right ) + V(\rho) \right ] \; \Phi( \vec{R} ,    \vec{\rho} ) = \; \varepsilon \; 
 \Phi (\vec{R} ,  \vec{\rho} )    \; ,
\end{equation}
This allows to write:
\begin{equation}
\label{3.8} 
 \Phi (\vec{R} ,  \vec{\rho} )   \; =  \;    \exp{ (i  \vec{K}  \cdot  \vec{R})} \; \varphi(  \vec{\rho} )    \; 
\end{equation}
and
\begin{equation}
\label{3.9} 
\varepsilon  \; =  \;   \frac{\hbar^2 \, \vec{K}^2}{4 \, m^*_h}    \; - \; \Delta       \; .
\end{equation}
It is, now, easy to find:
\begin{equation}
\label{3.10} 
 \left [  - \, \frac{ \hbar^2}{ m^*_h}    \nabla^2_{ \vec{\rho}} \;  + \; V(\rho) \right ] \; \varphi( \vec{\rho}  ) = - \; \Delta  \;   \varphi( \vec{\rho}  )    \; .
\end{equation}
As discussed in I, Sect.~(3.1), Eq.~(\ref{3.10}) admits a d-wave bound state solution with $  \Delta =  \Delta_2(\delta) > 0$.  
The hole ground-state wavefunction can be, now, obtained as in the BCS approximation.  Group the N conduction holes
 into $M = N/2$ pairs and let each pair be described by a bound-state wave function $\Phi(\vec{r}, \vec{r}^{'})$. 
Then consider the $N$-hole wavefunction that is just the antisymmetrized product of $M$ identical such two-hole wavefunctions:
\begin{eqnarray}
\label{3.11} 
\Omega(\vec{r}_1, ..., \vec{r}_N)  \; = \; \frac{1}{\sqrt{N!}}   \sum_P \; (-1)^P \;
\\ \nonumber
 P \bigg \{ \Phi(\vec{r}_1-\vec{r}_2)  \cdot \, \cdot \, \cdot   \Phi(\vec{r}_{N-1}-\vec{r}_N) \bigg \} \; ,
\end{eqnarray}
in which the sum is over all permutations $P$ of the N holes. Since we are interested in the ground-state wavefunction in the
underdoped region, we may simplify considerably the BCS-like wavefunction    Eq.~(\ref{3.11}). In fact,  we already known~\cite{Cea:2013}
 that  the spatial range  of the pair wave function is  $\xi_0 \simeq  2 a_0$ (see I, Fig.~4).  So that, in the underdoped region  $\xi_0$ is smaller
 than the average distance between pairs:
\begin{equation}
\label{3.12} 
d_0  \;  \simeq  \; \frac{1}{\sqrt{n_s}}  \; \simeq \; \frac{\sqrt{2} a_0}{\sqrt{\delta}}  \; .
\end{equation}
Therefore, in this case wavefunctions in the sum in   Eq.~(\ref{3.11}) differing by the exchange of single members of two or more pairs
do not overlap very much. This means that we may keep in   Eq.~(\ref{3.11}) only terms where there are exchanges of two or more
pairs. Thus, to a good approximation we can rewrite the ground state wavefunction as:
\begin{eqnarray}
\label{3.13} 
\Omega(\vec{R}_1, ..., \vec{R}_M, \vec{\rho}_1, ..., \vec{\rho}_M)  \; \simeq \; \frac{1}{\sqrt{M!}}  \;  \sum_P \; 
\\ \nonumber
P \bigg \{ e^{i \left [ \theta( \vec{R}_1) +  ... +  \theta( \vec{R}_M) \right ]} \; 
\varphi(\vec{\rho}_1)  \cdot \, \cdot \, \cdot  
 \varphi(\vec{\rho}_{M}) \bigg \} \; ,
\end{eqnarray}
where $ \theta( \vec{R}_i) =  \vec{K}  \cdot  \vec{R}_i$. With these approximations, it is quite easy to check that the wavefunction
 Eq.~(\ref{3.13}) satisfy the time-independent Schr\"odinger equation    Eq.~(\ref{3.4}) with eigenvalue:
\begin{equation}
\label{3.14} 
E  \;  \simeq  \;  M \; \left (  \frac{\hbar^2 \, \vec{K}^2}{4 \, m^*_h}    \; - \; \Delta_2(\delta) \right )    \; .
\end{equation}
Of course, the ground state wavefunction is obtained when the center of mass of the pairs is at rest, $\vec{P}= \hbar \, \vec{K} = 0$:
\begin{equation}
\label{3.15} 
E_0 \; = \; E_c  \; \; \; , \; \;  E_c \; \simeq  \;  - \; M \;  \Delta_2(\delta)    \; ,
\end{equation}
where $E_c$ is the condensation energy.  \\
For later convenience, let us rewrite   Eq.~(\ref{3.13}) in the following form:
\begin{eqnarray}
\label{3.16} 
\Omega(\vec{R}_1, ..., \vec{R}_M, \vec{\rho}_1, ..., \vec{\rho}_M)  \; \simeq \; \frac{1}{\sqrt{M!}}  \;  \sum_P \; 
\\ \nonumber
P \bigg \{  \Theta( \vec{R}_1)   \cdot \, \cdot \, \cdot    \Theta( \vec{R}_M)  \; 
\varphi(\vec{\rho}_1)  \cdot \, \cdot \, \cdot  
 \varphi(\vec{\rho}_{M}) \bigg \} \; .
\end{eqnarray}
If we normalize the wavefunctions as:
\begin{equation}
\label{3.17} 
\int \; d \vec{R}_i \;   |\Theta( \vec{R}_i)|^2  \;  =  \; 1   \; ,
\end{equation}
\begin{equation}
\label{3.18} 
\int \; d \vec{\rho}_i \;   |\varphi( \vec{\rho}_i)|^2  \;  =  \; 1   \; ,
\end{equation}
then, it easy to find:
\begin{equation}
\label{3.19} 
\int \; \prod_{i} \; d \vec{R}_i \; d \vec{\rho}_i \;  \left | \Omega(\vec{R}_1, ..., \vec{R}_M, \vec{\rho}_1, ..., \vec{\rho}_M) \right |^2  \;  =  \; 1   \; .
\end{equation}
It is worthwhile to note that:
\begin{equation}
\label{3.20} 
\Theta( \vec{R})  \;  =  \;  \frac{1}{\sqrt{V}} \;     \exp{ (i  \vec{K}  \cdot  \vec{R})}   \; ,  
\end{equation}
so that for the ground state:
\begin{equation}
\label{3.21} 
\Theta( \vec{R})  \;  =  \; \Theta_0( \vec{R})  \;  =  \; \frac{1}{\sqrt{V}}  \; .     
\end{equation}
Following Ref.~\cite{Leggett:2012b} we introduce the reduced  wavefunction  $\Psi(\vec{R}_1, ..., \vec{R}_M)$ defined by:
\begin{eqnarray}
\label{3.22} 
\Psi^*(\vec{R }_1\, ', ..., \vec{R}_M \, ') \;  \Psi(\vec{R}_1, ..., \vec{R}_M) \; = \; 
  \int \; \prod_{i} \; d \vec{\rho}_i \hspace{1.0cm} 
\\ \nonumber  
   \Omega^*(\vec{R}_1 \, ', ..., \vec{R}_M \, ', \vec{\rho}_1, ..., \vec{\rho}_M) 
  \Omega(\vec{R}_1, ..., \vec{R}_M, \vec{\rho}_1, ..., \vec{\rho}_M)   \; .
\end{eqnarray}
Using   Eq.~(\ref{3.16}), a straightforward calculation gives: 
\begin{equation}
\label{3.23} 
\Psi(\vec{R}_1, ..., \vec{R}_M)  \; \simeq \; \frac{1}{\sqrt{M!}}  \;  \sum_P \; 
\; P \left \{  \Theta( \vec{R}_1)   \cdot \, \cdot \, \cdot    \Theta( \vec{R}_M)  \;  \right \} \; .
\end{equation}
We introduce, now, the single-particle density matrix~\cite{Leggett:2012b}:
\begin{eqnarray}
\label{3.24} 
  \rho(\vec{R },  \vec{R} \, ')  \; = \; M \;  
  \int \; \prod_{i=2}^{M} \; d \vec{R}_i \;  
\\ \nonumber  
   \Psi^*(\vec{R } ,\vec{R }_2 \,..., \vec{R}_M ) \;  \Psi(\vec{R} \, ' ,\vec{R }_2 \,..., \vec{R}_M )    \; .
\end{eqnarray}
The physical meaning of    $\rho(\vec{R },  \vec{R} \, ')$ is that the density matrix gives the probability amplitude to find a particular
hole pair at $ \vec{R}$ times the amplitude to find it at  $\vec{R} \, '$ taking into account the correlations due to the remaining
$M-1$ hole pairs. The factor $M$ in the definition of the density matrix is due to the fact that we have $M$ identical pairs.
Equivalently, in terms of  the second quantization formalism one may introduce the creation operator of a pair with center of mass coordinate 
$ \vec{R}$:  
\begin{equation}
\label{3.25} 
{ \hat{\it{b}}^{\dagger}(\vec{R})}  =  \int d \vec{\rho}   \;  \Phi^*(\vec{R}, \vec{\rho} ) \;   
 \hat{\psi}^{\dagger}_{h,\uparrow}(\vec{R} - \frac{\vec{\rho}}{2})   \; 
 \hat{\psi}^{\dagger}_{h,\downarrow}(\vec{R} + \frac{\vec{\rho}}{2})   \; .
\end{equation}
Then, we have:
\begin{equation}
\label{3.26} 
  \rho(\vec{R },  \vec{R} \, ')  \; = \    < \Omega | \;  { \hat{\it{b}}^{\dagger}(\vec{R})}  \, { \hat{\it{b}}(\vec{R} \, ')} \;  |\Omega >  \; .
\end{equation}
From   Eq.~(\ref{3.24}), after taking into account   Eq.~(\ref{3.23}), we get easily: 
\begin{equation}
\label{3.27} 
  \rho(\vec{R },  \vec{R} \, ')  \; \simeq \; M \;   \Theta^*(\vec{R})  \;   \Theta(\vec{R} \, ')    \; .
\end{equation}
We may, now, introduce the condensate wavefunction  $\chi(\vec{R})$ which plays the role of the order parameter:
\begin{equation}
\label{3.28} 
  \rho(\vec{R },  \vec{R} \, ')  \; \equiv \;  \chi^*(\vec{R})  \;   \chi(\vec{R} \, ')    \; .
\end{equation}
This last equation, combined with previous equations, leads to:
\begin{equation}
\label{3.29} 
  \chi(\vec{R} )    \;  \simeq \;    \sqrt{M} \;  \Theta(\vec{R} ) \; = \; \sqrt{\frac{M}{V}} \;  e^{i  \theta(\vec{R} ) } \; = \; 
   \sqrt{n_s}   \;  e^{i  \theta(\vec{R}) } \; ,
 \end{equation}
 where $n_s$ is given by  Eq.~(\ref{2.18}). In particular, the ground-state condensate wavefunction is simply:
\begin{equation}
\label{3.30} 
 \chi_0(\vec{R} )    \;  \simeq \;    \sqrt{n_s}   \; .
 \end{equation}
From our discussion it follows that the low-lying condensate excitations can be obtained by solving the
Schr\"odinger-like eigenvalue equation: 
\begin{equation}
\label{3.31} 
  - \, \frac{ \hbar^2}{4 m^*_h}  \;   \nabla^2_{ \vec{R}} \;    \chi(\vec{R} )    \;  = \; \varepsilon \;   \chi(\vec{R} )  \;  , 
\end{equation}
where $\varepsilon$ is the energy per particle in excess to the condensation energy.  Eq.~(\ref{3.31}) shows that the condensate wavefunction
satisfies the Schr\"odinger equation of a free particle with mass $2 m^*_h$.  In fact, this is due to our approximations which neglect
the Coulomb interactions, always present in metal, between hole pairs. Since the hole pairs have charge $+2 e$, then the pairs
must avoid to be too close due to the Coulomb repulsion. As a consequence, in the ground state the pairs fill the system with
a density almost uniform over distance $R \gg d_0$.  If one takes care of the Coulomb interactions, then, in general, one finds that
the condensate wavefunction satisfies a non-linear Schr\"odiger equation, the time-independent Gross-Pitaevskii equation.
In that case the energy per particle must be replaced by the chemical potential~\footnote{A good account can
be found, for example, in Ref.~\cite{Leggett:2012b}.}. \\
In general, the condensate excitations are characterized by a non-zero planar current density:
\begin{equation}
\label{3.32} 
\vec{j}(\vec{R}) \; =   \; \frac{ - i \hbar}{4 m^*_h}  \;  \left [ \chi^*(\vec{R} ) \,  \nabla  \,   \chi(\vec{R})    \;  - \; 
  \nabla  \,  \chi^*(\vec{R} ) \,   \chi(\vec{R})    \right ] \; .
\end{equation}
Using  Eq.~(\ref{3.29}) we obtain:
\begin{eqnarray}
\label{3.33} 
\vec{j}(\vec{R}) \; =   \; n_s \; \frac{  \hbar}{ 2 m^*_h}  \;    \nabla  \,   \theta(\vec{R})    \; =  \; 
 n_s \;  \vec{v}_s(\vec{R})    \;  \;  , 
\\ \nonumber 
  \vec{v}_s(\vec{R})  \; = \;   \frac{  \hbar}{ 2 m^*_h}  \;    \nabla  \,   \theta(\vec{R})  \; .
\end{eqnarray}
For slow-varying condensate phase  $\theta(\vec{R})$, i.e. $| \nabla^2 \theta(\vec{R})| \ll  | \nabla \theta(\vec{R})|^2$, the energy
of the condensate excitations can be written as:
\begin{equation}
\label{3.34} 
\varepsilon  \; \simeq  \;  \frac{1}{2} \;   2 m^*_h  \;  \vec{v}_s^{\,2}(\vec{R})  \; = \;  m^*_h  \;  \vec{v}_s^{\, 2}(\vec{R}) \; .
\end{equation}
Finally, it is worthwhile to remark once more that the superfluidity of the condensate is assured by phase coherence. We have already 
remarked that for $\delta > \delta_{min}$ the exponentially small overlap of the hole pair wavefunctions constrains the
pair wavefunctions to have the same phase. Therefore, if a current is established in the condensate, all the hole pairs must move together.
Let  $\vec{P}= \hbar \, \vec{K}$ be the momentum of the center of mass of pairs. It is easy to see that the condensate wavefunction is 
given by   Eq.~(\ref{3.29}) with  $ \theta(\vec{R})  =    \vec{K}  \cdot \vec{R}$. This corresponds to a condensate with velocity
$ \vec{v}_s =  \frac{  \hbar   \vec{K} }{ 2 m^*_h}$ and current  density $\vec{j} =   n_s  \vec{v}_s$.  One might expect that such a current could be
degraded by a single-pair collision in which the center of mass momentum is reduced back to zero. However, all the
other pairs are described by identical pair wavefunctions. Thus one cannot change the pair wavefunction individually without 
destroying the whole condensate, and this cost an enormous amount of energy. We see, therefore, that it is the phase coherence
which assures rigidity to the  wavefunction implying condensate superfluidity.
\subsection{Low-lying excitations of the condensate}
\label{s3.1}
Our previous discussion  summarizes the essential features of the ground state at zero temperature. To describe
the thermal or transport properties of the hole pair condensate we need to determine the low-lying excited states.
The most obvious  possibility is to excite the system by breaking a single pair. This requires an energy $\varepsilon \simeq \Delta_2$.
This kind of excitations, however, are relevant for temperatures exceeding the pseudogap temperature $T^*$. Therefore, in the 
pseudogap region, where  $T^*$ is greater than the superconductive critical temperature, these excitations cannot contribute
to  the dynamics of the superfluid condensate.  We are led, thus, to inquire if there are excitations with energies 
$\varepsilon \lesssim \Delta_2$. According to our previous discussion, the reduced wavefunctions of these excitations
can be written as:
\begin{equation}
\label{3.35} 
\Psi_1(\vec{R}_1, ..., \vec{R}_M)  \; =  \;  e^{ i \;   \Theta( \vec{R}_1 , ... , \vec{R}_M) }  \; 
\Psi_0(\vec{R}_1, ..., \vec{R}_M)  \; , \hspace{0.5cm}
\end{equation}
where $ \Psi_0(\vec{R}_1, ..., \vec{R}_M)$ is the ground-state wavefunction, and   $\Theta( \vec{R}_1 , ... , \vec{R}_M)$ is 
a totally symmetric function of  $\vec{R}_1$, ... , $\vec{R}_M$. To determine the wavefunction  $ \Psi_1$ we shall follow
quite closely the Feynman's superb discussion on the excited states in liquid $^4He$~\cite{Feynman:1955}.
The simplest choice for the excited state wavefunction would be:
\begin{equation}
\label{3.36} 
\Psi_1(\vec{R}_1, ..., \vec{R}_M)  \; =  \;  e^{ i \;  \vec{K} \cdot \vec{R}_i  }  \; 
\Psi_0(\vec{R}_1, ..., \vec{R}_M)  
\end{equation}
for some fixed  $\vec{R}_i$. However, since the wavefunction must be symmetric we must write:
\begin{equation}
\label{3.37} 
\Psi_1(\vec{R}_1, ..., \vec{R}_M)  \; =  \; \sum_{i=1}^{M} \;  e^{ i \;  \vec{K} \cdot \vec{R}_i  }  \; 
\Psi_0(\vec{R}_1, ..., \vec{R}_M)  
\end{equation}
According to the results of the previous Section, the corresponding excitation energy is $\varepsilon =  \frac{ \hbar^2}{4 m^*_h}  \vec{K}^2$.
Then, to have low-energy excitations we are forced to consider very small wavenumber  $|\vec{K}|$.
However, since the wavefunction  $\Psi_1$ must be orthogonal to $ \Psi_0$, i.e.
\begin{equation}
\label{3.38} 
\int   d\vec{R}_1 ....  d\vec{R}_M  \;   \Psi_0^*(\vec{R}_1, ..., \vec{R}_M)  \Psi_1(\vec{R}_1, ..., \vec{R}_M)  \; =  \;  0  \;, 
\end{equation}
we need  configurations where  $\vec{K} \cdot \vec{R}_i$ is appreciably different from zero. Due to the
symmetry for exchanges of pairs, these configurations can be realized only by changing the density of the condensate. In fact, Feynman
argued that  in liquid $^4He$ these configurations  are the only available low-energy excitations and they  give rise to the phonon spectrum. 
In fact, these sound-wave modes are present also in traditional BCS superconductors~\cite{Bogoliubov:1958,Bogoliubov:1959}.
However, due to the Coulomb interactions the sound-wave mode is pushed up to high energy and  becomes the plasma mode.
As a result, the putative low-energy excitations are  realized by density distributions that oscillate at the plasma frequency $\omega_{pl}$.
We may estimate the plasma frequency by the well-known expression:
\begin{equation}
\label{3.39} 
\omega_{pl} \; =  \;  \sqrt{\frac{4 \pi \rho q^2}{m}}   \;, 
\end{equation}
where $\rho$ is the volumetric density of particles with charge $q$ and mass $m$. Accordingly,  taking into account that:
\begin{equation}
\label{3.40} 
  m \; = \; 2 \, m^*_h \; \; , \; \; q \; = \; 2 \, e \;\; , \; \; \rho \; \simeq \; \frac{n_s}{c_0} \; \simeq \; \frac{\delta}{2 \,  a_0^2 \, c_0}    \;, 
\end{equation}
we obtain:
\begin{equation}
\label{3.41} 
\omega_{pl} \; \simeq  \;  \sqrt{\frac{4 \pi e^2 \delta}{ m^*_h a_0^2 \, c_0}}   \;.
\end{equation}
After taking into account  Eq.~(\ref{2.9}) we get:
\begin{equation}
\label{3.42} 
\varepsilon_{pl} \; \simeq \;  \hbar \; \omega_{pl} \; \simeq  \;  \sqrt{32 \,  \pi \, \frac{e^2 }{c_0} \frac{t^2}{U} \delta}   \; .
\end{equation}
Using the numerical values of the parameters,  Eq.~(\ref{2.12}), we find that in the region of interest
 $\varepsilon_{pl} \, \simeq  \, 1.11 \, ev \, \sqrt{\delta} \; \gg \Delta_2$. Therefore, we are led to conclude that the only
 low-lying excitations of the pair condensate are analogous the the rotons in $^4He$. These kind of excitations are
 described by the condensate wavefunction $\chi(\vec{R})$,  Eq.~(\ref{3.29}), where $\theta(\vec{R})$ is a rapidly varying
 function over a distance of order $d_0$. We may estimate the roton energy by assuming that the phase   $\theta(\vec{R})$ 
 is subject to a variation of $ 2 \pi$ over  distance of $d_0$. In fact this allows to localize the condensate disturbance in a region
 of linear size $d_0$ around a given $\vec{R}$. To see this, we note that  $\theta $  equals  $0$   or  $2 \pi$ for $\vec{R}^{ \, '}$
 such that    $ |\vec{R}^{\, '}  -    \vec{R}|  > d_0$. So that, since $e^{ i \theta}=1$, we have    $\chi(\vec{R}) \simeq  \chi_0$
 outside the disturbance region. Now, taking into account that:
\begin{equation}
\label{3.43} 
| \nabla \, \theta_{rot}(\vec{R}) | \; \simeq \;   \frac{2 \, \pi }{d_0 }    \; ,
\end{equation}
we readily obtain:
\begin{equation}
\label{3.44} 
\varepsilon_{rot} \; \simeq \;    m^*_h  \;  \vec{v}_s^{\, 2}   \;  \simeq  \; \frac{ \hbar^2 \pi^2}{m^*_h}  \;  n_s  \; .
\end{equation}
Actually, one could consider rotons corresponding to a phase jump of $ 2 \pi n$, $n$ integer. However,  Eq.~(\ref{3.44})
shows that the energy of these rotons is higher by a factor $n^2$. So that, only rotons with $n=1$ are of interest.  
Using the numerical values of the model parameters, we find:
\begin{equation}
\label{3.45} 
\varepsilon_{rot} \; \simeq \;   434 \;  mev \; \delta \;  .
\end{equation}
For $\delta \simeq 0.1$ we have  $\varepsilon_{rot} \; \simeq \;   43 \;  mev \; \simeq \Delta_2(0)$.  Therefore, we may conclude
that below the critical superconductive temperature the roton-like condensate excitations could play a role in the dynamics
of the system. \\
We said that the low-lying quasi-particle excitations, described by the condensate wavefunction:
\begin{equation}
\label{3.46} 
  \chi_{rot}(\vec{R} )    \;  \simeq \;    \sqrt{n_s}   \;  e^{i  \theta_{rot}(\vec{R}) } \; ,
\end{equation}
are analogous to the rotons in liquid  $^4He$. However, in liquid helium Feynman and Cohen~\cite{Feynman:1956,Cohen:1957}
argued that the roton wavefunction could not provide a correct description because it does not offer a proper account of 
the motion of an excitation through the condensate. In fact the quasiparticle excitations are characterized by a non-zero current
given by  Eq.~(\ref{3.33}). Such a description cannot be appropriate for stationary excitations since the corresponding
current must necessarily vanish. In fact, Feynman and Cohen pointed out that one must take care of the backflow of the condensate
as the excitation moves through it. It turns out that the backflow corresponds to a slow drift of the condensate outside the region
of the excitation which can be described by a vortex-antivortex distribution of the condensate phase. As a result, the backflow acts
to cancel the excitation current and to increase the effective mass of the excitation~\footnote{For a very clear discussion,
see Ref.~\cite{Pines:1994}.}.
Therefore, the roton excitation energy can be written as:
\begin{equation}
\label{3.47} 
\varepsilon_{rot} \; \simeq \;    \alpha \; m^*_h  \;  \vec{v}_s^{\, 2}   \;   \; ,
\end{equation}
where $\alpha$ is some constant expected to be greater  than $1$, $\alpha > 1$. Actually, the precise numerical value of
this constant is not important for our purposes. Microscopic calculations in liquid $^4He$ indicated that 
$\alpha \sim 1.5$~\cite{Pines:1994}, nevertheless, to be conservative, in the following we shall assume  $\alpha \simeq 1.0$. \\
It is worthwhile to stress that these roton-like elementary excitations rely on the phase coherence of the hole pair condensate.
According to our model, in the pseudogap region the superconducting transition is described by the 
Berezinskii-Kosterlitz-Thouless  order-disorder transition. It is well known that the  Berezinskii-Kosterlitz-Thouless  transition
is driven by  vortex-antivortex unbinding which destroys the phase coherence above the critical temperature, $ T > T_c$.
In other words,  for  $ T > T_c$ configurations corresponding to an uniform condensate velocity $v_s$ are unstable to the
decay into vortex excitations. Kosterlitz~\cite{Kosterlitz:1974} introduced  the correlation or screening length:
\begin{equation}
\label{3.48} 
\xi_{+}(T) \; = \;  \xi_{V} \;  e^{ \frac{ b \; \pi}{\sqrt{T/T_c - 1}}}  \; \; , \; \;  T    \;  \rightarrow  \; T_c^+ \;  ,
\end{equation}
where $\xi_V$ is vortex core linear size and $b$ is a non-universal constant, such that the number density of free
vortices is proportional to $\left[ \xi_{+}(T) \right ]^{-2}$~\cite{Ambegaokar:1978,Ambegaokar:1980}. For temperatures near 
the critical temperature $T_c$, configurations leading to a condensate velocity $v_s$ are expected to be be screened
by free vortices. Moreover the screening should depend on the free vortex density. Therefore, we expect that for  
 $T     \rightarrow   T_c^+$:
\begin{equation}
\label{3.49} 
v_s(T) \; \sim \; \left [ \xi_{+}(T) \right ]^{-2}  \; \sim \;   e^{ \frac{2 \, \pi \,  b}{\sqrt{T/T_c - 1}}}  \; \; .
\end{equation}
In fact, similar arguments has been adopted to determine the temperature dependence of the conductivity in the resistive transition of
superconducting films~\cite{Halperin:1979}. On the other hand, we are mainly interested in the temperature dependence of the condensate
velocity below the critical temperature,   $T     \rightarrow   T_c^-$. To this end,  we may employ the Kosterlitz's recursion
relation~\cite{Kosterlitz:1974}:
\begin{equation}
\label{3.50} 
\xi_{+}(T) \; \sim \;  \xi_{V} \;  \left [ \frac{ \xi_{-}(T)}{\xi_{V}} \right ]^{2 \, \pi}   \; \; ,
\end{equation}
where $ \xi_{-}(T)$ is the screening length for   $T     \rightarrow   T_c^-$. Combining Eqs.~(\ref{3.48})  and (\ref{3.50}) we get:
\begin{equation}
\label{3.51} 
\xi_{-}(T) \; = \;  \xi_{V} \;  e^{ \frac{ b}{2 \, \sqrt{1 - T/T_c}}}  \; \; , \; \;  T    \;  \rightarrow  \; T_c^- \;  .
\end{equation}
Therefore, we obtain the remarkable result that superfluid velocities are screened to zero for     $T     \rightarrow   T_c^-$ according
to:
\begin{equation}
\label{3.52} 
v_s(T) \; \sim \;  \xi_{V}^{-2} \;   e^{  - \; \frac{b}{\sqrt{1 \, - \, T/T_c}}}  \; \; , \; \;  T    \;  \rightarrow  \; T_c^- \;  .
\end{equation}
 Eq.~(\ref{3.52}) is valid for temperatures quite close to $T_c$. However, due to the exponential dependence on the temperature,
 we may extrapolate   Eq.~(\ref{3.52}) down to very low temperatures according to:
\begin{equation}
\label{3.53} 
v_s(T) \;  \simeq \;  v_s(T=0)  \;   e^{  - \, b \; \left [ \frac{1}{\sqrt{1 \, - \, T/T_c}} \; - \; 1 \right ] } \; \; , \; \;  T    \; \le  \; T_c \;  .
\end{equation}
The previous arguments are valid also for the hole pair condensate fraction $n_s(T)$ which shares phase coherence. In fact, 
as discussed in Sect.~\ref{s3}, at $T=0$ all the hole pairs are condensed so that  $n_s(T=0) = \frac{\delta}{2 a_0^2}$ .
However, at finite temperatures the thermal activation of vortex-antivortex pairs tend to disorder the system such that
 $n_s(T) <  n_s(0)$. Now, as for $v_s(T)$  the depletion of the phase-coherent condensate is proportional to the density of
 vortices. Proceeding as before, we obtain the analogous of  Eq.~(\ref{3.53}), namely:
\begin{eqnarray}
\label{3.54} 
n_s(T)  \simeq  n_s(T=0)  \;   e^{  - \, b' \; \left [ \frac{1}{\sqrt{1 \, - \, T/T_c}} \; - \; 1 \right ] } \;  ,  \;   T    \le  T_c  \hspace{0.5cm}
\\ \nonumber 
n_s(T=0) \; \simeq \;   \frac{\delta}{2 a_0^2} \;  \hspace{1.0cm}
\end{eqnarray}
 where $b'$ is a non-universal constant, in principle different from $b$.  As we will discuss later,  Eqs.~(\ref{3.53}) and 
Eq.~(\ref{3.54})  allow to track the temperature dependence of various physical quantities. In fact, we will fix the
values of the non-universal constants $b$ and $b'$ by fitting to the experimental data. Indeed, we anticipate that these constants
are seen to assume quite  different values, $b \sim  \mathcal{O}(10^{-1})$ and $b' \sim \mathcal{O}(10^{0})$.
\subsection{Condensate in external magnetic fields}
\label{s3.2}
In this Section we are interested in the dynamics of the  hole pair condensate  in presence of an external magnetic field perpendicular 
to the copper-oxide plane:
\begin{equation}
\label{3.55}  
\vec{h}(\vec{r}) \; = \; \; \nabla \times \vec{A}(\vec{r}) \; \; \;  , \;  \; \; \vec{A}(\vec{r})  \; = \; \left ( A_1(\vec{r}),  A_2(\vec{r}), 0 \right ) \; ,
\end{equation}
where $\vec{h}(\vec{r})$ is the microscopic magnetic field, and we adopt the London gauge  $ \nabla  \cdot  \vec{A} = 0 $.
The reduced Hamiltonian in the Schr\"odinger representation is still given by  Eq.~(\ref{3.1}), but now: 
\begin{equation}
\label{3.56} 
 \hat{H}_0  \; = \;      \sum_{i=1}^{N}   \;  \frac{ 1}{2 m^*_h} \left [  - \, i \,  \hbar  \nabla_{ \vec{r}_i} \; - \; \frac{e}{c} \,   \vec{A}(\vec{r}_i) 
 \right ]^2  \; .
\end{equation}
In this case the Schr\"odinger equation for two holes, Eq.~(\ref{3.5}), becomes:
\begin{eqnarray}
\label{3.57} 
  \frac{ 1}{2 m^*_h} \left (  - i \hbar \nabla_{ \vec{r}}  -  \frac{e}{c} \vec{A}( \vec{r})  \right )^2    \Phi( \vec{r} , \vec{r} \,') 
 \; + 
  \\ \nonumber 
  \frac{ 1}{2 m^*_h}  \left ( - i \hbar \nabla_{ \vec{r} \, '}  -  \frac{e}{c} \,\vec{A}( \vec{r} \, ')  \right )^2    
  \Phi( \vec{r} , \vec{r} \,')  \; +
\\ \nonumber  
   + \; V(\vec{r} - \vec{r} \, ' )  \;   \Phi( \vec{r} , \vec{r} \, ' ) \;  =   \;  \varepsilon \;   \Phi (\vec{r} , \vec{r} \, ' )  \;   \; .  
\end{eqnarray}
Changing variables as in Eq.~(\ref{3.6}), we recast  Eq.~(\ref{3.57}) into:
\begin{eqnarray}
\label{3.58} 
  \frac{ 1}{4 m^*_h} \bigg (  - i \hbar \nabla_{ \vec{R}}  -  \frac{2e}{c} \vec{A}( \vec{R})  \bigg )^2    \Phi( \vec{R} , \vec{\rho})  + 
\\ \nonumber
  \frac{ 1}{ m^*_h}  \bigg (  - i \hbar \nabla_{ \vec{\rho}}  -  \frac{e}{2c} \,\vec{A}( \vec{\rho})  \bigg )^2    \Phi( \vec{R} , \vec{\rho}) 
\\ \nonumber
   + \; V(\vec{\rho} )  \;  \Phi( \vec{R} , \vec{\rho} ) \;  =   \;  \varepsilon  \; \Phi (\vec{R} , \vec{\rho})  \;   \; .  
\end{eqnarray}
Again, this allows to  write:
\begin{equation}
\label{3.59}  
\Phi (\vec{R} , \vec{\rho})  \; = \;  \Psi (\vec{R} ) \; \varphi( \vec{\rho})  \; ,  
\end{equation}
and 
\begin{equation}
\label{3.60}  
\varepsilon \; = \;  \varepsilon_{cm} \; - \; \Delta  \; .
\end{equation}
So that: 
\begin{equation}
\label{3.61} 
 \frac{ 1}{4 m^*_h} \left [  - i \hbar \nabla_{ \vec{R}}  -  \frac{2e}{c} \vec{A}( \vec{R})  \right ]^2   \Psi (\vec{R} )  
\;  =   \;  \varepsilon_{cm} \;  \Psi (\vec{R})  \;  , 
\end{equation}
\begin{equation}
\label{3.62} 
\left [  \frac{ 1}{ m^*_h}  \left ( - i \hbar \nabla_{ \vec{\rho}}  -  \frac{e}{2c} \,\vec{A}( \vec{\rho})  \right )^2   + \; V(\vec{\rho} ) \right ]    
\varphi( \vec{\rho}) \;  =   \; - \; \Delta  \;  \varphi( \vec{\rho}) \;  .
\end{equation}
Eq.~(\ref{3.62}) has been already discussed in I, Sect.~4.1. It turns out that the bound-state wavefunction  $\varphi( \vec{\rho})$
is practically unaffected by the magnetic field for applied magnetic field strengths  employed in experiments,
$H \lesssim$ $100 \,T$~\footnote{Even though we are using CGS units, it is customary
 in the literature to express the applied magnetic field in Tesla, $1 T = 10^4 G$.}. 
Moreover, the external magnetic field lifts the degeneracy with respect to the magnetic quantum number. However, 
the resulting Zeeman splitting is completely negligible such  that $\Delta  \simeq \Delta_2(\delta)$. \\
To determine the condensate wavefunction  we proceed as in Sect.~\ref{s3}. Obviously, one finds  that the  wavefunction of the low-lying condensate 
excitation has the form given by  Eq.~(\ref{3.29}) and it satisfies the   Schr\"odinger-like eigenvalue equation:
\begin{equation}
\label{3.63} 
 \frac{ 1}{4 m^*_h} \left [  - \, i \,  \hbar  \nabla  \; - \; \frac{2e}{c}  \,   \vec{A}(\vec{R})  \right ]^2  \,
  \chi(\vec{R} )    \;  = \; \varepsilon \;   \chi(\vec{R} )  \;  , 
\end{equation}
where  $\varepsilon$ is the excitation energy with respect to the condensation energy.  As expected,   Eq.~(\ref{3.63}) shows that
low-lying excitations of the condensate behave like quasi- particles with mass  $2 m^*_h$ and positive charge $+ 2e$.
Interestingly enough, we may write down the electromagnetic current for low-lying condensate excitations. In fact, we use
the quantum-mechanics expression for the current density when a charged particle moves in a magnetic 
field~\footnote{See, for example, Ref.~\cite{Landau:1977}.} to get:
\begin{eqnarray}
\label{3.64} 
\vec{j}_{em}(\vec{R}) =  2 e \, \vec{j}(\vec{R}) \; = \; 
\frac{ - i \, e \; \hbar}{2 m^*_h}  \;  \bigg [ \chi^*(\vec{R} ) \,  \nabla  \,   \chi(\vec{R})    \;  \; \; 
\\ \nonumber
- \;    \nabla  \,  \chi^*(\vec{R} ) \,   \chi(\vec{R})    \bigg ]  \; - \; \frac{ 2 \,  e^2 }{m^*_h \, c}  
   \chi^*(\vec{R} ) \,   \chi(\vec{R})  \vec{A}(\vec{R}) \; .
\end{eqnarray}
 Using Eq.~(\ref{3.29}), and introducing  the superfluid velocity:
\begin{equation}
\label{3.65}  
 \vec{v}_s (\vec{R}, \vec{A}) \; = \; \frac{\hbar}{  2   m^*_h}  \;   \nabla  \; \Theta(\vec{R}) \; - \;  \frac{e}{ m^*_h c} \; \vec{A}( \vec{R})  \; ,
\end{equation}
it is easy to check that:
\begin{equation}
\label{3.66} 
\vec{j}_{em}(\vec{R})  \; = \;  2  e \, n_s \,   \vec{v}_s (\vec{R}, \vec{A})  \; .
\end{equation}
\subsection{The roton gas and the specific heat anomaly }
\label{s3.3}
In this Section we address the problem of the specific heat anomaly  close to the transition temperature $T_c$.  
The specific heat is one of the few bulk thermodynamic probe of the superconducting state.
In high temperature cuprate superconductors the  normal-superconducting transition is substantially broadened
relative to that in many conventional superconductors.  In classic superconductors there is  a specific heat jump
at the critical temperature. On the other hand, cuprate high temperature superconductors show both  a pronounced
peak  or only a broad hump  in the specific heat at the critical temperature $T_c$.
In general, the  broadened specific heat anomaly has been attributed to superconducting fluctuations
(see, for example, the  reviews in Refs.~\cite{Junod:1998,Fisher:2007} and references therein). 
For sake of definiteness, we will focus on  the Yttrium compound  YBa$_2$Cu$_3$O$_{6+x}$  (YBCO) with
superconducting critical temperatures up to $T_c \simeq 93 K$~\cite{Wu:1987}. Indeed,
the measurements of specific heat on YBCO are more complete than those for any other high
temperature superconductor.  The specific heat anomaly in the YBCO sample shows a sharp peak structure.
Customarily  the specific heat anomaly is characterized by   $\Delta c(T_c)$, the difference of the specific heat
between the peak value with respect to the background  lattice specific heat. Furthermore, it should be keep in mind 
 that the  analysis of the shape of the anomaly  is always complicated by uncertainties in the subtraction of the background.
It turns out that   the specific heat jump is clearly seen at the superconducting transition, even though 
the specific heat  anomaly at  the critical temperature is only about a few percent of the total. 
For YBCO in the optimal doping region with $T_c \simeq 87.5 \, K$, form Fig.~5 of Ref.~\cite{Junod:1998} we
infer a specific heat anomaly:
\begin{equation}
\label{3.67} 
\frac{\Delta c(T_c)}{T_c}  \; \simeq \;  5  \; 10^{-3} \; \frac{J}{K^2 \, gat}  \; ,
\end{equation}
where $gat=gram-atom$ is the volume occupied by $N_A$ unit cell of the crystal, $N_A$ being the Avogadro's number.
Let $V_u$ be the volume of the unit cell, then from Eq.~(\ref{3.67}) we may estimate the dimensionless specific heat anomaly:
\begin{equation}
\label{3.68} 
V_u \; \frac{\Delta c(T_c)}{k_B}  \; \simeq \;  5.26  \; 10^{-2} \;  \; ,
\end{equation}
where $k_B$ is the Boltzmann constant. The main advantage of the dimensionless specific heat anomaly resides in the 
fact that it can be directly compared with theoretical calculations within our model. In fact, we now show that 
in our approach the specific heat anomaly can be accounted for by the specific heat of the roton thermal gas. \\
The thermodynamics of the superconductive condensate is determined by the low-lying excitations. The previous Section
showed that the condensate low-lying excitations are the rotons which behave like quasi-particle with mass $2 m^*_h$, 
charge $2 e$, and temperature-dependent energy:
\begin{equation}
\label{3.69}  
\varepsilon_{rot}(T)  \; \simeq \;    \alpha \; m^*_h  \;  \vec{v}_s^{\, 2}(T)   \;  ,
\end{equation}
where:
\begin{equation}
\label{3.70}  
 |\vec{v}_s(T)| \; = \;  v_s(T) \;  \simeq \;  v_s(0)  \;   e^{  - \, b \; \left [ \frac{1}{\sqrt{1 \, - \, T/T_c}} \; - \; 1 \right ] } \; \; , 
 \; \;  T    \; \le  \; T_c \;  .
\end{equation}
Eqs.~(\ref{3.69}) and  (\ref{3.70}) imply that the roton energy vanishes continuously at the critical temperature. Therefore, near
$T_c$  there is a proliferation of thermally activated rotons which, therefore, dominate the thermodynamic potential. On the other hand,
at low temperatures the excitation of rotons is exponentially suppressed since $ \varepsilon_{rot}(0) \sim \Delta_2(\delta)$.
Since rotons are bosons (like the phonons), the thermal distribution function is given by the Bose-Einstein distribution:
\begin{equation}
\label{3.71}  
f(\varepsilon_{rot})  \; =  \;  \frac{1}{e^{\frac{\varepsilon_{rot}}{k_BT} } \; - \;1 }   \; \;  .
\end{equation}
Our approximation amounts to deal with  the roton gas as  an ideal gas.  We may, then, easily evaluate the roton energy density:
\begin{equation}
\label{3.72}  
\mathfrak{u}_{rot}(T)  \; \simeq   \; \int  \; \frac{d \vec{k}}{(2 \pi)^2} \; \varepsilon_{rot}(T)  \; \; \frac{1}{e^{\frac{\varepsilon_{rot}(T)}{k_BT} } \; - \;1 }  
\; \delta\left( \frac{|\vec{k}|}{k_0} - 1\right )  \;   .
\end{equation}
In  Eq.~(\ref{3.72}) the Dirac $\delta$-function takes care of the fact that the  roton wavenumber is constrained 
according to    Eq.~(\ref{3.43}):
\begin{equation}
\label{3.73}  
|\vec{k}_{rot}|   \; \simeq  \;  k_0 \; \simeq \;   \frac{2 \, \pi }{d_0 }   \; \simeq  \; 2 \pi \, \sqrt{\frac{\delta}{2  a_0^2}} \;   .
\end{equation}
 A straightforward calculation gives:
\begin{equation}
\label{3.74}  
\mathfrak{u}_{rot}(T)   \simeq     \frac{\pi \delta}{a_0^2} \; \varepsilon_{rot}(0)  \; f_{KT}(2b,T) \; 
\frac{1}{e^{\frac{\varepsilon_{rot}(0) f_{KT}(2b,T) }{k_BT} } \; - \;1 }   \; \; 
\end{equation}
 where, for convenience, we introduced the function:
\begin{equation}
\label{3.75}  
 f_{KT}(b,T)  \; \equiv  \;     e^{  - \, b \; \left [ \frac{1}{\sqrt{1 \, - \, T/T_c}} \; - \; 1 \right ] }   \; \;  .
\end{equation}
From the roton internal energy density we may evaluate the specific heat at constant volume:
\begin{equation}
\label{3.76}  
c_{rot}(T) \; = \; \frac{\partial \, \mathfrak{u}_{rot}(T)}{\partial T}  \; \;  .
\end{equation}
\begin{figure}
\vspace{0.8cm}
\resizebox{0.5\textwidth}{!}{%
\includegraphics{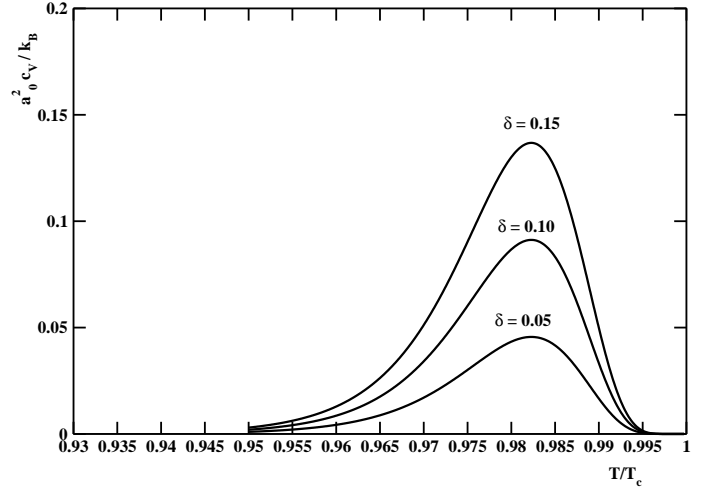}
}
\caption{\label{Fig3} Dimensionless roton specific heat versus the reduced temperature for different hole doping levels.}
\end{figure}
To compare the roton specific heat with experimental data we must take into account that the rotons
are the low-lying excitations of the superfluid condensate. Therefore the rotons contribute to
the specific heat only within the condensate fraction which share phase coherence.  According to our previous
discussion, at zero temperature all the hole pairs are condensed with phase coherence. However, at finite temperatures
the phase-coherent condensate fraction is given by:
\begin{equation}
\label{3.77}  
 \frac{n_s(T)}{n_s(0)} \; = \; f_{KT}(b',T)    \; \;  ,
\end{equation}
according to  Eqs.~(\ref{3.54}) and  (\ref{3.75}).  Thus,  the contribution of rotons to the specific heat at constant volume 
can be written as:
\begin{equation}
\label{3.78}  
c_{V}(T) \; \simeq  \; \frac{n_s(T)}{n_s(0)} \; c_{rot}(T) \; \simeq \;  \frac{n_s(T)}{n_s(0)} \;  \frac{\partial \, \mathfrak{u}_{rot}(T)}{\partial T}  \; \;  .
\end{equation}
In Fig.~\ref{Fig3} we display the temperature dependence of the dimensionless specific heat at constant volume for
different hole doping levels.  For our numerical estimates, we assumed $b \simeq 0.193$. As we will discuss later on, this
value has been fixed by fitting the temperature dependence of the excitation energy of the nodal quasielectron liquid 
to the experimental data. As concern
the parameter $b'$, we choose the value $b' \simeq 0.80$ that it is relevant for the London penetration
length in optimally doped YBCO (see Sect.~\ref{s3.4}). From Fig.~\ref{Fig3}  we see that, as expected, the roton specific heat is sizable
in the critical region only. Moreover, $c_{V}(T)$ displays a rather sharp peak at temperatures very near the critical temperature, with
a well developed curvature below $T_c$. Note that this feature has no equivalent in classic superconductors and it is 
in fair agreement with observations (see, for instance,  Fig.~6 in Ref.~\cite{Junod:1998}). Finally, we see that the  roton specific heat 
scales almost linearly with the hole doping $\delta$ in agreement with several measurements in the pseudogap region. \\
From   Fig.~\ref{Fig3} we estimate the dimensionless specific heat anomaly near the optimal doping:
\begin{equation}
\label{3.79} 
a_0^2  \; \frac{\Delta c_V(T_c)}{k_B}  \; \sim \;  10  \; 10^{-2} \;  \; .
\end{equation}
Comparing this estimate with   Eq.~(\ref{3.68}), we may safely conclude that the thermal roton gas can account, both qualitatively
and quantitatively, for the observed specific heat anomaly in hole doped cuprate superconductors in the pseudogap region.\\
Let us conclude this Section by briefly discussing the effects of an applied magnetic field on the specific heat anomaly.
Experimentally, it results that a magnetic field applied along the normal direction to the $CuO_2$ planes strongly suppresses 
the specific heat anomaly at $T_c$, while the effect of magnetic fields applied parallel to the $CuO_2$  planes is much less pronounced.
Moreover, the magnetic field shifts the specific heat anomaly to lower temperatures without strongly affecting the critical
temperature $T_c$. In Sect.~\ref{s3.2} we have seen that the electromagnetic current can be written in term of the field-dependent
superfluid velocity   $\vec{v}_s (\vec{R}, \vec{A})$, Eq.~(\ref{3.65}). Evidently, at finite temperatures we have:
\begin{equation}
\label{3.80}  
 \vec{v}_s (T, \vec{A}) \; = \;  \vec{v}_s (T)  \; - \;  \frac{e}{ m^*_h c} \; \vec{A}( \vec{R})  \; ,
\end{equation}
with:
\begin{equation}
\label{3.81}  
 \vec{v}_s(T)  \; = \; \vec{v}_s(0) \;  \;   e^{  - \, b \; \left [ \frac{1}{\sqrt{1 \, - \, T/T_c}} \; - \; 1 \right ] } \; \;   .
\end{equation}
In the mixed state the magnetic field penetrates into the superconductor with an array of Abrikosov vortices (see Sect.~\ref{s3.5}). 
The roton excitations are present outside the Abrikosov vortices. The effects of the magnetic field on these roton excitations
amount to shifting the superfluid velocity  $ \vec{v}_s$ according to Eq.~(\ref{3.80}). To estimate the magnetic field at the roton site,
we recall that rotons are disturbances of the superfluid condensate of size $\sim d_0$. Within the roton core the condensate losses 
the phase coherence, so that the roton core constitutes a region of normal fluid. Moreover, the screening of the roton
velocity  $\vec{v}_s (\vec{R})$ is due to Kosterlitz-Thouless vortex-antivortex pairs. Evidently, the screening of the magnetic field
due to a given  Kosterlitz-Thouless vortex is compensated by the corresponding antivortex. Then, we may safely assume that
the magnetic field in the roton region coincides with the external magnetic field $H$. This means that in the vortex core
$\nabla \times \vec{A}(\vec{R}) \simeq  \vec{H}(\vec{R})$. 
Since the size of the roton is very small, we may consider the magnetic field  $\vec{H}(\vec{R})$ almost constant within the roton
core and obtain  the roughly estimate:
\begin{equation}
\label{3.82}  
|\vec{A}(\vec{R})|  \; \simeq \; H \;  d_0  \;  \simeq  \;  H \; \sqrt{\frac{2 a_0^2}{\delta}} \;   .
\end{equation}
Accordingly, we find for the roton energy:
\begin{equation}
\label{3.83}  
\varepsilon_{rot}(T,H)   \simeq   \alpha \, m^*_h  \,  \vec{v}_s^{\, 2}(T, \vec{A})   \simeq 
  \alpha \, m^*_h   \left [ v_s(T) \; - \; \frac{eH}{m^*_h c} \, d_0 \right ]^2 .
\end{equation}
Now, we proceed as before and obtain the roton energy density by simply replacing  in Eq.~(\ref{3.72}) the roton energy
$\varepsilon_{rot}(T)$ with  $\varepsilon_{rot}(T,H)$ given by   Eq.~(\ref{3.83}). After that, the roton specific heat is given
by Eq.~(\ref{3.76}). \\
\begin{figure}
\vspace{0.8cm}
\hspace{-0.2cm}
\resizebox{0.5\textwidth}{!}{%
\includegraphics{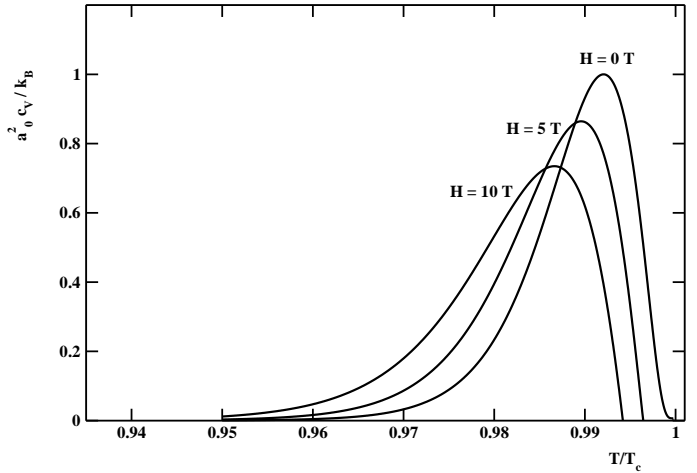}
}
\caption{\label{Fig4}  Dimensionless roton specific heat versus the reduced temperature at hole doping $\delta=0.10$
for different magnetic field strengths. The curves are normalized such that at $H=0$ the maximum of $a_0^2 \, \frac{c_V}{k_B}$
is set to $1.0$.}
\end{figure}
In Fig.~\ref{Fig4} we show the dimensionless roton specific heat at hole doping $\delta=0.10$ for different magnetic
field strengths, and  using the same values of the parameters as before. Qualitatively  the effect of the magnetic field is to shift 
the peak in the roton specific heat and to reduce the peak value. In fact, these  features are in  qualitative agreement
with experimental observations.
\subsection{The London penetration length }
\label{s3.4}
In this Section we discuss the magnetic properties of our ideal planar superconductor. Let us assume
that the superconductor is immersed in an external constant magnetic field of strength $H_0$ perpendicular
to the $CuO_2$ plane. If the magnetic field does not exceed the lower critical field $H_{c1}$, then the external
field does not penetrate into the superconductor (Meissner effect) so that the magnetic induction $B$ vanishes,
$B=0$. However, it should be mentioned that, in fact, the magnetic field penetrates into the superconductor to a depth
given by the London penetration length $\lambda(T)$ which depends on the temperature. Let $\vec{h}(\vec{R})$ be
the microscopic  magnetic field. The macroscopic field is defined as the spatial average of  $\vec{h}(\vec{R})$:
\begin{equation}
\label{3.84}  
\left <  \vec{h}(\vec{R}) \right  >_{\text{vol}} \;  = \; \vec{B}(\vec{R})  \; .
\end{equation}
Let us consider, now, a homogeneous superconductor in thermodynamic equilibrium. In this case there is no normal current,
so that the electromagnetic current is given by Eq.~(\ref{3.65}) that we rewrite as:
\begin{equation}
\label{3.85} 
\vec{j}_{em}(\vec{R})  \; = \;  \frac{e \hbar}{m^*_h}  \,  n_s \;  \nabla  \; \Theta(\vec{R}) \;  - \; 
 \frac{2 e^2}{ m^*_h c} \, n_s  \; \vec{A}( \vec{R})  \; ,
\end{equation}
where:
\begin{equation}
\label{3.86}  
\vec{h}(\vec{R}) \; = \; \; \nabla \times \vec{A}(\vec{R}) \; \; .
\end{equation}
Moreover, within our approximations  to obtain gauge invariant results we need to adopt the physical London gauge:
\begin{equation}
\label{3.87}  
 \nabla   \vec{A}(\vec{R}) \; =  \;  0 \; \; .
\end{equation}
In fact our reduced Hamiltonian approximation is similar to the reduced BCS Hamiltonian.  It is known since long time that
the simple BCS pairing approximation gives an accurate account of the response of the system to transverse electromagnetic
fields, but it does not give the correct response to longitudinal fields. Indeed, it was soon 
realized~\cite{Bardeen:1957b,Anderson:1958,Rickayzen:1959} that this  difficulty can be overcome once one realizes that
the longitudinal gauge potential couples primarily to the collective density fluctuation mode.  The density fluctuation modes
correspond to collective plasma oscillations of the charged condensate. Since there are no low-lying collective oscillations
due to the Coulomb interaction, the contributions of the longitudinal vector potential to the electromagnetic current are
negligible. In other words, when the gauge is chosen so that   $\nabla  \vec{A}(\vec{R})  =    0 $  one is guaranteed  
that gauge invariance is maintained.\\
From the Maxwell equation:
\begin{equation}
\label{3.88}  
 \nabla \,  \times \,  \vec{h}(\vec{R}) \; =  \;  \frac{4 \pi}{c}  \; \,   \vec{j}_{em}(\vec{R})    \;  \; ,
\end{equation}
we get:
\begin{equation}
\label{3.89}  
 \nabla \, \times  \, \nabla  \, \times \,  \vec{h}(\vec{R}) \; =  \;  \frac{4 \pi}{c}  \;  \nabla  \times   \vec{j}_{em}(\vec{R})  \; \; .
\end{equation}
Using  Eq.~(\ref{3.85}) we obtain readily:
\begin{equation}
\label{3.90}  
 \nabla \, \times  \, \nabla  \, \times \,  \vec{h}(\vec{R}) \; =  \; -  \; \frac{1}{\lambda^2(T)}  \;  \vec{h}(\vec{R})  \; \; ,
\end{equation}
where:
\begin{equation}
\label{3.91}  
 \frac{1}{\lambda^2(T)}  \; = \frac{ 8 \pi e^2}{m^*_h c}   \; n_s(T)  \; .
\end{equation}
Since     $\nabla \cdot \vec{h}(\vec{R})  =    0 $,  Eq.~(\ref{3.90}) leads to:
\begin{equation}
\label{3.92}  
 \nabla^2 \, \vec{h}(\vec{R}) \; =  \;  \frac{1}{\lambda^2(T)}  \;  \vec{h}(\vec{R})  \; \; ,
\end{equation}
which shows that $\lambda(T)$ is the London penetration length.  To obtain  Eq.~(\ref{3.90}) we assumed
 $ \nabla \,  \times \,  \nabla   \theta(\vec{R}) = 0$, which corresponds to irrotational superfluid flow. As discussed in Sect.~\ref{s3.5}, if
 the external magnetic field exceeds the lower critical field $H_{c1}$, then it is thermodynamically favored the formation of Abrikosov 
 vortices where   $\nabla  \theta(\vec{R})$ develops a singularity. The temperature dependence of the London penetration length results
 from the temperature behavior of the superfluid condensate,  Eq.~(\ref{3.54}). Accordingly, we can write:
\begin{eqnarray}
\label{3.93}  
 \frac{1}{\lambda^2(T)}  \; =   \frac{1}{\lambda^2(0)}  \;   e^{  - \, b' \; \left [ \frac{1}{\sqrt{1 \, - \, T/T_c}} \; - \; 1 \right ] } \;  ,  
 \\ \nonumber  
    \frac{1}{\lambda^2(0)} =  \frac{ 8 \pi e^2}{m^*_h c}   \; n_s(0)  \; , \;  n_s(0) \; \simeq \;   \frac{\delta}{2 a_0^2} \; .
\end{eqnarray}
%
%
\begin{figure}
\vspace{0.8cm}
\resizebox{0.5\textwidth}{!}{%
\includegraphics{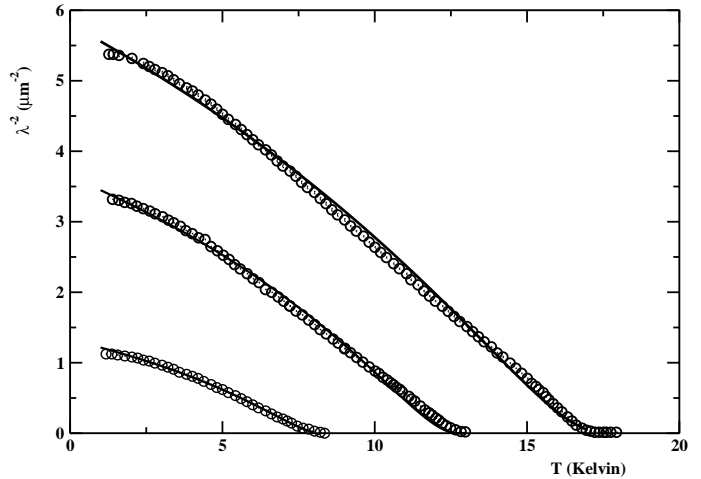}
}
\caption{\label{Fig5}  The London penetration length versus the temperature for highly underdoped
 YBCO superconductor. The data has been taken from  Fig.~2 of Ref.~\cite{Broun:2007}. The solid curves 
 are the fits of the data to  Eq.~(\ref{3.93}).}
\end{figure}
%
The peculiar temperature behavior of the London penetration length implied by   Eq.~(\ref{3.93}) can be contrasted with
experimental observations in cuprate superconductors in the pseudogap region. In fact, we checked that our results are in satisfying
agreement with observations. In the remainder of the present Section we present some  representative  examples in the highly
underdoped  and  optimal doped regions for different compound of copper-oxide superconductors. In Fig.~\ref{Fig5}
 we display the penetration length at finite temperatures for highly underdoped $YBa_2Cu_3O_{6+x}$ superconductors
 with critical temperatures  $T_c = 8.4 \, K , 13.0 \, K, 17.9 \, K$ respectively. The corresponding doping level $\delta$
can be inferred from the phenomenological parabolic relationship~\cite{Takagi:1989,Torrance:1989,Presland:1991}:
\begin{equation}
\label{3.94}  
1 \; - \frac{T_c(\delta)}{T_c^{max}} \; = \;  82.6 \, (\delta - 0.16)^2  \; \; , \; \;  T_c^{max} \; \approx \; 98 \, K \;.
\end{equation}
The data have been extracted from Fig.~2 of Ref.~\cite{Broun:2007} where it is displayed the London penetration length for
YBCO with 20 different doping levels and critical temperatures ranging from $T_c \simeq  3 \, K$ up to   $T_c \simeq  17 \, K$.
We fitted the data to our  Eq.~(\ref{3.90})   leaving as free parameters the non-universal constant $b'$ and   $\frac{1}{\lambda^2(0)}$.
To implement  the fits  we used  the program Minuit~\cite{James:1975} which is conceived as a tool to find the 
minimum value of a multi-parameter function and analyze the shape of the function around the minimum.  The principal application 
is to compute the best-fit parameter values and uncertainties by minimizing the total chi-square $\chi^2$. 
 As  rule of thumb,  a sensible fit results in $\chi_r^2 \sim 1$, where  $\chi_r^2$ is the reduced chi-square,
namely the total chi-square divided by the number of degree of freedom. Remarkably, we found that  Eq.~(\ref{3.90}),  
with $b' \, \simeq 1.40$, gives a fit with  $\chi_r^2 \sim 1$, in quite good agreements with the experimental data (see Fig.~\ref{Fig5}). 
Moreover, the fits returned values of  $\frac{1}{\lambda^2(0)}$ which were quite consistent with the ones reported in Ref.~\cite{Broun:2007},
 Fig.~3. Note, however, that  to compare our theoretical estimate for $\lambda(0)$ to measurements, we  must slightly 
 modify our result which is relevant for a planar superconductor:
\begin{equation}
\label{3.95}  
 \frac{1}{\lambda^2(0)}  \;  =  \; \frac{ 8 \pi e^2}{m^*_h c}   \; n_s(0)  \; 
  \simeq  \;  \frac{ 8 \pi e^2}{m^*_h c} \;  \frac{\delta}{2 a_0^2c_0}  \; .
\end{equation}
%
%
\begin{figure}
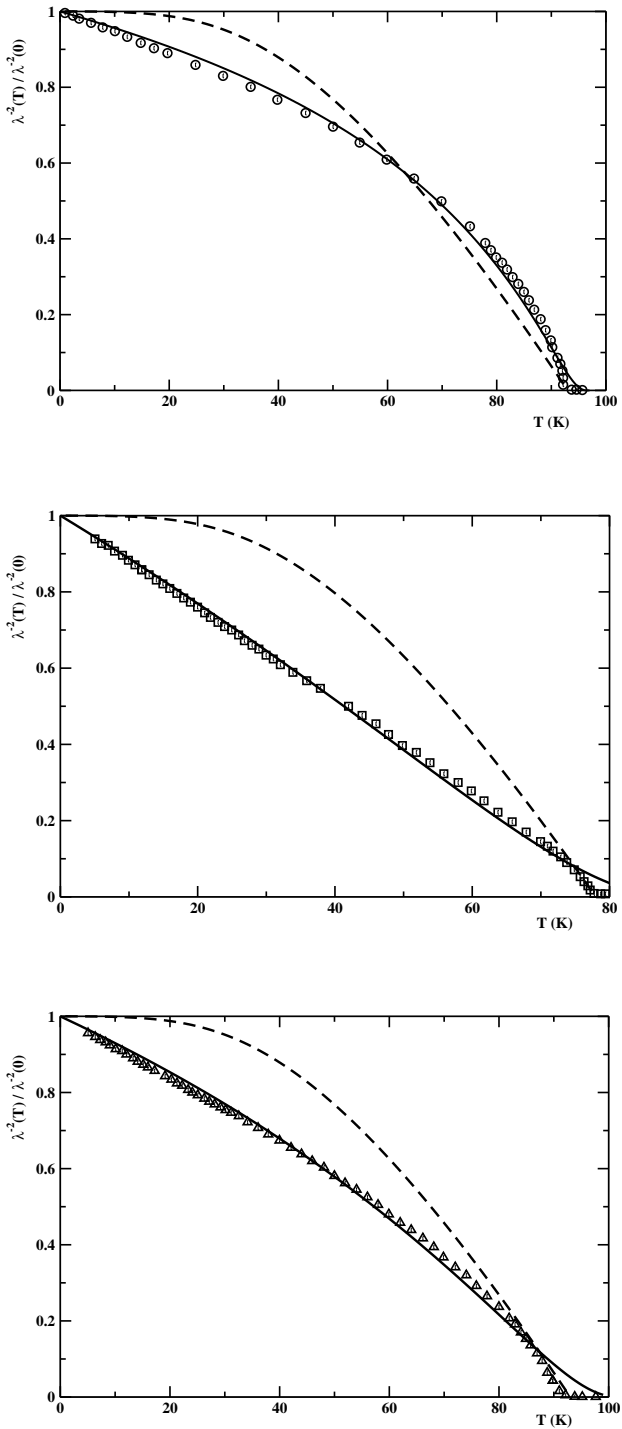

\vspace{0.8cm}
\centering
\resizebox{0.45\textwidth}{!}{%
\includegraphics{Fig6a.eps} }

\vspace{1.0cm}

\resizebox{0.45\textwidth}{!}{%
\includegraphics{Fig6b.eps}  }

\vspace{1.0cm}

\resizebox{0.45\textwidth}{!}{%
\includegraphics{Fig6c.eps}} 
\caption{\label{Fig6} (Top) London penetration length versus the temperature in the optimal doping region for YBCO, 
 $T_c=93 \, K$,  (Middle) Tl-2201 $T_c = 78 \, K$,   (Bottom) BSCCO,   $T_c =  93 \, K$. The solid lines are the fits of the
 data to  Eq.~(\ref{3.93}). The dashed lines are the London penetration length in the  weak-coupling d-wave BCS theory, Eq.~(\ref{A.34}).}
\end{figure}
%
We found that, by using the numerical values of the model parameters,   Eq.~(\ref{3.95}) gives the correct order of magnitude
for the zero-temperature London penetration length. In addition, the linear dependence of    $\frac{1}{\lambda^2(0)}$ on the doping
$\delta$ is consistent with experimental observations (see Fig.~3. in  Ref.~\cite{Broun:2007}). \\
We have also compared our peculiar temperature dependence of the penetration length with experimental observations
for three different class of cuprate superconductors in the optimal doping region. In Fig.~\ref{Fig6} we display the observed
 temperature behavior of the normalized London penetration length for very high quality single crystal optimal doped 
 $YBa_2Cu_3O_{6.95}$, $T_c=93 \, K$, 
single crystal $Tl_2Ba_2CuO_{6+\delta}$ (Tl-2201), $T_c=78 \, K$, and high-quality single crystal $Bi_2Sr_2CaCu_2O_6$ 
(BSCCO),   $T_c=93 \, K$.  
The temperature dependence of the London penetration length in YBCO has been obtained
 in Ref.~\cite{Hardy:1993} by microwave techniques which allow to track the deviations of $\lambda(T)$ from its zero
 temperature value. We extracted the data from Fig.~12 in Ref.~\cite{Hardy:1998}.  
 For Tl-2201, Ref.~\cite{Broun:1997} reports the measurements of the in-plane microwave conductivity which allowed
 to extract the variations of the London penetration length with the temperature. The data has been taken from Fig.~3 in 
Ref.~\cite{Broun:1997}.
Finally, for BSCCO the data have been taken from Fig.~3 in Ref.~\cite{Lee:1996} where the temperature dependence of
$\lambda(T)$ has been obtained from the in-plane microwave surface impedance. \\
We fitted the data to  Eq.~(\ref{3.93}) leaving $b'$ and $T_c$ as free parameters. The results of our fits are displayed
as full  lines in Fig.~\ref{Fig6}.  Concerning the fitted values of the parameters, we obtained $b' \simeq 0.80$, 
$T_c \simeq  97.0 \, K$ (YBCO),   $b' \simeq 2.05$,  $T_c \simeq  93.7 \, K$ (Tl-2201), and  $b' \simeq 1.41$, 
$T_c \simeq  103.8 \, K$ (BSCCO).
Fig.~\ref{Fig6} shows that our proposal for the temperature dependence of the London penetration length, Eq.~(\ref{3.95}),
seems to be in reasonable agreements with experimental measurements, at least for temperatures not too close
to the critical temperature. To appreciate better this point, in  Fig.~\ref{Fig6}  we also compare the data with the
weak-coupling d-wave BCS  prediction (see Eq.~(\ref{A.34}) in Appendix~\ref{AppendixA}). It seems evident that
our best fits to  Eq.~(\ref{3.95}) compare much better with  experimental data. Near the critical temperature $T_c$
there are deviations of the data with respect to  Eq.~(\ref{3.95}).  Moreover, the best-fit values for the critical temperatures
are systematically slightly higher than the observed $T_c$. These features, however, are to be expected. In fact, we already noticed
that our mean field approximation could be questionable in the optimal doping region due to the enhanced role of phase
fluctuations in the crossover from the pseudogap to the d-wave BCS gap. In fact, it is known that the mean field critical
temperature tends to overestimate the actual value of $T_c$ due to sizable fluctuations  in the critical region.
As regard the parameter $b'$, we already remarked that this parameter is not universal, so that it could, in principle,
depends on the  crystal structure, on the presence of disorder and defects, and also on the 
hole doping fraction.  In any case,  as anticipated, we found that $b' \, \sim 1$.
\subsection{Vortex structure and critical magnetic fields}
\label{s3.5}
We obtained the London equation   Eq.~(\ref{3.90}) assuming an irrotational superfluid flow. Let, now, relax this assumption
by letting   $\nabla   \theta(\vec{R})$ to develop a singularity.  In this case, in Appendix~\ref{AppendixB} we show that
the magnetic field obeys the following equation:
\begin{equation}
\label{3.104-bis}  
 h(\vec{R}) \; - \;  \lambda^2(T) \;   \nabla^2  \;  h(\vec{R}) \; =  \; 
 \phi_0  \;      \delta(\vec{R}) \; ,
\end{equation}
which, indeed, correspond to  the well-known Abrikosov vortex field distribution. As is well known, in type II superconductors
when the external magnetic field exceed a minimal magnetic field strength, the lower critical
magnetic field, it is energetically favored the production of Abrikosov vortices. In fact, we obtain for the lower critical
magnetic field (see Appendix~\ref{AppendixB}):
\begin{eqnarray}
\label{3.130-bis}  
H_{c1}(T) \; =  \;  H_{c1}^{em} (T) \; + \;   H_{c1}^{core}(T)  \; \; ,  \hspace{2.0cm}
\\ \nonumber
H_{c1}^{em}(T)   \simeq    \frac{\phi_0}{4 \pi  \lambda^2(T)}  \;   \ln (\kappa)  \; \; ,  \; \;  
H_{c1}^{core}(T)  \simeq     \frac{4 \pi}{c_0 \phi_0} \;   \varepsilon_{rot}(T)   \; .
\end{eqnarray}
As concern the upper  critical field $H_{c2}$, we recall that in BCS superconductors  $H_{c2}$ is given by the Cooper pair-breaking
critical field. Moreover, the pair-breaking critical field is of the same order of the depairing field and of the Ginzburg-Landau
critical field, defined as the magnetic field strength such that  the Abrikosov vortices became to overlap. In our model,
however, it turns out that both the pair-breaking and Ginzburg-Landau critical fields are much higher than the depairing
field (see I, Sect.~4.2). In fact we found~\cite{Cea:2013}  that the upper critical magnetic field is given by the depairing
field:
\begin{equation}
\label{3.133}  
H_{c2}(T) \; \simeq  \; \kappa^2  \frac{8 \pi e}{c} \; n_s(T) \;  \sqrt{\frac{\Delta_2(\delta)}{m^*_h}}   \; .
\end{equation}
As the applied magnetic field exceeds $H_{c1}$, more and more Abrikosov vortices are found. Increasing the strength
of the external magnetic field these vortices become more dense and, eventually, their cores overlap  so that the medium
becomes normal. In conventional superconductors, the field at which this happens is the upper critical field. Within
the Ginzburg-Landau formulation one introduces the Ginzburg-Landau coherence length  $\xi_{GL}$ such that:
\begin{equation}
\label{3.134}  
H_{c2}(T) \; \simeq  \;  \frac{\phi_0}{\pi \, \xi_{GL}^2(T)}  \;   \; .
\end{equation}
 Interestingly enough, if we define the   Ginzburg-Landau coherence length  $\xi_{GL}$  using   Eq.~(\ref{3.134})
 with the upper critical field given by   Eq.~(\ref{3.133}), then the usual   Ginzburg-Landau $\kappa$-parameter is:
\begin{equation}
\label{3.135}  
\kappa_{GL} \; \simeq  \;  \frac{\lambda(T)}{ \xi_{GL}(T)}  \;   \; .
\end{equation}
After some manipulations, we find:
\begin{equation}
\label{3.136}  
\kappa_{GL} \; \simeq  \;  \kappa \; \left [ \Delta_2(\delta) \;  \frac{m^*_h c_0^2}{\hbar^2} \right ]^{\frac{1}{4}}  \;   \; .
\end{equation}
Using   Eq.~(\ref{2.14}) we readily get: 
\begin{equation}
\label{3.137}  
\kappa_{GL} \; \simeq  \;  1.44 \; \kappa \; \left [ 1 \;  - \; (\frac{\delta}{\delta^*})^{1.5} \right ]^{\frac{1}{4}}  \; .
\end{equation}
For hole doping not to close to $\delta^*$, where our mean field approximation is anyway  questionable, we see
that  $\kappa_{GL} $ is almost independent on the hole doping and, in addition, $\kappa_{GL}  \sim \kappa$. \\
 In the subsequent  Section we shall compare the temperature dependence of the critical magnetic fields,   Eqs.~(\ref{3.130-bis}) 
 and  (\ref{3.133}), with experimental data. In the remained  of the present Section we intend to compare to experimental 
 measurements the peculiar doping  dependence implied by Eq.~(\ref{3.133}) for the upper critical magnetic field 
 at zero temperature.
 %
 %
\begin{figure}
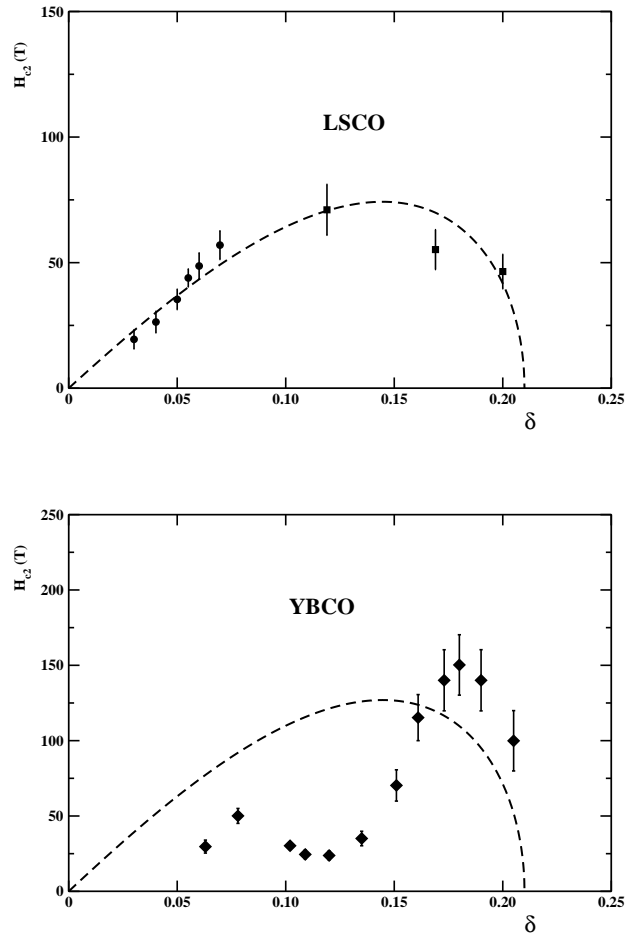

\vspace{0.8cm}
\centering
\resizebox{0.45\textwidth}{!}{%
\includegraphics{Fig7a.eps} }

\vspace{1.0cm}

\resizebox{0.45\textwidth}{!}{%
\includegraphics{Fig7b.eps}  }
\caption{\label{Fig7} (Top) Upper critical magnetic field at low temperatures for LSCO versus the hole doping fraction. 
Data have been extracted from Ref.~\cite{Li:2007} (full circles) and Ref.~\cite{Wang:2006} (full squares). 
The dashed line is Eq.~(\ref{3.133})  with $\kappa \simeq 130$.  (Bottom)  Upper critical magnetic field at low temperatures for YBCO versus the hole doping fraction.  
Data have been taken  from Table~1 in  Ref.~\cite{Grissonnanche:2014}. The dashed line is Eq.~(\ref{3.133}) with $\kappa \simeq 170$.}
\end{figure}
%
In Fig.~\ref{Fig7} we report the low-temperature upper critical magnetic field for the hole doped cuprates $La_{2-x}Sr_xCuO_4$ (LSCO)
 and YBCO in the pseudogap region $\delta \lesssim \delta^*$.  For LSCO in lightly doped region the estimate of the critical depairing
 field has been obtained in Ref.~\cite{Li:2007} from field suppression of the purely diamagnetic term in the effective magnetization
 measured by torque magnetometry.  It is interesting to point out that the authors of  Ref.~\cite{Li:2007} also reported that torque
 magnetometry measurements indicated that phase-disordered condensate survived up to $\delta = 0.03 < \delta_{min}$, while, in
 zero magnetic field, quantum phase fluctuations were  seen to destroy superconductivity at $\delta \simeq 0.055$. These experimental observations  
 give  support to our theory where, in the pseudogap region,   the superconductive  transition is driven by  phase fluctuations.
In the optimally doped LSCO, the data have been extracted from Fig.~18 in Ref.~\cite{Wang:2006}. The upper critical magnetic
field has been estimate from Nernst signal at the lowest available temperature $T=4.2 K$. In   Fig.~\ref{Fig7}, top panel, we compare
our theoretical result  Eq.~(\ref{3.133}) by assuming  $\kappa \simeq 130$.  Indeed, it seems that our model calculations compare
quite well  with measurements.  In the bottom panel we report the upper critical magnetic field for YBCO in the pseudogap
region. The data have been taken from Table~1 in  Ref.~\cite{Grissonnanche:2014}.  The upper critical magnetic field was defined
as the zero-temperature limit of $H_{vs}(T)$ within the theory of vortex-lattice melting. The critical magnetic field  $H_{vs}(T)$
was determined from high-field resistivity data as the critical field below which the resistance is zero. These data are compared
to  Eq.~(\ref{3.133}) with  $\kappa \simeq 170$.  In this case we see that, for $\delta \lesssim 0.15$,  the upper critical 
magnetic field $H_{c2}$ seems to be suppressed with respect to the theoretical expectations. We believe that this sudden drop in
$H_{c2}$ is revealing the presence of a competing phase which weakens the superconductivity. In fact, it is already known the existence
of competing order due to presumably the onset of incommensurate spin modulations detected by neutron scattering
and muon spectroscopy~\cite{Hang:2010,Coneri:2010,Drachuck:2014}.
\subsection{Temperature dependence of the critical magnetic fields }
\label{s3.6}
In this Section we would like to check  the temperature dependence of the critical magnetic fields by comparing
to available experimental data. For reasons of space, we have made a selection of representative examples. 
According to our previous discussion the dependence of the critical fields,  Eq.~(\ref{3.130-bis}) and  Eq.~(\ref{3.133}), on the
 temperature is basically due to  $n_s(T)$ and $\varepsilon_{rot}(T)$. Let us consider, firstly, the lower critical magnetic field.
 We recall that:
\begin{equation}
\label{3.138}  
H_{c1}(T) \; =  \;  H_{c1}^{em} (T) \; + \;   H_{c1}^{core}(T)  \; ,
\end{equation}
where, after taking into account Eqs. (\ref{3.93}),  (\ref{3.69}) and (\ref{3.70}), we can write:
\begin{equation}
\label{3.139}  
H_{c1}^{em}(T)  \; \simeq \;    H_{c1}^{em}(0) \;  e^{  - \, b' \; \left [ \frac{1}{\sqrt{1 \, - \, T/T_c}} \; - \; 1 \right ] } \; , 
\end{equation}
and
\begin{equation}
\label{3.140}  
H_{c1}^{core}(T)  \; \simeq  \;  H_{c1}^{core}(0)   \;  e^{  - \, 2 \, b \; \left [ \frac{1}{\sqrt{1 \, - \, T/T_c}} \; - \; 1 \right ] } \; .
\end{equation}
%
%
\begin{figure}
\vspace{0.8cm}
\resizebox{0.5\textwidth}{!}{%
\includegraphics{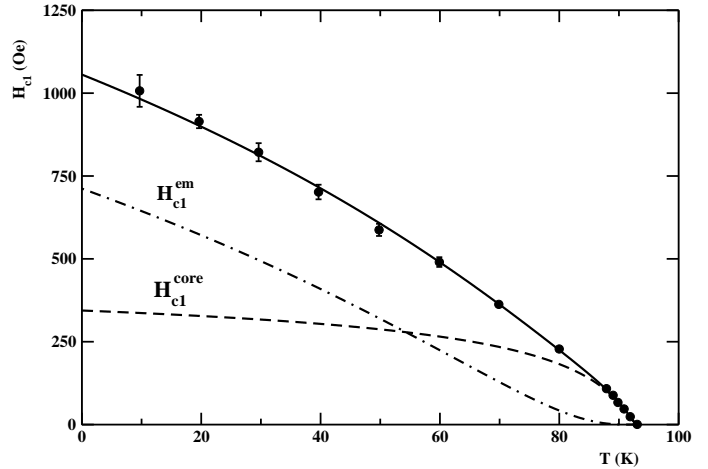}
}
\caption{\label{Fig8} Temperature dependence of the lower critical magnetic field for YBCO.
The data  have been extracted from Fig.~4  of Ref.~\cite{Liang:1994}. The solid line is the best fit of data to Eq.~(\ref{3.138}).
Dashed and dot-dashed lines correspond to   $H_{c1}^{core}(T)$  and  $H_{c1}^{em} (T)$  respectively.}
\end{figure}
In Ref.~\cite{Liang:1994} the lower critical fields in optimal doped YBCO ($T_c \simeq 93.1 K$) have been determined by magnetization measurements  
using a thin platelet  crystal for external magnetic fields perpendicular to the $CuO_2$ planes. In  Fig.~\ref{Fig8}
we report the relevant experimental data extracted from Fig.~4 in   Ref.~\cite{Liang:1994}. As one can see, at temperature well
below the critical temperature $T_c$ the lower critical magnetic field behaves essentially with a linear temperature dependence.
This behavior is qualitatively different from the characteristic saturation at low temperatures for conventional BCS superconductors.
Moreover, it turned out that the observed $H_{c1}(0)$ is considerably larger with respect to what one might expect within the 
Ginzburg-Landau theory. Indeed, the authors of  Ref.~\cite{Liang:1994} correctly argued that the core energy of an isolated 
Abrikosov vortex gives a non-negligible contribution to the lower critical magnetic field.  
%
%
\begin{figure}
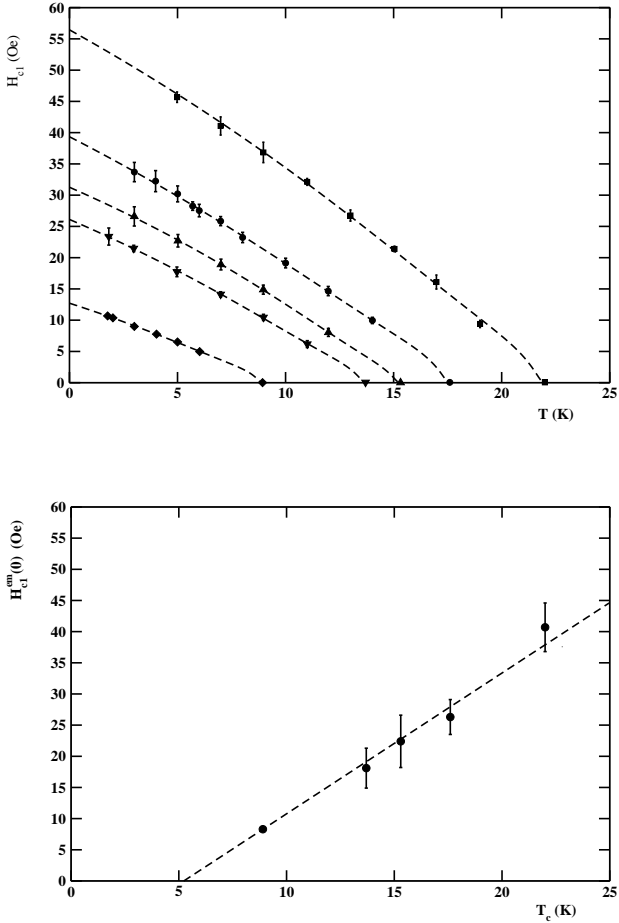

\vspace{0.8cm}
\centering
\resizebox{0.45\textwidth}{!}{%
\includegraphics{Fig9a.eps} }

\vspace{1.0cm}

\resizebox{0.45\textwidth}{!}{%
\includegraphics{Fig9b.eps}  }
\caption{\label{Fig9} (Top) Lower critical magnetic field versus the temperature for highly  underdoped YBCO.
The corresponding hole doping fraction are $\delta \simeq 0.055, 0.058, 0.059, 0.060, 0.064$ respectively.
Data have been extracted from Fig.~2 in  Ref.~\cite{Liang:2005}. The dashed lines are the best fits of the experimental data to
  Eqs.~(\ref{3.138}) - (\ref{3.140}).  (Bottom) Fitted values of the zero-temperature lower critical magnetic field  $H_{c1}^{em}(0)$ 
versus the superconductive critical temperature. The statistical errors correspond to the  68 \% confidence level.
 The dashed line is the best fit to the assumed linear relation between $H_{c1}^{em}(0)$ versus $T_c$.}
\end{figure}
We, now, show that this is indeed the case.
To this end, we fitted the experimental data for the temperature dependence of the lower critical magnetic field, displayed in
Fig.~\ref{Fig8}, to our theoretical results Eqs.~(\ref{3.138}) - (\ref{3.140}). In the fitting procedure we found the there is degeneracy
in the parameters $b$ and $b'$. So that we fixed $b \simeq 0.193$, as we did before, and taken $b'$, $T_c$, $H_{c1}^{em}(0)$,
 and $H_{c1}^{core}(0)$ as free parameters. We found  $b' \simeq 1.7$, $T_c \simeq 93.3$,   $H_{c1}^{em}(0) \simeq 710 \; Oe$, and
 $H_{c1}^{core}(0) \simeq 350 \; Oe$. These results confirm that the contribution of the vortex core energy to the lower critical magnetic
 field is, indeed, sizable. The results of our fit are displayed in  Fig.~\ref{Fig8}. Evidently, the agreement
between theory and experiment is rather good.  For completeness, we also show  the different temperature behavior of
 $H_{c1}^{em}(T)$ and $H_{c1}^{core}(T)$.  Note that the almost linear dependence on the temperature in $H_{c1}(T)$ is
 due to  $H_{c1}^{em}(T)$, which dominates at low enough temperatures. \\ 
Next, we looked at the data for the lower critical magnetic field in underdoped YBCO.  Ref.~\cite{Liang:2005} reports the lower critical magnetic
field for underdoped YBCO with critical temperatures varying between $8.9 \, K$ and $22 \, K$. The critical fields were determined by
measurements of magnetization in an applied magnetic field perpendicular to the copper-oxide planes. In Fig.~\ref{Fig9}, top panel,
we display the temperature dependence of the lower critical magnetic field for underdoped YBCO with critical temperatures
$T_c \simeq 8.9 \, K$, $13.7 \, K$, $15.3 \, K$, $17.6 \, K$,  $22.0 \, K$ respectively. The corresponding hole doping $\delta$ can be inferred from the
phenomenological relation Eq.~(\ref{3.94}). The hole doping level ranges from $\delta \simeq 0.055$ up to $\delta \simeq 0.064$.
The data have been extracted from Fig.~2 in  Ref.~\cite{Liang:2005}.
Even in the highly underdoped region the lower critical magnetic field seems to vary linearly with temperature, at least for temperatures
below $\sim 0.6 \, T_c$. In fact, the authors of   Ref.~\cite{Liang:2005}, performing  a linear fit in this temperature range, was able
to extract $H_{c1}(0)$.   They found that    $H_{c1}(0)$ was not a linear function of $T_c$. From the best fits  they found a non-linear
relation between the zero-temperature lower critical  magnetic field versus the critical temperature, 
$H_{c1}(0) = 0.366 \, T_c^{1.64 \pm 0.004}$~\cite{Liang:2005}. This last result looks puzzling. Indeed,  in conventional BCS
superconductors the effect of the vortex core energy is to slightly modify the logarithmic term $\ln \kappa$ in $H_{c1}(0)$. Therefore,
one expects that  $H_{c1}(0) \sim \lambda^{-2}(0) \sim n_s(0)$. Since $T_c \sim n_s(0)$ is the expected behavior if the critical temperature
is governed only by phase fluctuations in two dimensions, then  the above results would imply   $T_c \sim n_s(0)^{0.61}$, that is 
inconsistent with the observed linear relation between the critical temperature and the superfluid condensate 
density~\cite{Uemura:1989,Corson:1999}. However, we already remarked that the vortex core energy contribution to the lower
critical magnetic field cannot be neglected. On the other hand, it is evident that only the electromagnetic term $H_{c1}^{em}(0)$
needs to scale with   $\lambda^{-2}(0)$, while  $H_{c1}^{core}(0)$ is almost independent on the London penetration length.
To determine   $H_{c1}^{em}(0)$ we fitted the available data to our Eqs.~(\ref{3.138}) - (\ref{3.140}). We fixed the critical temperatures
to the measured values and let   $H_{c1}^{em}(0)$, $H_{c1}^{core}(0)$, and $b'$ be free fitting parameters. Moreover,
as in previous analysis, we fixed $b \simeq 0.193$. In Fig.~\ref{Fig9}, bottom panel, the dashed lines are the best-fitted curves.
We see that, in fact, the experimental data are in good agreement with our theoretical expectations.  We found that the parameter
$b'$ did not showed statistically significant dependence on the hole doping $\delta$, at least  in the rather narrow 
explored range of $\delta$, $b' \simeq 2.0$. Even in this case, we confirm that the vortex core energy contribution to
the lower critical magnetic field is not negligible. In fact, it turns out that    $H_{c1}^{core}(0) \sim   0.5 \, H_{c1}^{em}(0)$.
In Fig.~\ref{Fig9}, right  panel, we report the best-fitted values for   $H_{c1}^{em}(0)$ as a function of the critical superconductive 
temperature. Remarkably,  we found that, as expected,   $H_{c1}^{em}(0)$ scales linearly with $T_c$. The dashed line in
Fig.~\ref{Fig9}, right  panel, is the best fit straight line   $H_{c1}^{em}(0) \propto T_c$. Note that the best-fit straight line 
 does not extrapolate to the origin. This is due to the fact that the superconductive transition set in abruptly  for 
 $\delta \gtrsim \delta_{min} \simeq 0.05$, so that the critical temperature $T_c$ is not strictly proportional to $\delta$
 for hole doping too close to $\delta_{min}$. \\
 %
\begin{figure}
\vspace{0.8cm}
\centering
\resizebox{0.48\textwidth}{!}{%
\includegraphics{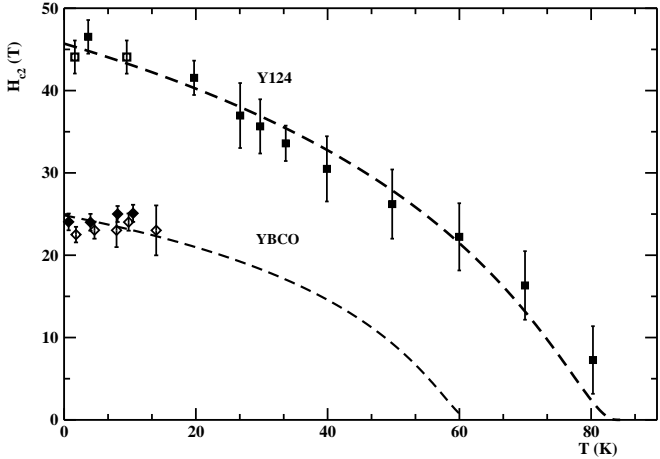} }
\caption{\label{Fig10} Temperature dependence of the upper critical magnetic field for two different compounds 
of cuprate superconductors. Full and open squares correspond to $H_{c2}(T)$ for Y123 ($\delta \simeq 0.14$).
The data  have been extracted from Fig.~3, panel b, in  Ref.~\cite{Grissonnanche:2014}. Full squares correspond roughly to
the upper boundary of the vortex-liquid phase. Open squares correspond to the rapid drop in the longitudinal thermal conductivity
versus the applied magnetic field. Full and open diamonds  correspond to $H_{c2}(T)$ for YBCO ($\delta \simeq 0.11$). 
The data  have been extracted from Fig.~1, panel a, in  Ref.~\cite{Grissonnanche:2015}. Full and open diamonds corresponds to
$H_{c2}(T)$  detected in the  Hall and longitudinal thermal conductivity respectively. The dashed  lines are the  fits  to 
Eq.~(\ref{3.141}) of the experimental measurements.}
\end{figure}
 Finally, we turn on the temperature dependence of the upper critical magnetic field. The upper critical magnetic field is essential to
 identify the factor that limit the strength of superconductivity. However, in hole doped cuprates the direct measurement
of the upper critical field is difficult for $H_{c2}(0)$ can attain values as high as $\sim 10^{2} \, T$. In 
Refs.~\cite{Grissonnanche:2014,Grissonnanche:2015}  it is suggested that the thermal conductivity can be used to directly detect
the upper critical magnetic field in YBCO, $YBa_2Cu_4O_8$ (Y124) and Tl-2201. In  Fig.~\ref{Fig10} we report the temperature
dependence of  $H_{c2}$ for Y124 and YBCO in the pseudogap region.
In our theory the temperature dependence of the upper critical magnetic field is given by   Eq.~(\ref{3.133}), which it is
useful to rewrite as:
\begin{equation}
\label{3.141}  
H_{c2}(T)  \; \simeq \;    H_{c2}(0) \;  e^{  - \, b' \; \left [ \frac{1}{\sqrt{1 \, - \, T/T_c}} \; - \; 1 \right ] } \; . 
\end{equation}
Indeed, we have fitted the experimental data to   Eq.~(\ref{3.141}).  It turned out that our   Eq.~(\ref{3.141})
allows to track the temperature dependence of the upper critical magnetic field   in a satisfying way. 
The resulting best-fit curves are shown in  Fig.~\ref{Fig10} as dashed lines. We
found $T_c \simeq 84.5 \; K, H_{c2}(0) \simeq 45.7 \; T, b' \simeq 0.88$ for Y124 and 
 $T_c \simeq 62.1 \; K, H_{c2}(0) \simeq 24.8 \; T, b' \simeq 0.80$ for YBCO. 
\subsection{Temperature dependence of the  critical current}
\label{s3.7}
The critical current density $j_c$ represents a fundamental quantity for the superconductor applications. Usually, the maximum 
limit  for the critical current is set by the depairing current. In high temperature superconductors the depairing critical
current density is considerably higher than in conventional superconductors. For instance, in YBCO films one finds 
for the  critical current density extrapolated to zero temperature the extraordinarily high value 
 $j_c(0) \sim 10 \, MA/cm^2$~\footnote{See Table 6.3 in Ref.~\cite{Wesche:2015}. Even though we are using cgs units, 
 it is widespread consuetude to measure the current density in $A/cm^2$.}. \\
For zero applied field the critical current densities in high temperature superconductors are
typically well described by the scaling law;
\begin{equation}
\label{3.142}  
j_{c}(T)  \; \simeq \;    j_{c}(0) \;  \left [  1 \, - \, \frac{T}{T_c}  \right ]^{\gamma}  \; . 
\end{equation}
The values of the exponent are typically  $\gamma \simeq  1  -  2$. According to the Ginzburg-Landau theory, the
depairing current density depends on the London penetration length $\lambda(T)$ and the Ginzurg-Landau
coherence length $\xi_{GL}(T)$. Using the empirical temperature dependence of the two characteristic lengths on finds
that near the critical temperature  $j_{c}(T)$ satisfies   Eq.~(\ref{3.142}) with~\cite{Tinkham:1996}
\begin{equation}
\label{3.143}  
\gamma_{GL} \; \simeq   \; \frac{3}{2} \; \; \; . 
\end{equation}
As we have already discussed, the electrical  current density is given by  Eq.~(\ref{3.66}):
\begin{equation}
\label{3.144} 
\vec{j}_{em}(\vec{R},T)  \; = \;  2  e \, n_s(T) \, \left [ 
 \vec{v}_s (\vec{R},T)  \; - \;  \frac{e}{ m^*_h c} \; \vec{A}( \vec{R}) \right ] \; ,
\end{equation}
where, we recall that $\vec{h}(\vec{R})  =  \nabla \times \vec{A}(\vec{R})$ and 
 $\left <  \vec{h}(\vec{R}) \right  >_{\text{vol}}   =  \vec{B}(\vec{R})$.
For zero microscopic magnetic field $\vec{h}(\vec{R})$ and using  Eq.~(\ref{3.53}) and  Eq.~(\ref{3.54}), we get:
\begin{eqnarray}
\label{3.145} 
\vec{j}_{em}(\vec{R},T)  \; \simeq \;  2  e \, n_s(0) \,   \vec{v}_s (\vec{R},0)  \;
\\ \nonumber
e^{  - \, b" \; \left [ \frac{1}{\sqrt{1 \, - \, T/T_c}} \; - \; 1 \right ] } \; \; , \; \;  b" \; = \; b  +  b' \; \;.
\end{eqnarray}
The critical depairing current density is attained when $v_s(0)$ equals the critical velocity $v_c$ given by  Eq.~(\ref{3.109}).
Accordingly, we obtain the depairing critical current density:
\begin{eqnarray}
\label{3.146} 
j_{c}(T)  \; \simeq \;  j_c(0) \;  e^{  - \, b" \; \left [ \frac{1}{\sqrt{1 \, - \, T/T_c}} \; - \; 1 \right ] } \; , 
\\ \nonumber
j_c(0) \; = \; 2  e \, n_s(0) \,  v_c   \; .
\end{eqnarray}
It is worthwhile to estimate the zero-temperature critical current density in terms of the model parameters. Explicitly, we
have:
\begin{equation}
\label{3.147} 
j_{c}(0)  \; \simeq \; \frac{e}{c_0 a_0^2} \;  \sqrt{\frac{\Delta_2(0)}{m^*_h}}   \; \;  \delta \,
 \left [ 1 \;  - \; (\frac{\delta}{\delta^*})^{1.5} \right ]^{\frac{1}{2}}  \; . 
\end{equation}
Using the numerical values of the parameters we  obtain the estimate:
\begin{equation}
\label{3.148} 
j_{c}(0)  \; \simeq \;  2.84 \; 10^3 \; \frac{MA}{cm^2}    \; \; \; \delta \,
 \left [ 1 \;  - \; (\frac{\delta}{\delta^*})^{1.5} \right ]^{\frac{1}{2}}  \; . 
\end{equation}
Eq.~(\ref{3.148}) shows that the zero-temperature critical current density has the same doping dependence as the upper
critical magnetic field. In the pseudogap region $\delta \lesssim \delta^*$ one finds  $j_{c}(0) \sim 10^2 \; MA/cm^2$, confirming
that  the critical depairing current density  in hole doped cuprate superconductors  can reach very high values. \\
%
%
\begin{figure}
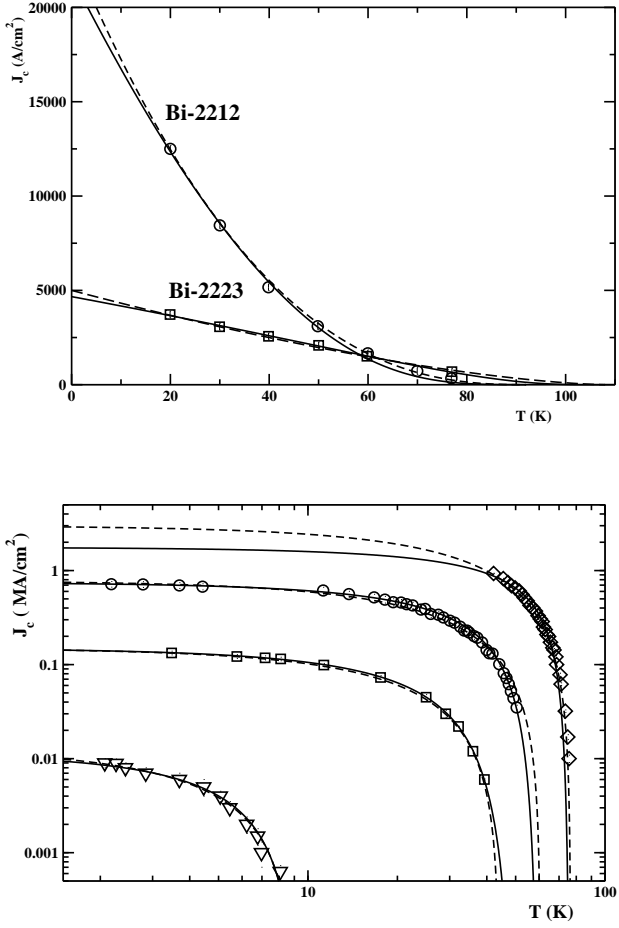

\vspace{0.8cm}
\centering
\resizebox{0.45\textwidth}{!}{%
\includegraphics{Fig11a.eps} }

\vspace{1.0cm}

\resizebox{0.45\textwidth}{!}{%
\includegraphics{Fig11b.eps}  }
\caption{\label{Fig11} (Top) Critical current density versus the temperature  for optimal doped
Bi-2212 (opne circles) and Bi-2223 (open squares). Data have been extracted from Fig.~3 in
Ref.~\cite{Fagnard:2010}. The dashed and solid lines are the fits to Eq.~(\ref{3.142}) and  Eq.~(\ref{3.146})
respectively.   (Bottom) Critical current density versus the temperature  for Bi-2212 with different hole doping fraction
 in the pseudogap region. Data have been taken  from Fig.~5, panel a,  in Ref.~\cite{Naamneh:2014}. 
The dashed and solid lines are the fits to Eq.~(\ref{3.142}) and  Eq.~(\ref{3.146}) respectively. }
\end{figure}
To extract a quantitative estimate of the parameter $b"$ we need to compare the temperature dependence of
the critical current density to observations. In Fig.~\ref{Fig11}  we display the temperature dependence of the
critical current density for the hole doped cuprate superconductors $Bi_2Sr_2CaCu_2O_8$ (Bi-2212) and
$Bi_{1.8}Pb_{0.26}Sr_2Ca_2Cu_3O_{10+x}$ (Bi-2223). In the top panel, the data correspond to the average
critical current density of Bi-2212 hollow cylinder ($T_c = 92 \, K$) and a tabular polycrystalline sample of 
Pb-doped Bi-2223 ($T_c = 108 \, K$)~\cite{Fagnard:2010}. By fitting the data to the Ginzburg-Landau
scaling law  Eq.~(\ref{3.142}), the authors of Ref.~\cite{Fagnard:2010} found $j_c(0) \simeq 2.3 \, 10^4 \, A/cm^2$,
$\gamma \simeq 2.5$  for Bi-2212, and   $j_c(0) \simeq 5.0 \, 10^3  \, A/cm^2$, $\gamma \simeq 1.5$  for Bi-2223.
The resulting best-fit curves are displayed in  Fig.~\ref{Fig11} as dashed lines. We performed the fit of data to
our  Eq.~(\ref{3.146}). The resulting fits, displayed as full lines, are practically indistinguishable from the 
Ginzburg-Landau power law. As concern the parameter $b"$, we found $b" \simeq 4.7 \; , 2.3$ for  Bi-2212
and   Bi-2223 respectively, while for the  zero-temperature critical current densities we found values consistent
with the Ginzburg-Landau fits.\\
In the bottom panel of   Fig.~\ref{Fig11}, we report the critical current density measurements for high quality
Bi-2212 thin films for different hole doping fraction reported in Ref.~\cite{Naamneh:2014}.  The data correspond
to critical temperature $T_c \simeq 9 \, K$, $44 \, K$, $60.5 \, K$, $76.5 \, K$. The corresponding hole doped fraction
$\delta$ has been determined by using the phenomenological relationship  Eq.~(\ref{3.94}) by assuming
$T_c^{max} \simeq  76.5 \, K$.  The authors of   Ref.~\cite{Naamneh:2014} found that the temperature dependence
of the critical current density can be reproduced by the Ginzburg-Landau  phenomenological power law
with $\gamma \simeq 1.5$ for all doping fraction $\delta$. Indeed, we have fitted the data to  Eq.~(\ref{3.142}) with
fixed  $\gamma = 1.5$ (see dashed lines in  Fig.~\ref{Fig11}, bottom panel). On the other hand, the fits of the data
to  Eq.~(\ref{3.146}), displayed as solid lines in   Fig.~\ref{Fig11}, bottom panel, track quite closely the Ginzurg-Landau
phenomenological pawer law. Moreover, we found   $b" \simeq  2.7$ almost independently on the doping fraction. 
It is worthwhile to observe that the best-fit parameter   $b"$  assumes quite different values for the same
compound. In fact, for  Bi-2212 hollow cylinder we obtained  $b" \simeq  4.7$ , while for  Bi-2212 thin films 
 $b" \simeq  2.7$. This can be easily understood within our model, for  $b"$ results from both the screening of the
 superfluid velocity and the depletion of the superconductive condensate fraction due to thermal proliferation
 of vortex-antivortex excitations. It is natural, therefore, to expect that this screening mechanism is more
 efficient in bulk material with respect to thin films. Note that this also explains naturally why critical
 current densities in superconducting films are much larger than in bulk material.  
\section{The  Nodal  Quasielectron  Liquid}
\label{s4}
It is now well established  that  hole doped high temperature cuprate superconductors in the  pseudogap region have 
 an electron-like Fermi surface occupying a small fraction of the Brillouin zone.
Indeed,  angle resolved photoemission studies (see  Refs.~\cite{Lynch:1999,Damascelli:2003,Campuzano:2004} and references therein) 
showed that low-energy excitations are characterized by Fermi arcs, namely truncate segments of a Fermi surface. Moreover, several recent
studies (see Refs.~\cite{Sebastian:2011,Sebastian:2012,Vignolle:2013,Sebastian:2015} and references therein)
reported unambiguous identification of quantum oscillations in high magnetic fields. Interestingly enough,
the measured low oscillation frequencies reveals a Fermi surface made of small pockets. In fact, from the Luttinger's 
 theorem~\cite{Luttinger:1960}   and the Onsager relation~\cite{Onsager:1952,Lifshitz:1956}  between the frequency and the 
 cross-sectional area of the  orbit (see, for example,  Refs.~\cite{Abrikosov:1972,Ashcroft:1976,Shoenberg:1984}), 
 it results that the area of the pocket correspond to about a few percent of the first Brillouin zone area
in sharp contrast to that of overdoped cuprates where the frequency corresponds to a large hole Fermi surface. 
 In addition, there is convincing evidence  of  negative Hall and Seebeck effects  which reveals that these pockets 
 are electron-like rather than hole-like.  Moreover, it turns out that these pockets are  associated with states near the nodal 
 region of the Brillouin zone.
 In I we provided some theoretical arguments to justify the occurrence of the nodal quasielectron liquid.
 Let us, briefly, recapitulate the main arguments presented in I, Sect.~4.4.  \\
 From the geometry of the $CuO_2$ planes we argued:
\begin{equation}
\label{4.1}  
\varphi(\vec{\rho}) \; = \;  \varphi( \rho,\theta) \; = \;   \varphi(\rho) \cdot \cos{(2 \theta)}    \; , 
\end{equation}
where the coordinate axis are directed along the $Cu-O$ bond directions. This is the most natural choice since the wavefunction is sizable along 
the $Cu-O$ bonds and vanishes at $\theta = \pm \frac{\pi}{4}$. Let us consider the Fourier transform of the wavefunction: 
\begin{equation}
\label{4.2}  
\tilde{\varphi}( \vec{k}) \; = \;  \int d \vec{\rho} \;  \varphi( \rho,\theta) \; \exp{ (i \vec{k} \cdot \vec{\rho}) }   \; . 
\end{equation}
Assuming that the $k_x$, $k_y$ axis are oriented along the copper-oxygen bond directions, one readily obtains:
\begin{equation}
\label{4.3}  
\tilde{\varphi}(k, \theta_k ) \; = \; \tilde{\varphi}(k) \,  \cos{(2 \theta_k)} 
\end{equation}
where:
\begin{equation}
\label{4.4}  
\tilde{\varphi}(k) \; = \; -  2 \pi  \int_0^{\infty} d \rho  \, \rho \,   \varphi(\rho) \, J_2(k\rho) \; . 
\end{equation}
We see, thus, that the Fourier transform of the wavefunction vanishes along the {\it nodal} directions:
\begin{equation}
\label{4.5}  
 \theta_k \; = \;  \pm \; \frac{\pi}{4} \; , 
\end{equation}
while it is sizable along the {\it antinodal} directions:
\begin{equation}
\label{4.6}  
 \theta_k \; = \;  0 \; , \; \pm \; \frac{\pi}{2}  \; . 
\end{equation}
 Even though  the pair wavefunction vanishes along the nodal directions $k_x = \pm k_y$ ($\theta_k = \pm \frac{\pi}{4}$), 
 there are not nodal low-lying hole excitations. This is due to the fact that the pairing of the holes is in the real space and not in momentum 
 space~\cite{Sakai:2013}. On the other hand, we may freely perform rotations of the pairs without spending energy since this  
 modify only the phase of pair wavefunction.  The rigid rotations of pairs is equivalent to hopping of electrons according to the 
 hopping term in the Hamiltonian Eq.~(\ref{1.1}): 
\begin{equation}
\label{4.7}
\hat{H}_0^{(e)}  \; = \; -t \; \sum_{<i,j>,\sigma}   \left [  \hat{c}^{\dagger}_{i,\sigma}  \, \hat{c}_{j,\sigma}  \,  + \,   \hat{c}^{\dagger}_{j,\sigma}  \, \hat{c}_{i,\sigma} \right]  \; .
\end{equation}
We may diagonalize this Hamiltonian obtaining:
\begin{equation}
\label{4.8} 
\hat{H}_0^{(e)}    \; = \;    \; \sum_{\vec{k},\sigma} \;  \varepsilon_{\vec{k}}^{(e)}   \;  \;  \hat{\psi}^{\dagger}_{e}(\vec{k},\sigma)  \;   \hat{\psi}_{e}(\vec{k},\sigma) \; .
 \end{equation}
In the small-k limit we have:
\begin{equation}
\label{4.9} 
\varepsilon_{\vec{k}}^{(e)}  \; \simeq  \;   \frac{\hbar^2 \, \vec{k}^2}{2 \, m^*_e}   \; ,
 \end{equation}
where:
\begin{equation}
\label{4.10} 
m^*_e \; = \;  \frac{\hbar^2}{2  \, t  \, a_0^2} \; \simeq \; 2.17 \; m_e \;  . 
\end{equation}
Since there are $1 - \delta$ electrons per $Cu$ atoms, from the Hamiltonian Eq.~(\ref{4.8}) we may determine the electron Fermi energy:
\begin{equation}
\label{4.11} 
\varepsilon_{F}^{(e)}  \; = \;  \frac{\hbar^2 \, (\vec{k}_F^{(e)})^2}{2 \, m^*_e}  \;  \; \; , \; \; \;  a_0 \; k_F^{(e)} \; \simeq \;\sqrt{ 2 \pi (1 - \delta)}   \;  . 
\end{equation}
%
\begin{figure}
\vspace{0.7cm}
\hspace{0.62cm} 
\resizebox{0.44\textwidth}{!}{%
\includegraphics{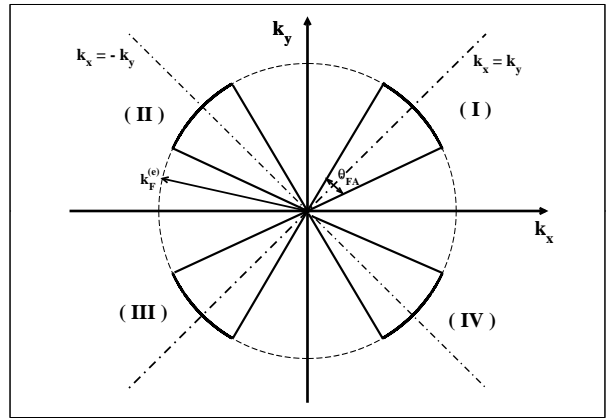} }
\caption{\label{Fig12} The quasielectron Fermi sector and Fermi arcs  in the first Brillouin zone. For later convenience, the
we have labelled the four quadrant of the Brillouin zone.}
\end{figure}
%
At first glance, one expects that the quasielectrons fill in momentum space the circle with radius $ k_F^{(e)}$ (the electron Fermi surface). 
However, one should keep in mind that the hopping of electrons is possible thanks to the paired holes. Since in momentum space the wavefunction of
 a given pair is spread over a region around  $k \sim \frac{1}{\xi_0}$, we see that the wavefunction of quasielectrons is likewise localized on a region in k-space 
 around $\frac{1}{\xi_0}$. Thus, the quasielectrons do not have the needed coherence to propagate over large distances with a well defined momentum. 
 However, this argument does not apply along the nodal directions where the 
momentum-space hole pair wavefunction vanishes. Therefore we are led to the conclusion  that there are coherent quasielectrons that fill small circular 
sectors of the electron Fermi circle around the nodal directions  $k_x = \pm k_y$. Thanks to the rotational symmetry,
we have four circular sectors with the same area. Since the number of coherent quasielectrons is determined by the doping fraction
of holes  (assuming that all the holes are paired), we obtain (see Fig.~\ref{Fig12}):
\begin{equation}
\label{4.12} 
\frac{\delta}{a_0^2} \;  \simeq \;  \frac{4 \times 2}{(2 \pi)^2 } \; \frac{1}{2} \;  (k_F^{(e)})^2 \; \theta_{FA} \; ,
\end{equation}
namely
\begin{equation}
\label{4.13} 
 \theta_{FA} \; \simeq \; \frac{\pi}{2} \; \frac{\delta}{1 - \delta}  \;  \; .
\end{equation}
We see, thus, that the Fermi surface is made by four Fermi arcs in qualitative  agreement with the angle resolved photoemission data. 
Moreover, the area of the Fermi sector with respect to the area of the Fermi circle in overdoped region turns out to be:
\begin{equation}
\label{4.14} 
\frac{{\cal{A}}_{FA}}{ {\cal{A}}_{overdoped}}\; \simeq \; \frac{1}{4} \; \frac{\delta}{1 + \delta}  \;  \; .
\end{equation}
For the typical value of hole doping fraction $\delta \simeq 0.1$, we infer from Eq.~(\ref{4.14}) that, indeed, 
${\cal{A}}_{FA}$ is about $ 2.3 \; \%$ of the first Brillouin zone area in the overdoped region, in satisfying agreement with quantum oscillation experiments. 
\\
It is useful to  emphasize that the nodal quasielectron low-lying excitations are basically controlled by the pseudogap. Therefore,
we expect that these excitations would be present up to the pseudogap temperature $T^*$. However, our previous arguments
rely on the possibility to freely vary the phase of the hole pair wavefunction without spending energy. Therefore, we see that
nodal quasielectron excitations are present only in the disordered phase of the pair condensate. Finally, we must
admit that  the discussion on the origin of the nodal quasielectron is somewhat qualitative and cannot be considered
a truly microscopic explanation. However, in the spirit of our approach,  the roughness of these arguments
are nevertheless corroborated  by the clear experimental evidence of the nodal quasielectron
liquid in the pseudogap region of hole doped cuprate superconductors.
\subsection{The nodal gap}
\label{s4.1}
\begin{figure}
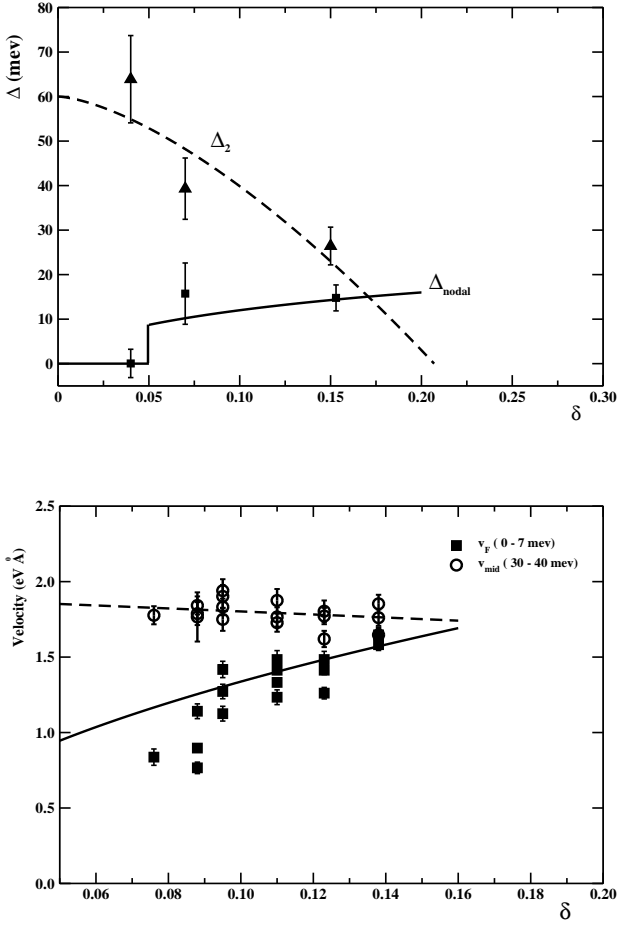

\vspace{0.8cm}
\centering
\resizebox{0.45\textwidth}{!}{%
\includegraphics{Fig13a.eps} }

\vspace{1.0cm}

\resizebox{0.45\textwidth}{!}{%
\includegraphics{Fig13b.eps}  }
\caption{\label{Fig13} (Top) Doping dependence of the pseudogap and nodal gap. The data refer to  lightly to optimally doped LSCO and have
been extracted from Fig.~2, panel c, in Ref.~\cite{Yoshida:2009}. The solid line is Eq.~(\ref{4.25}), the dashed line is  Eq.~(\ref{2.14}) assuming
$\Delta_2(0) \simeq 60 \, mev$.
  (Bottom) Doping  dependence of the low- and high-energy nodal Fermi velocities extracted from momentum distribution analysis in 
  underdoped Bi-2212.  The data have been taken form Fig.~3, panel a) of  Ref.~\cite{Vishik:2010a}. The solid and dashed lines are our
 Eq.~(\ref{4.27}) and Eq.~(\ref{4.29}) respectively.}
\end{figure}

The most extensive investigation of excitation gaps in high temperature cuprate superconductors has been done
by angle resolved photoemission spectroscopy (ARPES)~\cite{Lynch:1999,Damascelli:2003,Campuzano:2004}.
This technique  generally gives  informations on the electronic structure of a material.
 In the high temperature cuprate superconductors, due to the quasi-two dimensionality of the electronic structure,
ARPES studies  permit  to unambiguously determine the momentum of the initial state from the measured final state momentum,
since the component parallel to the surface is conserved in photoemission (note that the photon momentum can be
neglected at the low photon energies typically used in experiments). From the detailed momentum dependence of
the excitation gap along the Fermi surface contour, these studies suggested the coexistence of two distinct spectral
gap components, namely the pseudogap and another gap which was identified with the superconductive 
gap~\cite{Tanaka:2006,He:2009,Pushp:2009,Yoshida:2009,Hashimoto:2010,Reber:2012,Yoshida:2012,Hashimoto:2012,Hashimoto:2014,Hashimoto:2015}
(see, also, Ref.~\cite{Hufner:2008} and references therein). There are two main motivations which led to identify the second gap detected in ARPES
experiments with the superconductive gap. First, the gap were sizable in the nodal regions in momentum space. Second, the
gap closes at the critical temperature $T_c$. Therefore, it was natural to consider the gap as the d-wave BCS gap.
However, this point of view is in striking contrast with the substantial experimental evidence that the superconductive transition
is driven by the B-K-T transition, for in this case there is not any superconductive gap to play with. In addition, we have already seen
that the temperature dependence of the London penetration length within the weak coupling d-wave BCS theory is in
undeniable disagreement with the experimental data. Indeed, now we shall show that the gap detected in ARPES
studies, which will be referred to as nodal gap, is not the superconductive gap but, nevertheless, it is related   intimately
to the low-lying excitations of the hole pair superfluid condensate. \\
In ARPES experiments the superfluid condensate is excited by the incoming photons. We saw in Sect.~\ref{s3.1} that
the low-lying condensate excitations are rotons which behave like elementary quasiparticle with mass $2 m_h^*$ and charge
$2e$. Since the in-plane photon momentum is negligible, naively one expects that the lowest energy excitations would be
two rotons with opposite superfluid velocity. However, we known that  $\varepsilon_{roton} \sim \Delta_2$, so that
these kind of configurations are lying  at very high energies. Moreover, the roton-antiroton pairs do not give rise directly to
photoelectrons and, therefore, they can be hardly detected in ARPES experiments.  On the other hand, if we take into account
that in the vortex core region the condensate is phase disordered, then we see that, according to the discussion in Sect.~\ref{s4},
a moving roton can excite nodal quasielectrons in such a way that the roton momentum can be compensated. In other words,
we are led to consider condensate configurations corresponding to a moving roton and  a few of nodal quasielectrons with
opposite velocities.  We have seen that the energy of the nodal quasielectrons is:
\begin{equation}
\label{4.15} 
\varepsilon_{\vec{k}}^{(e)}  \; \simeq  \;   \frac{\hbar^2 \, \vec{k}^2}{2 \, m^*_e}   \; .
 \end{equation}
Eq.~(\ref{4.15}) is relevant in the superfluid condensate rest frame. If the condensate has velocity $\vec{v}_s$, then,
following the well-known Landau's arguments (see, eg, Ref.~\cite{Lifshitz:1980}), we have:
\begin{equation}
\label{4.16} 
\varepsilon_{\vec{k}}^{(e)}  \; \simeq  \;  \frac{1}{2} \, m^*_e \left (\vec{v}_e \; + \; \vec{v}_s \right )^2 \; \simeq \;
  \frac{\hbar^2 \, \vec{k}^2}{2 \, m^*_e}  \; + \;  \hbar \,  \vec{k} \cdot  \vec{v}_s \; + \;    
  \frac{1}{2} \, m^*_e \vec{v}_s^{\,2}  \; .
 \end{equation}
The low-lying excitation spectrum of the nodal quasielectrons is determined by:
\begin{equation}
\label{4.17} 
\varepsilon_{\vec{k}}^{(e)}  \; -  \varepsilon_{F}^{(e)} \; \ge \; 0  \; \; .
 \end{equation}
Combining Eqs.~(\ref{4.16}) and (\ref{4.17}), and observing that $\cos \theta_{\vec{k}\vec{v}_s}$ $\simeq - 1$ since the total
momentum must vanish, we obtain readily:
\begin{equation}
\label{4.18} 
\varepsilon_{\vec{k}}^{(e)}  \; -  \varepsilon_{F}^{(e)} \; \gtrsim  \;   \hbar \,  k \; v_s \; - \;    
  \frac{1}{2} \, m^*_e v_s^{\,2}    \; \; .
 \end{equation}
For $ k \, \approx \, k_F^{(e)}$ and using  Eq.~(\ref{3.43}) one can check that  the second term on the right hand side of
Eq.~(\ref{4.18}) is negligible with respect to the first term. Therefore we obtain that near the Fermi surface the nodal quasielectron
spectrum is gapped:
\begin{equation}
\label{4.19} 
\varepsilon_{\vec{k}}^{(e)}  \; -  \varepsilon_{F}^{(e)} \; \gtrsim  \;   \hbar \,  k \; v_s  \; \; , \; \; 
k \;  \gtrsim   k_F^{(e)} \; \; .
 \end{equation}
This allow to introduce the nodal gap:
\begin{equation}
\label{4.20} 
\Delta_{nodal} \; \simeq \;   \hbar \,  k_F^{(e)} \; v_s  \; \; .
\end{equation}
Evidently, the nodal gap determines the energy gap size at the Fermi surface. Nevertheless, from Eq.~(\ref{4.19}) we see that
the quasiparticle excitation energy dispersion  is linear in the wavenumber $k$ with slope  $\hbar \,  v_s $.   The temperature 
dependence of the nodal gap is mainly due to the thermal screening of the superfluid velocity as implied by  Eq.~(\ref{3.53}): 
\begin{equation}
\label{4.21} 
\Delta_{nodal}(T) \;  \simeq \;  \Delta_{nodal}(0)  \;   e^{  - \, b \; \left [ \frac{1}{\sqrt{1 \, - \, T/T_c}} \; - \; 1 \right ] } \; \; , \; \; 
 T    \; \lesssim  \; T_c \;  \;  ,
\end{equation}
where:
\begin{equation}
\label{4.22} 
\Delta_{nodal}(0) \; \simeq  \;  \hbar \,  k_F^{(e)} \; v_s(0)  \; \; .
\end{equation}
Before comparing  our results to experimental data, it is useful  to determine the doping dependence of the nodal gap. This can be easily
accomplished using our previous results on the superfluid velocity of rotons, Eq.~(\ref{3.43}), and on the Fermi wavenumber of
nodal quasielectrons,  Eq.~(\ref{4.11}):
\begin{eqnarray}
\label{4.23} 
\Delta_{nodal}(0) \; \simeq  \;  \frac{\hbar}{a_0} \, \sqrt{ 2 \pi (1 - \delta)}  \; \frac{\hbar}{2 m_h^*}  \frac{2 \pi}{d_0}  \; 
\\ \nonumber
\simeq \;    \pi^{\frac{3}{2}} \; \frac{\hbar^2}{ m_h^* a_0^2} \; \sqrt{\delta (1 \, - \, \delta)} \; .
\end{eqnarray}
A few comments are in order. Firstly, the nodal gap is sizable in the nodal regions in momentum space  $k_x \simeq  \pm k_y$, for
the nodal quasielectron excitations are present only  there. The doping dependence of the nodal gap is quite different from
the pseudogap and the critical temperature $T_c$. In fact, the pseudogap decreases with increasing doping according
to  Eq.~(\ref{2.14}), while the critical temperature scales roughly linearly with the hole doping $\delta$ (see   Eq.~(\ref{2.19})).
This has interesting consequences. In fact, let us consider the ratio between the nodal gap and the critical temperature:
\begin{equation}
\label{4.24} 
\frac{\Delta_{nodal}(0)}{k_B T_c}  \; \sim  \; \sqrt{\frac{1 \, - \, \delta}{\delta}} \; .
\end{equation}
If one identifies the nodal gap with the weak coupling d-wave BCS gap, then this ratio should be $\sim 2.14$ 
(see Eq.~(\ref{A.16}))  independently on the hole doping.  In contrast,  Eq.~(\ref{4.24}) implies that that ratio is doping dependent
and it increases roughly as $\sim 1/\sqrt{\delta}$ in the highly underdoped region, in qualitative agreement with observations. \\
It should be evident that our interpretation of the nodal gap being highly unconventional  needs to be compared
both qualitative and quantitatively with observations. We attempted several checks  to test both the  dependence 
on temperature and on hole doping of the nodal gap by comparing  to available  ARPES studies in literature. Here we present
some illustrative examples of such comparisons. In Ref.~\cite{Yoshida:2009} it was investigated the doping dependence of the
pseudogap and nodal gap in high-quality single crystals of  LSCO by angle resolved photoemission spectroscopy. The hole doping fractions
were $\delta = 0.03, 0.07, 0.15$. For $\delta = 0.07, 0.15$ the critical temperatures were $T_c \simeq 14 \, K, 39 \, K$ respectively.
The sample with hole doping $\delta = 0.03$ was not superconducting. In Fig.~\ref{Fig13}, top panel,  we display the doping dependence of
the pseudogap $\Delta_2(\delta)$ and the nodal gap $\Delta_{nodal}(\delta)$.  The data  have been extracted from Fig.~2, panel c, 
in Ref.~\cite{Yoshida:2009}. As regard the pseudogap, we compare the data to  Eq.~(\ref{2.14}) by taking $\Delta_2(0) \simeq 60 \, mev$.
Note that this value for  $\Delta_2(0)$ is in reasonable agreement with the estimate obtained with the numerical
values of the model parameters. The doping dependence of the nodal gap is given by Eq.~(\ref{4.23}). Accordingly, in  Fig.~\ref{Fig13},
top panel, we display 
\begin{equation}
\label{4.25} 
\Delta_{nodal} \; \simeq \;    40 \; mev \;  \; \sqrt{\delta (1 \, - \, \delta)} \; \; , \; \; \delta \; \gtrsim  \; \delta_{min} \; \simeq \; 0.05 \; ,
\end{equation}
where the constant has been fixed to match the data. For  $\delta  < \delta_{min}$ the nodal gap vanishes since, as we said,
it is related to the low-lying excitations of the superfluid condensate. We see that the data are in satisfying agreement with
our theoretical expectations. Nevertheless, it must be mentioned that the estimate of the constant in  Eq.~(\ref{4.25})
with the model parameters gave a value higher by about one order of magnitude. This could be due to the fact that
in LSCO the $T_c^{max} \sim 40 \, K$, while in our model  $T_c^{max} \sim 100 \, K$ (see Fig.~\ref{Fig2}). Moreover,
the data for the nodal gap are taken at finite temperature where the gaps are somewhat smaller with respect
to the zero-temperature values. \\
We have seen that the nodal quasielectron excitation energy dispersion  is linear in the wavenumber $k$ with slope  $\hbar \,  v_s $. 
Following the practice common among experimental physicists this slope is called nodal Fermi velocity $v_F$. Note that
$v_F$ does not have  the dimension of a velocity but its dimension is energy $\times$ length.  According to our previous
discussion, we find that the nodal Fermi velocity depends on hole doping according to:
\begin{equation}
\label{4.26} 
v_F  \; \simeq  \;   \hbar \,  v_s \; \simeq \;  \frac{\pi}{\sqrt{2}} \;  \frac{\hbar^2}{m_h^* a_0} \, \sqrt{\delta}   \; .
\end{equation}
In fact, a doping dependent nodal Fermi velocity has been revealed by high-resolution ARPES~\cite{Vishik:2010a}. 
In  Fig.~\ref{Fig13}, bottom panel, we report the doping dependence of the nodal Fermi velocity  obtained in   Ref.~\cite{Vishik:2010a}
by high-resolution ARPES spectra for Bi-2212 with doping in the range $0.076$  $< \delta < 0.14$.  The nodal Fermi velocity
was obtained from the slope of the excitation energy spectra in the range $0 - 7 \, mev$. The data for $v_F$ are contrasted to
Eq.~(\ref{4.26}). More precisely, we compare the data with:
\begin{equation}
\label{4.27} 
v_F  \; \simeq  \;   4.22 \; mev \,  {\text \AA } \; \sqrt{\delta}   \; .
\end{equation}
Even in this case, we found that the estimate of the constant  with the model parameters produces a value
about a factor $\sim 5$ smaller than  in   Eq.~(\ref{4.27}). Nevertheless, we see that the puzzling doping dependence of
the nodal Fermi velocity is in fair agreement with our model calculations. Interestingly, the authors of   Ref.~\cite{Vishik:2010a}
reported that to fully characterize the ARPES spectra it was needed two different velocities. In fact, they introduced the
nodal Fermi velocity $v_{mid}$ defined by the linear fits of the spectra in the energy range  $30 - 40 \, mev$. Looking
at  Fig.~\ref{Fig13}, bottom panel, it is evident that  $v_{mid}$ does not display a strong doping dependence, while $v_F$
has a pronounced  dependence on doping. Since the nodal Fermi velocity is extracted from the excitation spectrum
in an energy range well above the the nodal gap, we suggest that the high-energy spectrum is due to  unconstrained 
nodal quasielectrons. If this is the case, then the relevant Fermi velocity would be:
\begin{equation}
\label{4.28} 
v_{mid}  \; \simeq  \;  \frac{\hbar^2 k_F}{m_e^*} \; \simeq \;  \frac{\hbar^2}{m_e^* a_0} \;    \sqrt{1 \, - \, \delta}   \; .
\end{equation}
%
%
%
In fact, in Fig.~\ref{Fig13}, bottom panel,  we see that the data can be accounted for quite accurately by:
\begin{equation}
\label{4.29} 
v_{mild}  \; \simeq  \;   1.90 \; mev \,  {\text \AA } \;  \sqrt{\delta}   \; .
\end{equation}
We would like to note that, this time,  the estimate of the constant  with the model parameters produces a value in fair 
agreement with the value in Eq.~(\ref{4.29}). \\
%
%
\begin{figure}
\vspace{0.8cm}
\centering
\resizebox{0.5\textwidth}{!}{%
\includegraphics{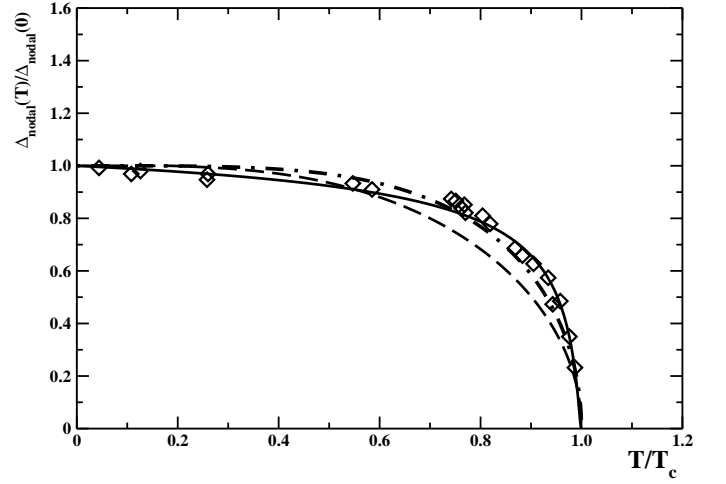} }
\caption{\label{Fig14} Temperature dependence of the nodal gap for Bi-2223  as extracted from Andreev reflection spectroscopy. The data 
have been taken from Fig.~2 of Ref.~\cite{Svistunov:2000}. The solid line is our Eq.~(\ref{4.21})  with $b \simeq 0.193$, the dashed 
line is the  weak coupling d-wave BCS gap, Eq.~(\ref{A.33}), and the dot-dashed line is the phenomenological relation, 
Eq.~(\ref{4.30}). }
\end{figure}
%
We turn, now, on the temperature dependence of the nodal gap. In Ref.~\cite{Svistunov:2000} it was presented spectroscopic
studies of the cuprate superconductor compound Bi-2223  in the optimal doped region. These authors employed both
scanning tunneling spectroscopy~\cite{Fischer:2007} and Andreev-Saint-James reflection spectroscopy~\cite{Deutscher:2005}.
The tunneling spectroscopy, that in principle is sensitive to any gap in the excitation spectra, detected a rather large gap
which, however, did not reveal any appreciable temperature dependence. On the contrary,  it was reported a qualitatively different temperature
dependence of the energy gap observed in superconductor-normal-superconductor junctions that unequivocally exhibited
Andreev reflections. Remarkably,  the  data for the temperature dependence of the gap detected in Andreev spectroscopy were in reasonable 
agreement with the phenomenological BCS-like relation:
\begin{equation}
\label{4.30} 
\frac{\Delta_{nodal}(T)}{\Delta_{nodal}(0)}  \;  \simeq   \;  \sqrt{ 1 \; - \; \left ( \frac{T}{T_c}  \right )^4 }  \;  \;  .
\end{equation}
Indeed, in Fig.~\ref{Fig14} we report the temperature dependence of the energy gap detected with the  Andreev spectroscopy.
The data has been extracted from  Fig.~2 of Ref.~\cite{Svistunov:2000}.  We compare the data with the phenomenological
relation  Eq.~(\ref{4.30}) (dot-dashed line). We see, in fact, that this relation accounts quite well the temperature dependence
of the energy gap. On the other hand, the temperature behavior of the the weak-coupling BCS gap (see  Eq.~(\ref{A.33})
in Appendix~\ref{AppendixA}) is seen to deviate from the data. In fact, the BCS d-wave gap   underestimates 
 the gap for $T/T_c \gtrsim 0.7$ (see the dashed line in Fig.~\ref{Fig14}). \\
 Our interpretation of the puzzling results presented in   Ref.~\cite{Svistunov:2000} is as follows.
The tunnel data are sensitive to the pseudogap, but cannot detect easily the nodal gap since one does not have excitations 
of the condensate in tunneling spectroscopy experiments. On the other hand, Andreev reflection spectroscopy
is sensitive to the low-lying excitations. This implies that the Andreev gap must be identify with the nodal gap.
According to our theory, the pseudogap does not display a pronounced temperature dependence  for 
temperatures within the critical temperature since $T_c < T^*$. Moreover, the temperature dependence
of the nodal gap is given by  Eq.~(\ref{4.21}). Accordingly, we fitted the  Andreev gap data to our   Eq.~(\ref{4.21})
leaving $b$ as free fitting parameter. Remarkably, we found a very good fit with $b \simeq 0.193$. Indeed, the fitted curve,
displayed as solid line in    Fig.~\ref{Fig14},  follows quite closely the phenomenological law Eq.~(\ref{4.30}).
Nevertheless,  it seems that our best fit compare with the data slightly better  than  Eq.~(\ref{4.30}). \\
We, also,  tried several comparisons between   Eq.~(\ref{4.21}) and more recent data in the literature. Here, we merely present 
our results  for two different ARPES studies of the nodal energy gap. In  Ref.~\cite{Lee:2007} it was presented the temperature
dependence of the nodal energy gap extracted from spectra taken by angle resolved photoemission spectroscopy  for
underdoped Bi-2212 ($T_c \simeq 92 \, K$).
%
\begin{figure}
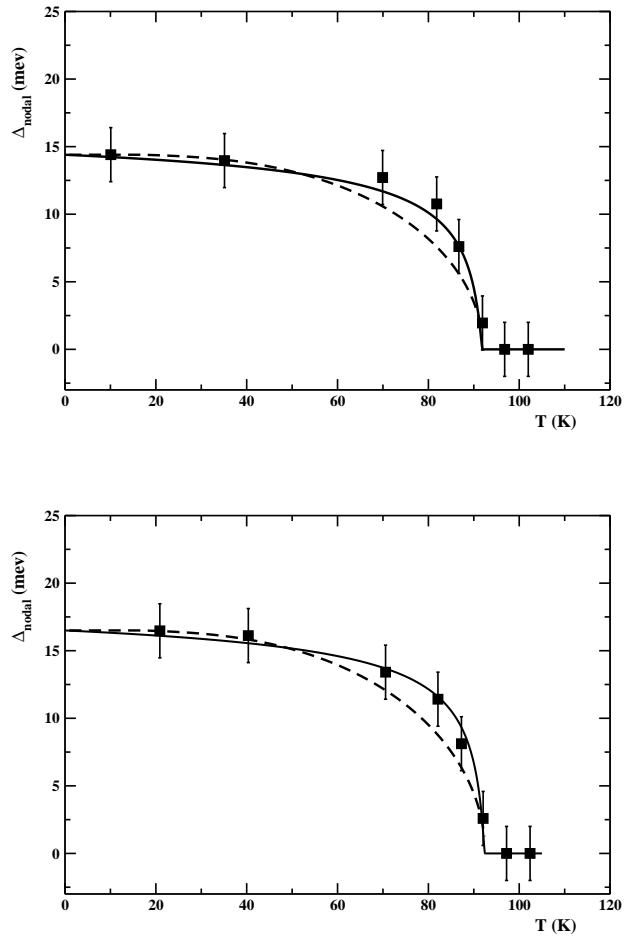

\vspace{0.8cm}
\centering
\resizebox{0.45\textwidth}{!}{%
\includegraphics{Fig15a.eps} }

\vspace{1.0cm}

\resizebox{0.45\textwidth}{!}{%
\includegraphics{Fig15b.eps}  }
\caption{\label{Fig15} (Top) Temperature dependence of the nodal gap for Bi-2212 as extracted from angle resolved photoemission spectroscopy.
 The data have been taken from Fig.~2, panel d, in Ref.~\cite{Lee:2007}. The solid and dashed curves are Eq.~(\ref{4.21})  with $b \simeq 0.193$ 
 and the BCS d-wave gap respectively,  assuming $\Delta_{nodal}(T=0) \simeq 14.4 \; mev$ and $T_c \simeq 92.0 \; K$.
 (Bottom) Temperature dependence of the nodal gap for Bi-2212 as extracted from angle resolved photoemission spectroscopy. The data have 
 been taken from Fig.~7, panel d, in Ref.~\cite{Vishik:2010b}. The solid and dashed curves are  Eq.~(\ref{4.21})  with $b \simeq 0.193$  and 
 the BCS d-wave gap respectively  assuming $\Delta_{nodal}(T=0) \simeq 16.5 \; mev$ and $T_c \simeq 92.5 \; K$.}
\end{figure}
 In Fig.~\ref{Fig15}, top panel, we display the nodal energy gap versus the temperature. The
data correspond to those  in  Ref.~\cite{Lee:2007},  Fig.~2, panel d. We compare the data to the weak coupling d-wave BCS
relation by taking  $\Delta_{nodal}(T=0) \simeq 14.4 \; mev$ and $T_c \simeq 92.0 \; K$. Fig.~\ref{Fig15} show that the BCS
relation is in fair agreement with the data mainly due to  the large experimental uncertainties.  However, one sees that
Eq.~(\ref{4.21}),  with $b \simeq 0.193$, compares better to the data.  Finally, we considered the angle resolved photoemission
spectroscopy study for the same compound   Bi-2212 with a slightly different critical temperature ($T_c \simeq 92.5 \, K$).
The data for the nodal energy gap, displayed in Fig.~\ref{Fig15},  bottom panel, have been extracted from  Fig.~7, panel d, in 
Ref.~\cite{Vishik:2010b}.  Again, we see that our nodal gap  Eq.~(\ref{4.21}) compares moderately better than 
 the BCS d-wave gap.
\subsection{Thermodynamics of the nodal quasielectron  liquid }
\label{s4.2}
We intend to calculate the thermal properties of the nodal quasielectron liquid. Throughout this Section we shall take
the nodal liquid as a free and independent  gas of quasiparticles. Here, we are explicitly neglecting corrections
due to Coulomb electron-electron or electron-ion interactions. The effects of the interactions with the lattice ions
will be discussed in Sect.~\ref{s5}.  Coulomb  interactions  are the main mechanism for electron-electron Umklapp scatterings.
These processes are relevant mainly in transport phenomena which, however, are beyond the aims of the present paper. \\
When the temperature is not zero the thermal distribution of nodal quasielectron is the Fermi-Dirac distribution:
\begin{equation}
\label{4.31}  
f(\varepsilon^{(e)}_{\vec{k}})  \; =  \;  \frac{1}{e^{\frac{\varepsilon^{(e)}_{\vec{k}} - \mu^{(e)}(T)}{k_BT} } \; + \;1 }   \; \;  ,
\end{equation}
where $\mu^{(e)}(T)$ is the quasielectron chemical potential. At zero temperature   $\mu^{(e)}(0) = \varepsilon_{F}^{(e)} $.
The temperature corrections to the quasielectron chemical potential can be obtained from:
\begin{equation}
\label{4.32}  
n^{(e)} \; \simeq \;    \frac{\delta}{a_0^2} \;      \; = \;   2 \; \int_{FA} \; \frac{d \vec{k}}{(2 \pi)^2} \; 
  \frac{1}{e^{\frac{\varepsilon^{(e)}_{\vec{k}}- \mu^{(e)}(T)}{k_BT} } \; + \;1 }   \; ,
\end{equation}
where the angular integration is performed over the four Fermi arcs (FA), see Fig.~\ref{Fig12}.
Introducing the density of state per spin at the Fermi energy:
\begin{equation}
\label{4.33} 
{\cal{N}}(0) \; \equiv \;  \int_{FA} \; \frac{d \vec{k}}{(2 \pi)^2} \;   \delta \left [ \varepsilon_{\vec{k}}^{(e)}   -   \varepsilon_F^{(e)} \right ] \; 
\simeq \; \frac{m^*_e}{\pi^2 \hslash^2} \, \theta_{FA}
 \; ,
\end{equation}
Eq.~(\ref{4.32}) can be rewritten as:
\begin{equation}
\label{4.34}  
n^{(e)}   \; = \;   2  \; {\cal{N}}(0)  \; \int_{0}^{\infty} \;  d \varepsilon \; 
  \frac{1}{e^{\frac{\varepsilon  - \mu^{(e)}(T)}{k_BT} } \; + \;1 }   \; \; .
\end{equation}
It is easy to check that:
\begin{equation}
\label{4.35}  
 \int_{0}^{\infty} \;  d \varepsilon \; 
  \frac{1}{e^{\frac{\varepsilon  - \mu^{(e)}(T)}{k_BT} } \; + \;1 }   \;  =  \;  \mu^{(e)}(T) \; + \; {\cal{O}}(e^{ - \frac{\mu^{(e)}(T)}{k_BT} })
 \; \;   .
\end{equation}
Therefore, neglecting the exponentially small thermal corrections, we find  $ \mu^{(e)}(T) \simeq  \varepsilon_{F}^{(e)} $.
We are interested in the calculation of the electronic contribution to the specific  heat. Actually, experimentally
one generally measures the specific heat at constant pressure. However, it is known that for temperatures
such that $k_B T \ll   \varepsilon_{F}^{(e)}$ the specific heat at constant pressure is practically coincident
with the specific heat at constant volume. The quasielectron contribution to the constant-volume specific heat is:
\begin{equation}
\label{4.36}  
c^{(e)}(T) \; = \; \frac{\partial \, \mathfrak{u}^{(e)}(T)}{\partial T}  \; \;  ,
\end{equation}
 $\mathfrak{u}^{(e)}(T)$ being the electronic energy density:
\begin{equation}
\label{4.37}  
\mathfrak{u}^{(e)}(T)  \;  =  \;   2 \; \int_{FA} \; \frac{d \vec{k}}{(2 \pi)^2}  \;  \varepsilon_{\vec{k}}^{(e)}  \; 
  \frac{1}{e^{\frac{\varepsilon_{\vec{k}}^{(e)} - \mu^{(e)}(T)}{k_BT} } \; + \;1 }   \; \; .
\end{equation}
Rewriting   Eq.~(\ref{4.37}) as:
\begin{equation}
\label{4.38}  
\mathfrak{u}^{(e)}(T)  \;  =  \;    2  \; {\cal{N}}(0)  \; \int_{0}^{\infty} \;  d \varepsilon \;  \;  \varepsilon  \; 
  \frac{1}{e^{\frac{\varepsilon - \mu^{(e)}(T)}{k_BT} } \; + \;1 }   \; \; ,
\end{equation}
and taking into account that~\cite{Gradshteyn:1980} 
\begin{equation}
\label{4.39}  
 \int_{0}^{\infty} \;  d \xi \;   \frac{\xi}{e^{\xi } \; + \;1 }   \;  = \;  \frac{\pi^2}{12} \; \; ,
\end{equation}
we readily obtain:
\begin{equation}
\label{4.40}  
\mathfrak{u}^{(e)}(T)  \;  \simeq  \;    {\cal{N}}(0)   \; (\varepsilon_{F}^{(e)})^2 \; \left [ 1 \; + \;  \frac{\pi^2}{3}  \;
\left ( \frac{k_BT}{\varepsilon_{F}^{(e)}} \right )^2  \right ]  \;  .
\end{equation}
Whereupon we obtain for the specific heat:
\begin{equation}
\label{4.41}  
c^{(e)}(T) \;  \simeq  \;  \frac{2}{3} \, \pi^2  {\cal{N}}(0) \; k_B^2  \;  T \; \;  .
\end{equation}
To compare with experiments it is convenient to introduce the molar specific heat:
\begin{equation}
\label{4.42}  
c^{(e)}_m(T) \;  =  \; a_0^2 \, N_A \,  c^{(e)}(T) \; \;  .
\end{equation}
Evidently we can write:
\begin{equation}
\label{4.43}  
c^{(e)}_m(T) \;  = \gamma_S \; T \; \;  ,
\end{equation}
where the Sommerfeld coefficient is:
\begin{equation}
\label{4.44} 
\gamma_S  \; \simeq \; \frac{\delta}{ 1  - \delta} \; \frac{\pi}{3} \;   \frac{m^*_e a_0^2}{ \hbar^2 }   \; N_A \; k_B^2 \; \; .
\end{equation}
Using the numerical values of the microscopic parameters, we find the estimate (in MKS units):
\begin{equation}
\label{4.45} 
  \gamma_S \; \simeq \; \frac{\delta}{ 1  - \delta} \; 6.82  \; \; \frac{mJ}{mol \,  K^2}  \; \; .
\end{equation}
This last equation implies that in the pseudogap region $\delta \sim 0.1$,   $ \gamma_S \; \sim \; 1.0 \;  mJ /mol K^2 $. \\
In the following Section we will deal with the effects of an applied magnetic field on the low-temperature specific heat
in hole doped cuprate superconductors in the pseudogap region. In fact, there are some controversial experimental evidences for
a field-induced   Sommerfeld coefficient in the electronic specific heat (see the general survey in Refs.~\cite{Junod:1998,Fisher:2007}). \\
we have already seen that at low temperatures the external magnetic field penetrates into the superconductor with an array of Abrikosov vortices. 
Long time ago, it was pointed out~\cite{Volovik:1993} that in d-wave superconductors, due to the line nodes of the superconducting gap,
there is a field-induced Sommerfeld coefficient in the electronic specific heat with  Sommerfeld coefficient $\gamma_S(H) \sim \sqrt{H}$.
Basically, this effect is due to the  supercurrent circulating around an Abrikosov vortex which shift the quasiparticle energy by Doppler
effect modifying, thereby, the density of states at the Fermi energy.  Nevertheless, within our model we found that 
(see Appendix~\ref{AppendixC}):
\begin{equation}
\label{4.53-bis} 
 < {\cal{N}}(0) >_{vor}   \; \simeq \; {\cal{N}}(0) \; .  
 \end{equation}
In other words, the Abrikosov vortices do not modify the  density of states at the Fermi energy of the nodal quasielectron liquid.
\subsection{Low temperature specific heat}
\label{s4.3}
The specific  heat is a bulk thermodynamic measurement that probes all excitations in a system. In
order to extract the excitations arising only from the electronic quasiparticle density of states, all
other contributions to the specific heat need to be subtracted out.
Specific heat data refer to the specific heat per mole that  are usually quoted in joules per mole per degree Kelvin.
 In the following, we will not distinguish between the specific heat at constant pressure and the specific heat at constant volume 
 for, at low temperatures, the two specific heats differ by a negligible amount. \\
In the simplest case, the specific heat of a metal is the sum of the lattice specific heat, the  phonon contribution,
and the contribution of the conduction electrons.  In the Debye model, the crystal lattice is considered as a
 continuous isotropic medium and thereby it turns out that  the phonon contribution to the specific heat 
 is proportional to $T^3$ at low temperatures.  At higher temperatures this Debye interpolation is
 insufficient to describe the phonon specific heat, so that one includes higher order terms. Usually, it
 turns out that it is enough to consider a term proportional to $T^5$. The resulting phonon contribution
 to the specific heat can be written as:
\begin{equation}
\label{4.54} 
c_{phonon} \; = \; \beta_1 \, T^3  \; + \; \beta_2 \, T^5 \; \; . 
\end{equation}
Alternatively, the phonon contribution can be parametrized as:
\begin{equation}
\label{4.55} 
c_{phonon} \; = \; \beta_1 \, T^3  \; + \; 
N_E \, \left ( \frac{T}{T_E} \right )^2 \frac{e^{\frac{T}{T_E}}}{\left [ 1 - e^{\frac{T}{T_E}} \right ]^2} \; .
\end{equation}
The second term represents an Einstein contribution associated to optical phonons and it  is characterized by the excitation 
of oscillators with a single frequency. In Eq.~(\ref{4.55}) $T_E$ is the Einstein temperature and $N_E$ is the Einstein constant. 
The electronic specific heat  depends linearly on temperature, and it is the dominating contribution to the
specific heat at sufficiently low temperatures:
\begin{equation}
\label{4.56} 
c^{(e)} \; =  \; \gamma \; T \; \; .
\end{equation}
The constant $\gamma$ can be easily obtained as the  the intercept at $T = 0$ in a plot of $c/T$ versus $T^2$. 
However, generally in hole doped cuprate superconductors an upturn of the specific heat has been found at 
very low  temperatures~\cite{Junod:1998,Fisher:2007}.  This upturn is interpreted as a  Schottky anomaly.
Indeed, it is known that in paramagnetic salts  a Schottky anomaly has been observed. 
The Schottky specific heat at low temperatures may be significantly larger than the electronic and the lattice specific heats. 
The nature of the Schottky term is not well understood and it is generally
ascribed to paramagnetic centers arising from disorder. The Schottky contribution to the specific heat is of the form:
\begin{equation}
\label{4.57} 
c_{Schottky} \; =   \; k_B \, N_A \, n \;  z^2 \; \frac{e^z}{\left [ 1 + e^z \right ]^2} \; \; , \; \; 
z \; = \frac{g \mu_B H_{eff}}{k_B T} \; ,
\end{equation}
where $\mu_B$ is the Bohr magneton, $g$ is the Land\`e g-factor, $n$ is the number of mole of the Schottky centers,
and $H_{eff}$ is a microscopic effective magnetic field. In the temperature regime of interest it results that
$k_B T \gg   \mu_B H_{eff}$,  so that the  Schottky specific heat falls off with the inverse square of the temperature:
\begin{equation}
\label{4.58} 
c_{Schottky} \; \simeq   \; \frac{a}{T^2} \; \; \; . 
\end{equation}
At low temperatures the specific heat comprises three contributions:
\begin{equation}
\label{4.59} 
c(T)  \; = \; c^{(e)}(T) \; + \;  c_{phonon}(T) \; +  \;  c_{Schottky}(T)  \;  \; ,
\end{equation}
which, according to our previous discussion, can be written as:
\begin{equation}
\label{4.60} 
c(T)  \; \simeq  \;  \frac{a}{T^2} \;  + \;   \gamma \, T  \; +  \;   \beta_1 \, T^3  \; + \; \beta_2 \, T^5  \; \; .
\end{equation}
%
%
\begin{figure}
\vspace{0.8cm}
\centering
\resizebox{0.5\textwidth}{!}{%
\includegraphics{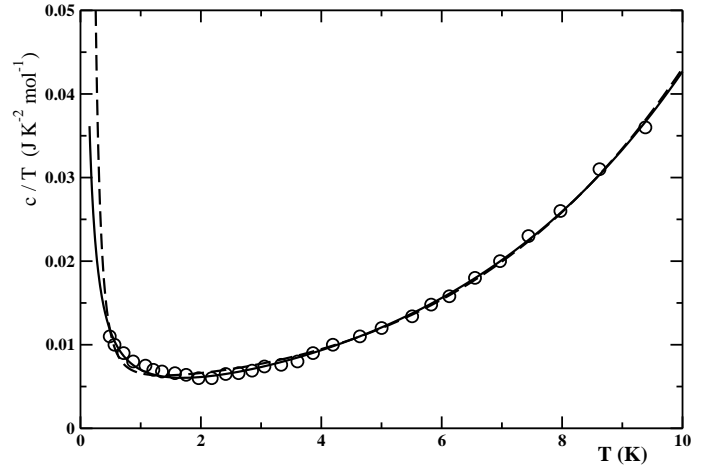} }
\caption{\label{Fig16} Temperature dependence of the specific heat divided by the temperature for underdoped YBCO, $T_c \simeq 89 \, K$. 
The data have been taken from Fig.~9.4, upper insert,  in Ref.~\cite{Fisher:2007}. The dashed  line is the best fit of data to  
Eq.~(\ref{4.60}).  The solid line is the best fit of the data to Eq.(\ref{4.69}). }
\end{figure}
%
For illustrative purposes, in Fig.~\ref{Fig16} we  shows the  low temperature specific heat data of  underdoped YBCO, $T_c \simeq 89 \, K$. 
The data have been taken from Fig.~9.4, upper insert,  in Ref.~\cite{Fisher:2007}. The low temperature upturn
 is  clearly displayed by the specific heat data. In fact, we tried to fit the data to  Eq.~(\ref{4.60}). As a result we found
 the following values for the fitting parameters (see the dashed line in  Fig.~\ref{Fig16}):
\begin{eqnarray}
\label{4.61} 
a  \, \simeq  \, 0.714 \, mJ \, K \, mole^{-1} \; \; \;  \; \;  ,  \hspace{1.2cm}  
\\ \nonumber
\gamma \,   \simeq  \, 5.66 \, mJ \, K^{-2} \, mole^{-1} \; \; \; \;  , \hspace{1.1cm}
\\  \nonumber 
\beta_1 \,  \simeq  \, 0.212 \, mJ \, K^{-4} \, mole^{-1} \;   ,  \hspace{1.0cm}
\\ \nonumber 
\; \beta_2 \,  \simeq  \, 0.0016 \, mJ \, K^{-6} \, mole^{-1} \;  . \hspace{0.9cm}
\end{eqnarray}
 An interesting feature resides in the fact that the fit leads to a non-zero value for $\gamma$. In fact, this  holds quite generally
 in any  hole doped cuprate superconductors with the observed values for $\gamma$  in the range 
 $ 1 - 10  \, mJ \, K^{-2} \, mole^{-1}$.  The presence of such a term looks puzzling. In fully gapped superconductors
 the electronic specific heat is exponentially suppressed at low temperatures. On  the other hand, if one identifies
 the nodal gap observed in angle resolved photoemission experiments with the d-wave BCS gap, then the density
 of states at the Fermi energy does not need to vanish along the gap nodes. In particular, the quasi two-dimensional
 d-wave gap proposed for cuprate superconductors  gives rise to a quadratic temperature term in the specific heat.
 Moreover, the coefficient of the quadratic term associated with the lines of gap nodes may be estimated to be 
 $a_{nodal} \approx \gamma_n/T_c$,  where $\gamma_n$ is the normal-state Sommerfeld coefficient. Therefore, one obtains
the  estimate  $a_{nodal} \approx 0.1  \, mJ \, K^{-3} \, mole^{-1} $.  Nevertheless, we added to the specific heat  Eq.~(\ref{4.60})  the
quadratic term $a_{nodal} \, T^2$ and fitted the data. We obtained for the best fitted  $a_{nodal}$ a value consistent with zero,
 $a_{nodal}   \lesssim 0.01$  $\, mJ \, K^{-3} \, mole^{-1} $. In general, these results are interpreted by assuming that the
 linear term $\gamma \, T$ is an additive contribution to the specific heat which is extrinsic to the superconductivity.
 As concern the quadratic term, it is believed that such term is masked by the phonon contribution so that
 it is difficult to extract its value by fitting   the low-temperature specific heat data. \\
 We discuss, now, the peculiar temperature dependence of the specific heat at low temperatures in cuprate superconductors
 within our phenomenological theory.  The specific heat is fully determined by the low-lying excitation of superfluid
 condensate. We have seen that  the low-lying condensate excitations are the roton quasiparticles.
 However, at  low temperatures the rotons contribution to the internal energy is exponentially suppressed.
 So that we are left with nodal quasielectrons as the only  relevant  low-energy excitations in the superconductive phase
 at low enough temperatures. 
 We showed in Sect.~\ref{s4.2} that the specific heat of the nodal quasielectron gas rises linearly with the temperature with
Sommerfeld coefficient given by Eq.~(\ref{4.44}).  However, we must take into account that nodal quasielectrons
can be excited only in the phase disordered hole pair condensate. At low temperatures the phase disordered condensate
is only a tiny fraction of the condensate. In fact, we may interpret the phase disordered condensate as the condensate  normal fraction
in an effective two fluid picture:
\begin{equation}
\label{4.62} 
n_n(T)  \;  =  \; n_s(0) \; - \; n_s(T)  \;  \; .
\end{equation}
So that the  fraction of phase disordered condensate is:
\begin{eqnarray}
\label{4.63} 
\frac{n_n(T)}{n_s(0)}   =    1  -  \frac{n_s(T)}{n_s(0)}   \simeq   
1 -  e^{  - \, b' \left [ \frac{1}{\sqrt{1 \, - \, T/T_c}}  -  1 \right ] } 
\\ \nonumber
 =  \frac{b'}{2} \, \frac{T}{T_c} \; + {\cal{O}}( \frac{T^{2}}{T_c^{2}}) \; \; , \; \;  
T \; \ll \; T_c  \; .
\end{eqnarray}
Therefore, at low temperatures the internal energy density of nodal quasielectrons is: 
\begin{eqnarray}
\label{4.64}  
\mathfrak{u}^{(ne)}(T) \; = \;  n_n(T) \, \mathfrak{u}^{(e)}(T)  \;  \simeq  \;    
\\ \nonumber 
 \frac{b'}{2} \, \frac{T}{T_c} \, {\cal{N}}(0)   \; (\varepsilon_{F}^{(e)})^2 \; 
  \left [ 1 \; + \;  \frac{\pi^2}{3}  \;
\left ( \frac{k_BT}{\varepsilon_{F}^{(e)}} \right )^2  \right ]  \;  .
\end{eqnarray}
Whereupon one obtains the nodal quasielectron specific heat at constant volume:
\begin{equation}
\label{4.65}  
c_{V}^{(ne)}(T) \; = \; \frac{\partial \, \mathfrak{u}^{(ne)}(T)}{\partial T}  \; \;  ,
\end{equation}
and  the molar specific heat:
\begin{eqnarray}
\label{4.66}  
c^{(ne)}(T)  =  a_0^2 \, N_A \,  c_{V}^{(ne)}(T)   \simeq \hspace{2.5cm}
\\ \nonumber
 \frac{b'}{2} \,  a_0^2 \, N_A \,  {\cal{N}}(0) \; \frac{(\varepsilon_{F}^{(e)})^2}{T_c}  +  
\frac{\pi^2}{2} \, b' \, a_0^2 \, N_A \,  {\cal{N}}(0) \;  \frac{k_B^2}{T_c}  \;  T^2  .
\end{eqnarray}
We see, then, that at low temperatures the nodal quasielectrons contribute to the specific heat with
a term of the form:
\begin{equation}
\label{4.67} 
c^{(ne)}(T) \; \simeq  \;  a_1 \; + \; a_2 \, T^2  \;  \; ,
\end{equation}
where:
\begin{eqnarray}
\label{4.68}  
a_1  \simeq  \frac{b'}{2} \,  a_0^2 \, N_A \,  {\cal{N}}(0) \; \frac{(\varepsilon_{F}^{(e)})^2}{T_c}    \; \; , 
\\ \nonumber
a_2  \simeq  \frac{\pi^2}{2} \, b' \, a_0^2 \, N_A \,  {\cal{N}}(0) \;  \frac{k_B^2}{T_c}  \;  \;   \; \; . 
\end{eqnarray}
Accordingly, we tried to fit  the specific heat data with the following functional form:
\begin{equation}
\label{4.69} 
c(T)  \; \simeq  \;  a_1  \;  + \;   a_2 \, T^2   \; +  \;   \beta_1 \, T^3  \; + \; \beta_2 \, T^5  \; \; .
\end{equation}
Remarkably, we found that  Eq.~(\ref{4.69}) is able to track quite well the data.  Indeed, we obtained the best fit 
 with the following values of the parameters (see the full curve in  Fig.~\ref{Fig16}):
\begin{eqnarray}
\label{4.70} 
a_1  \, \simeq  \, 5.307 \, mJ \, K^{-1} \, mole^{-1} \; , \hspace{1.0cm}
\\ \nonumber
a_2 \,   \simeq  \, 1.534  \, mJ \, K^{-3} \, mole^{-1}  \; , \hspace{1.0cm}
\\ \nonumber
\beta_1 \,  \simeq  \, 0.088 \, mJ \, K^{-4} \, mole^{-1} \; ,  \hspace{1.0cm}
\\ \nonumber
 \beta_2 \,  \simeq  \, 0.0018 \, mJ \, K^{-6} \, mole^{-1} \;  . \hspace{0.9cm}
\end{eqnarray}
Fig.~\ref{Fig16} shows that both  functional forms,  Eq.~(\ref{4.60}) and   Eq.~(\ref{4.69}), are quite consistent with measurements.
Note that our  Eq.~(\ref{4.69}) avoids to introduce a Schottky-like or a linear term to account for the temperature
dependence of the low temperature specific heat. It must be mentioned, however, that if one evaluates  the two coefficients
$a_1$ and $a_2$ by means of the parameters of our model, then it results in an  overestimate by order of magnitude
with respect to the best-fitted values. At moment, we have not found any means of explaining  this discrepancy.  \\
To conclude the present Section, we would like to discuss the low temperature specific heat in presence of an applied magnetic field.
At low temperatures and in magnetic fields of the order of a few Tesla it is well established that  hole doped superconductors
 in the pseudogap region  are in a vortex state. In fact, it results that $H > H_{c1}$, but the magnetic field strength is too small
to suppress the superconductivity, i.e. $H \ll H_{c2}$.  From the experimental point of view, it is established that the specific heat is of
the form:
\begin{equation}
\label{4.71} 
c(H,T)  \; \simeq  \;  c(T)  \;  + \;  \Delta c(H,T)  \; \; ,
\end{equation}
 where $c(T)$ is the low-temperature specific heat in absence of the external magnetic field. The specific heat of most high temperature
 superconductors in external magnetic field displays a broad low temperature structure which usually is ascribed to a Schottky anomaly.
 Even in this case, the origin of such anomaly is believed to arise from particles with spin whose concentration is, in general, dependent on
 the magnetic field strength. The Schottky contribution is given by  Eq.~(\ref{4.57}) where $H_{eff} \simeq H$ for strong enough
 applied magnetic fields. For temperatures not too low, $T \gtrsim 3 - 4 \, K$, the Schottky term is negligible and it turns out that
 $ \Delta c(H,T)$ increases linearly with the temperature. Therefore, we can write: 
\begin{equation}
\label{4.72} 
 \Delta c(H,T)  \; \simeq  \;  c_{Skottky}(H,T)  \;  + \;  \gamma(H) \;  T \; \; .
\end{equation}
Early measurements on optimal doping YBCO~\cite{Moler:1994,Moler:1997,Revaz:1998,Wright:1999,Wang:2001} performed at magnetic
fields up to $16 \, T$ evidenced  a coefficient of the low-temperature linear term growing with the magnetic field according
to $\gamma(H) \sim \sqrt{H}$. This dependence on the magnetic field is understood to be a clear signature of a d-wave BCS 
gap~\cite{Volovik:1993,Simon:1997}. More recently, this peculiar dependence of $\gamma(H)$ on the magnetic field strength has
been confirmed in Ref.~\cite{Riggs:2011} where it is reported the low temperature measurements of the specific heat in underdoped
YBCO with magnetic field up to $45 \, T$. However, in Ref.~\cite{Kemper:2015}  it is reported the low temperature specific heat of
underdoped YBCO ($\delta \simeq 0.076, 0.084$) in applied magnetic field up to $34.5 \, T$. These authors observed two regimes
in the low temperature limit. Namely, below a characteristic magnetic field $H' \sim 13 \, T$ the coefficient of the linear temperature
term in the specific heat behaves like  $\gamma(H) \sim \sqrt{H}$ as previously reported. However, near the above characteristic magnetic 
field there is a sharp inflection followed by a linear behavior of $\gamma(H)$ with $H$:
\begin{eqnarray}
\label{4.73} 
  \gamma(H) \; = \; A_c \, H \; \; , \; \;  \;  H \; \gtrsim \; H'  \; ,
\\ \nonumber
 A_c \;  \simeq  \;  0.16 \; mJ \, K^{-2} \, mole^{-1} \, T^{-1} \;  .
\end{eqnarray}
Indeed,  we shall see that  in our theory we can account quite easily for  $ \Delta c(H,T)$ of the form
given by Eq.~(\ref{4.72})  with  $\gamma(H) \sim H$, as well as with a  suitable  Schottky term.
In type-II superconductors the applied magnetic field exceeding the lower critical field penetrates into the system
with an array of Abrikosov vortices.  Since an Abrikosov vortex carries one magnetic flux quantum,
as the applied magnetic field increases  more and more Abrikosov vortices are formed.
As discussed in Sect.~\ref{3.5}, in our theory the upper critical magnetic field is much smaller than the Landau-Ginzburg
critical magnetic field where the Abrikosov vortices become to overlap. So that, the density of Abrikosov vortices is given by:
\begin{equation}
\label{4.74} 
n_V  \; \simeq \;  \frac{H}{\phi_0} \;   \; . 
\end{equation}
From Eq.~(\ref{4.74}) it follows that the average distance between vortices is  $d_H \simeq  \sqrt{\phi_0/H}$. Now, it is
easy to check that $d_H \gg \xi_V \simeq d_0$  for applied magnetic field strengths up to $40 \, T$. Therefore, we 
see that we may employ safely the dilute vortex approximation. Concerning  the specific heat at low temperatures,   
we already noticed that the only relevant low-energy excitations are the nodal quasielectrons which, however,  
can be excited only in the phase disordered condensate. Since we argued that in the vortex core the condensate
is phase disordered, we see that at low temperatures the specific heat  comprises two contributions due to
 nodal quasielectrons excitations arising from  to the disorder condensate fraction outside the vortices and
the disorder condensate in the vortex cores. Therefore, since the Abrikosov vortices are dilute, to a good approximation
the specific heat can be written as in  Eq.~(\ref{4.71}). The first term on the right hand side of  Eq.~(\ref{4.71}) is the
contribution due to the tiny disorder fraction of the hole condensate. Evidently this term is almost independent on the
strength of the applied magnetic field and it has been already discussed. Now we show that the other term can be written as in  
 Eq.~(\ref{4.72}). In fact,  we have written down the nodal quasielectron contribution to the constant volume specific heat,
 Eq.~(\ref{4.41}). Therefore we have:
\begin{equation}
\label{4.75} 
\Delta c^{(ne)}_V(H,T)  \; \simeq \;  \pi \, \xi_V^2 \, \frac{H}{\phi_0} \;  \frac{2}{3} \, \pi^2  <{\cal{N}}(0)>_{vor}  \; k_B^2  \;  T   \; ,
\end{equation}
and  the molar specific heat:
\begin{equation}
\label{4.76}  
\Delta c^{(ne)}(H,T) \;  =  \; a_0^2 \, N_A \,  \Delta c^{(ne)}_V(H,T)  \; \;  .
\end{equation}
Taking into account  Eq.~(\ref{4.53-bis}) and Eq.~(\ref{4.44})   we can write:
\begin{equation}
\label{4.77} 
\Delta c^{(ne)}(H,T) \;  =\;    \gamma(H) \;  T \; \; ,
\end{equation}
where:
\begin{equation}
\label{4.78} 
\gamma(H) \; \simeq \;  \pi \, \xi_V^2 \, \frac{H}{\phi_0} \;  a_0^2 \, N_A \,  \frac{2}{3} \, \pi^2  {\cal{N}}(0) \; k_B^2  \; = \; 
 \pi \, \xi_V^2 \, \frac{H}{\phi_0} \;   \gamma_S \; .
\end{equation}
Evidently, $\gamma(H)$ is conform to  Eq.~(\ref{4.73}) with
\begin{equation}
\label{4.79} 
A_s \; \simeq \;  \pi \, \xi_V^2 \, \frac{1}{\phi_0} \;   \gamma_S \; .
\end{equation}
The Schottky-like term in the specific heat arises as follows. In the Abrikosov vortex core there is a non-zero microscopic
magnetic field $h(0)$. Now, the spatial average of $h(0)$ is the magnetic induction $B$ not the applied magnetic field
$H$. For magnetic field strengths employed in the experiments and at low temperatures the microscopic magnetic field
is given by   Eq.~(\ref{3.113}):
\begin{equation}
\label{4.80}  
 h(0) \;  \simeq  \;   \frac{\phi_0}{2 \pi   \lambda^2(0)}  \;   \ln ( \kappa)   \; .
\end{equation}
This tiny magnetic field induces the Zeeman splitting of the energy of the hole bound states (see I, Sect.~4.1):
\begin{equation}
\label{4.81} 
\varepsilon_{Zeeman}  \; \simeq   \;  \pm \;  \frac{ \hslash \, e  h(0)}{m^*_h c} \; .
\end{equation}
The presence of energy levels separated by a very small energy difference gives rise to the specific heat at constant volume:
\begin{equation}
\label{4.82} 
c^{Zeeman}_V(H,T)  \; \simeq   \;  k_B \, z^2 \; \frac{e^z}{\left [ 1 + e^z \right ]^2} \; \; , \; \; 
z \; = \frac{2  \hslash \, e  h(0)}{m^*_h c \,  k_B T} \; .
\end{equation}
So that we find the following Schottky-like molar specific heat:
\begin{equation}
\label{4.83}  
c_{Schottky}(H,T)  \;  =  \;  \pi \, \xi_V^2 \, \frac{H}{\phi_0} \;  a_0^2 \, N_A \,   c^{Zeeman}_V(H,T)   \; \;  .
\end{equation}
Actually, since  $\varepsilon_{Zeeman}  \ll k_B T$, the Schottky-like specific heat can be written as:
\begin{equation}
\label{4.84} 
c_{Schottky}(H,T)  \; \simeq  \;  A \; \frac{H}{T^2}  \;  \; .
\end{equation}
In conclusion, we found that at low temperatures and in external magnetic field the specific heat acquires an additional
contribution which can be parametrized as:
\begin{equation}
\label{4.85} 
 \Delta c(H,T)  \; \simeq  \;  A \,  \frac{H}{T^2}   \;  + \;  A_s \, H \,  T \; \; .
\end{equation}
Even though Eq.~(\ref{4.85}) seems to be in qualitative agreement with observations, it would be of great interest
 to reanalyze  the experimental data following our theoretical suggestions.
\section{Charge Density Wave Instability}
\label{s5}
In this Section we will focus on charge order in hole doped cuprate superconductors in the pseudogap region~\footnote{For a recent review
see Ref.~\cite{Comin:2015a}.}. For charge order, in general,  it is intended an electronic phase which  breaks
translational symmetry through a self - organization   into periodic structures incompatible with the periodicity of the underlying lattice.
In fact, there is growing evidence of a charge order existing in the pseudogap state of several cuprate families 
(see Refs.~\cite{Wu:2011,Mounce:2011,Ghiringhelli:2012,Chang:2012,Achkar:2012,Wu:2013,LeBoeuf:2013,Blanco:2013,Croft:2014,Blanco:2014b,Hucker:2014,Fujita:2014,Forst:2014,Tabis:2014,Comin:2014,Silva:2014,Comin:2015b,Wu:2015,Gerber:2015,Forgan:2015,Hamidian:2015})  
which also coexists and competes with superconductivity at lower temperatures.
Recently, it seems  that there is a consensus on the presence of periodic modulations in the electronic density, 
a phenomenon called charge density wave (CDW).  Clearly, the ubiquitous    presence of charge density wave correlations 
is adding complexity to the physics of  high temperature superconductivity. \\
Charge density waves were discussed long time ago by Fr\"ohlich~{\cite{Frohlich:1954} and Peierls~\cite{Peierls:1955}
(see also Refs.~\cite{Gruner:1988,Gruner:1994} and references therein).
In fact,  it was  pointed out that a one dimensional metal coupled to the underlying lattice is not stable at low temperatures. 
Due to the peculiar topology of the Fermi surface in one dimensional metals with respect to
 three dimensional  systems,  the bare charge susceptibility diverges at wavevector $q = 2 k_F$, where $k_F$ is
 the Fermi wavenumber.  As a result, the formation of an electronic charge modulation with wavevector $q$ 
 is favored with respect to  homogeneous configurations. 
Since in conductors the  electrons are coupled to the ion lattices, such a deep reorganization of the electronic carrier distribution 
alters the ionic lattice resulting in  a small deformation to lower the total electrostatic energy of the electron-phonon system. In fact,
 the ions move towards new equilibrium positions leading to the so-called Peierls deformation.  The ionic displacements are limited in
extent not to exceed in the elastic energy cost. The resulting   ground state of the coupled electron-phonon system 
 is characterized by a gap in the single-particle excitation spectrum and by the presence of collective modes. 
Usually, a ground state with density waves is associated to a highly anisotropic conduction bands,
such as  organic and inorganic materials with  quasi-one dimensional  or quasi-two dimensional  
electronic structures~\cite{Gruner:1988,Gruner:1994}. \\
Besides the electronic spectrum, the formation of a charge density wave also affects other quasiparticle excitations.
Most prominently there appears the so-called Kohn anoma-ly~\cite{Khon:1959} which  correspond  to a softening of the 
phonon responsible for the electron-phonon instability. Remarkably, in hole doped cuprate superconductors in the pseudogap region
it seems that  giant phonon anomalies  are  common to the $CuO_2$ planes. In particular,  neutron and X-ray inelastic scattering
studies identified the anomalous behavior of longitudinal optical phonons arising from the  in-plane $Cu-O$ bond-stretching modes
(see the review Ref.~\cite{Reznik:2010} and references therein).  These anomalies  resemble those associated
 with the observed Kohn anomalies in quasi one-dimensional conductors with  nested Fermi surfaces.   This strongly suggests that the 
in-plane $Cu-O$ bond-stretching  modes  are deeply involved in  the observed  charge density wave formation in the underdoped cuprates. 
In our highly idealized model there are two degenerate  bond-stretching longitudinal optical phonon mo-des, as schematically
illustrated in Fig.~\ref{Fig17}. In the pseudogap region the only low-lying excitations which may interact with these longitudinal
phonons are the nodal quasielectrons. In fact, we have seen in Sect.~\ref{s5} that the nodal quasielectrons are intimately lied to the bound
hole pairs.  The Coulomb interactions between the nodal quasielectrons and the lattice ions give rise to an effective   
electron-phonon coupling. Now, we consider a model Hamiltonian  where  the coupling of the  nodal quasielectrons with the  
bond-stretching phonons is at the heart of the charge density wave instabilities.
 Evidently the planar lattice displacements are written as (see Fig.~\ref{Fig17}):
\begin{equation}
\label{5.1} 
 \vec{u}(\vec{r})  \; =  \;    u_1(\vec{r}) \, \hat{x}     \;  + \; u_2(\vec{r}) \, \hat{y} \; \; \; , \; \; \; \vec{r} \; = \; (x,y) \; \; .
\end{equation}
Adopting the quasi-harmonic approximation the in-plane lattice vibrations are described by the following Hamiltonian:
\begin{eqnarray}
\label{5.2} 
 \hat{H}^{(ph)}  \; =  \;   \sum_{i =1}^{2}  \; \sum_{\vec{q}} \; \bigg \{ \frac{ \hat{P}_{i}(\vec{q})  \hat{P}_{i}(-\vec{q})}{2 M} \; 
 \\ \nonumber 
 + \; \frac{M \Omega^2(\vec{q})}{2}  \hat{Q}_{i}(\vec{q})  \hat{Q}_{i}(-\vec{q}) \bigg \} \; ,
\end{eqnarray}
where $\hat{Q}_{i}(\vec{q})$ and  $\hat{P}_{i}(\vec{q})$, $i = 1, 2$, are the normal coordinates and conjugate momenta corresponding
to the lattice displacements  $u_1(\vec{r})$ and $u_2(\vec{r})$ respectively; $\Omega(\vec{q})$ is the normal mode frequency with wavevector 
$\vec{q}$, and  $M$ is the reduced atomic  mass of the $Cu$ and $O$ ions:
\begin{equation}
\label{5.3} 
 \frac{1}{M}  \; =  \;   \frac{1}{M_{Cu}}  \; +   \; \frac{1}{M_{O}}  \; \; \;  .
\end{equation}
We may introduce the phonon creation and annihilation operators as follows:
\begin{equation}
\label{5.4} 
 \hat{Q}_{i}(\vec{q}) \; = \;  \sqrt{ \frac{\hslash}{2 M \Omega(\vec{q})}} \;  \left [  \hat{b}_{i}(\vec{q}) \, + \,  \hat{b}^{\dagger}_{i}(-\vec{q})
\right ] \; ,
\end{equation}
\begin{equation}
\label{5.5} 
\hat{P}_{i}(\vec{q}) \; = \; i \;  \sqrt{ \frac{\hslash  M \Omega(\vec{q})}{2}} \;  \left [  \hat{b}^{\dagger}_{i}(\vec{q}) \, - \,  \hat{b}_{i}(-\vec{q})
\right ] \; .
\end{equation}
These operators satisfy the canonical commutation relations:
\begin{equation}
\label{5.6} 
 \left [ \hat{b}_{i}(\vec{q} ) \; , \; \hat{b}^{\dagger}_{j}(\vec{q} \, ')  \right ]  \;  = \; \delta_{i,j} \, \delta_{\vec{q}, \vec{q} \, '} \; , 
\end{equation}
while  all the other commutators are vanishing. In fact, it is easy to check that:
\begin{equation}
\label{5.7} 
 \left [ \hat{Q}_{i}(\vec{q}) \; , \; \hat{P}_{j}(\vec{q} \, ') \right ]  \;  = \; i \, \hslash \, \delta_{i,j} \, \delta_{\vec{q}, \vec{q} \, '} \; .
\end{equation}
Using Eqs.~(\ref{5.4}),  (\ref{5.5}) and (\ref{5.6}) the phonon Hamiltonian can be rewritten as:
\begin{equation}
\label{5.8} 
 \hat{H}^{(ph)}  \; =  \;   \sum_{i =1}^{2}  \; \sum_{\vec{q}} \;  \hslash \,  \Omega(\vec{q})  
 \left [ \hat{b}^{\dagger}_{i}(\vec{q} ) \,  \hat{b}_{i}(\vec{q} ) \;  + \; \frac{1}{2}  \right ]   \; .
\end{equation}
%
%
\begin{figure}
\vspace{0.7cm}
\hspace{0.65cm}
\resizebox{0.44\textwidth}{!}{%
\includegraphics{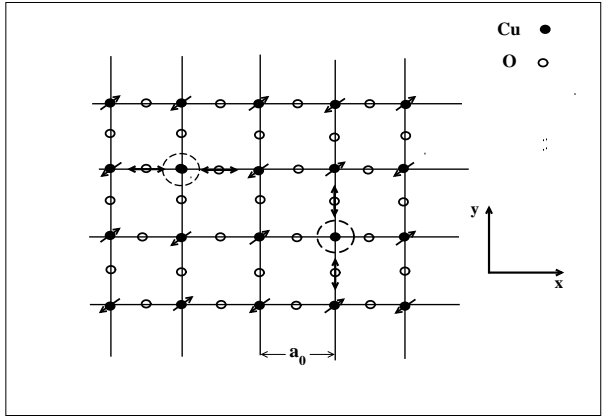} }
\caption{\label{Fig17} The idealized square copper-oxide plane with lattice spacing $a_0$. The dashed ellipses represent two holes in the
 antiferromagnetic background. The arrows indicate the longitudinal bond-stretching modes along the  planar $Cu-O$ bonds. }
\end{figure}
%
The planar lattice displacements can be expressed in terms of the creation and annihilation operators according to:
\begin{equation}
\label{5.9} 
 u_i(\vec{r})   \; =  \;  \frac{1}{\sqrt{N_u} }  \; \sum_{\vec{q} } \;   \sqrt{ \frac{\hslash}{  2 M \Omega(\vec{q})} } 
 \left [ \hat{b}_{i}(\vec{q} ) \; + \;  \hat{b}^{\dagger}_{i}( - \vec{q} ) \right ]   \;  e^{i \vec{q} \cdot \vec{r} } \; ,
\end{equation}
where $N_u$ is the number of unit cell such that the lattice volume is $V = N_u \, V_u$, $V_u = a_0^2$ being the unit cell volume.
As concern the nodal quasielectron Hamiltonian, according to Sect.~\ref{s4} we have:
\begin{equation}
\label{5.10} 
\hat{H}^{(e)}    \; = \;    \; \sum_{\vec{k},\sigma} \; \left ( \varepsilon_{\vec{k}}^{(e)} \, - \,    \varepsilon_{F}^{(e)}  \right )
 \;   \hat{\psi}^{\dagger}_{e}(\vec{k},\sigma)  \;   \hat{\psi}_{e}(\vec{k},\sigma) \; .
 \end{equation}
The electron-phonon interaction is usually obtained in the rigid ion approximation. In terms of the creation and annihilation
operators the interaction Hamiltonian reads:
\begin{eqnarray}
\label{5.11} 
\hat{H}^{(e-ph)}  \simeq  \frac{1}{V}   \sum_{\vec{q},\vec{k},\sigma} \,  \sum_{i =1}^{2}  
g_i(\vec{q})  \bigg( \hat{b}_{i}(\vec{q} ) \; + \;  \hat{b}^{\dagger}_{i}( - \vec{q} )   \bigg) \; \; \; 
\\ \nonumber
  \hat{\psi}^{\dagger}_{e}(\vec{k} + \vec{q},\sigma)  \;   \hat{\psi}_{e}(\vec{k},\sigma) \; \; \; , \; \; \;
 \end{eqnarray}
where~\cite{Gruner:1994}
\begin{equation}
\label{5.12} 
g_i(\vec{q}) \; \simeq \;   \sqrt{ \frac{\hslash}{  2 M \Omega(\vec{q})} }  \;  | q_i| \, V(\vec{q}) \; \; , \; \; \vec{q} \, = \, (q_1,q_2)  \;  .
\end{equation}
In Eq.~(\ref{5.12}) $ V(\vec{q})$ is the Fourier transform of the interaction potential between nodal quasielectrons and the lattice
ions. Accordingly, we assume:
\begin{equation}
\label{5.13} 
 V(\vec{q})  \; \simeq \;   \int \, d \vec{r} \;  e^{i \vec{q} \cdot \vec{r} } \; \frac{e^2}{| \vec{r} |} \;  =  \; \frac{ 2 \pi \, e^2}{| \vec{q} |} \; .
\end{equation}
Therefore, we are lead to the following effective Fr\"ohlich Hamiltonian:
\begin{equation}
\label{5.14} 
\hat{H}_{eff} \;  =  \;   \hat{H}^{(e)}    \; +   \;    \hat{H}^{(ph)}  \; + \;    \hat{H}^{(e-ph)}   \; .
\end{equation}
Note that the interaction Hamiltonian Eq.~(\ref{5.11}) can be rewritten as:
\begin{eqnarray}
\label{5.15} 
\hat{H}^{(e-ph)}   \simeq  \frac{1}{V}  \sum_{\vec{q},\vec{k},\sigma} \,  \sum_{i =1}^{2}  
g_i(\vec{q})  \;  \sqrt{ \frac{2 M \Omega(\vec{q})}{\hslash}  }  \; 
\\ \nonumber
  \hat{\psi}^{\dagger}_{e}(\vec{k} + \vec{q},\sigma)  \;   \hat{\psi}_{e}(\vec{k},\sigma) \; \hat{Q}_{i}(\vec{q}) \; .
 \end{eqnarray}
Moreover, we have:
\begin{equation}
\label{5.16} 
g_i(\vec{q}) \; \simeq \;   \sqrt{ \frac{\hslash}{  2 M \Omega(\vec{q})} }  \; \;  2  \pi  e^2 \;   \hat{q}_i   \;  .
\end{equation}
It is useful to consider the equation of motion of the normal coordinates:
\begin{equation}
\label{5.17} 
 \frac{\partial}{\partial t} \;   \hat{Q}_{i}(\vec{q},t)  \; = \; \frac{1}{i \hslash} \; 
 \left [  \hat{Q}_{i} \; , \;  \hat{H}_{eff} \right ] \; .   
\end{equation}
A straightforward calculation leads to:
\begin{equation}
\label{5.18} 
 \frac{\partial}{\partial t} \;   \hat{Q}_{i}(\vec{q},t)  \; = \; \frac{\hat{P}_{i}(\vec{q},t) }{M} \; .   
\end{equation}
Using Eq.~(\ref{5.18}) we obtain at once:
\begin{eqnarray}
\label{5.19} 
 \frac{\partial^2}{\partial t^2} \;   \hat{Q}_{i}(\vec{q},t)  \; = \; \frac{1}{i \hslash} \; 
 \left [ \frac{\hat{P}_{i}}{M} \; , \;  \hat{H}_{eff} \right ]   \; =  \;    - \;  \Omega^2(\vec{q}) \;  \hat{Q}_{i}(\vec{q},t) 
  \nonumber \\
  \; - \;
g_i(\vec{q})  \; \sqrt{ \frac{2  \Omega(\vec{q})}{\hslash M}  }  \; 
 \frac{1}{V}  \; \sum_{\vec{k},\sigma} \,  \hat{\psi}^{\dagger}_{e}(\vec{k} + \vec{q},\sigma)  \;   \hat{\psi}_{e}(\vec{k},\sigma) 
 \; . \hspace{1.0cm}
\end{eqnarray}
The last term on the right hand side of Eq.~(\ref{5.19}) is the effective force due to the modulation of the quasielectron density
induced by the lattice displacements. To evaluate this term we use the adiabatic approximation, namely we are supposing
that the ion and quasielectron motion can be decoupled. In fact, when a longitudinal wave progresses through the lattice,
causing local rarefactions and compressions in the ion density, we may suppose that the quasielectrons move so as to screen
out these fluctuations. Since the quasielectrons are able to respond to perturbations in times much shorter than the
inverse lattice frequencies, then they will effectively be following the motion of the lattice instantaneously. Thereby, the
presence of the quasielectrons results in a renormalization of the longitudinal frequencies of the ion vibrations. Within
the framework of the linear response theory a time independent external potential $\phi_{ex}(\vec{r})$ coupled to
the quasielectron gas leads to a rearrangement of the density which, expressed in Fourier space, 
is given by~\cite{Pines:1994b,Giuliani:2005}:
\begin{equation}
\label{5.20} 
\hat{\rho}(\vec{q})   = \left <    \frac{1}{V}  \; \sum_{\vec{k},\sigma}  \; 
  \hat{\psi}^{\dagger}_{e}(\vec{k} + \vec{q},\sigma)  \;   \hat{\psi}_{e}(\vec{k},\sigma)  \right >  
 =  \chi(\vec{q}) \; \hat{\phi}_{ex}(\vec{q}) \; , 
\end{equation}
where the brackets mean the quantum average over the fermion degrees of freedom. In Eq.~(\ref{5.20})  $\chi(\vec{q})$ is the so-called
static Lindhard response function:
\begin{equation}
\label{5.21} 
 \chi(\vec{q}) \;  = \;   \frac{1}{V}  \; \sum_{\vec{k},\sigma}  \; 
 \frac{n(\vec{k}) \; - \; n(\vec{k}+\vec{q})}{ \varepsilon_{\vec{k}}^{(e)} \, - \,   \varepsilon_{\vec{k}+\vec{q}}^{(e)}} \; ,
\end{equation}
where $ n(\vec{k})$ is the zero-temperature occupation number, $ n(\vec{k}) \, = \, 1$ if 
$  \varepsilon_{\vec{k}}^{(e)} \, < \, \varepsilon_{F}^{(e)}$ ,  $ n(\vec{k}) \, = \, 0$ otherwise. 
From Eq.~(\ref{5.15}) we infer that:
\begin{equation}
\label{5.22} 
 \hat{\phi}_{ex}(\vec{q})  \; = \;  \frac{g_i(\vec{q})}{V_u}  \; \sqrt{ \frac{2  M  \Omega(\vec{q})}{\hslash}  }  
 \;   \hat{Q}_{i}(\vec{q})  \; .
\end{equation}
So that, in the adiabatic approximation, one obtains: 
\begin{eqnarray}
\label{5.23} 
 \left <  \;  \frac{1}{V}  \; \sum_{\vec{k},\sigma}  \; 
  \hat{\psi}^{\dagger}_{e}(\vec{k} + \vec{q},\sigma)  \;   \hat{\psi}_{e}(\vec{k},\sigma) \; \right >  \; \simeq \; 
 \\ \nonumber
  \chi(\vec{q}) \;  \frac{g_i(\vec{q})}{V_u}  \; \sqrt{ \frac{2  M  \Omega(\vec{q})}{\hslash}  }  
 \;   \hat{Q}_{i}(\vec{q})  \; .
\end{eqnarray}
Inserting this last equation into  Eq.~(\ref{5.19}) we get:
\begin{equation}
\label{5.24} 
 \frac{\partial^2}{\partial t^2} \;   \hat{Q}_{i}(\vec{q},t)     \simeq    -  
 \left [ \Omega^2(\vec{q}) \; +  \frac{2 g_i^2(\vec{q})}{\hslash V_u}  \;   \chi(\vec{q}) \; \Omega(\vec{q})   \right ]
 \hat{Q}_{i}(\vec{q},t)  \; . 
\end{equation}
Eq.~(\ref{5.24}) shows that, indeed, the quasielectrons give rise to a renormalization of the longitudinal bond-stretching 
frequencies:
\begin{equation}
\label{5.25} 
 \tilde{\Omega}^2(\vec{q}) \; \simeq \;  \Omega^2(\vec{q})   \; + \;  \frac{2 g_i^2(\vec{q})}{\hslash V_u}  \;   \chi(\vec{q}) \; \Omega(\vec{q}) 
 \; .
\end{equation}
At finite temperatures one must perform the thermal average over the fermion degrees of freedom. As a result  Eq. (\ref{5.23}) 
still holds by replacing the zero-temperature Lindhard function with:
\begin{equation}
\label{5.26} 
 \chi(\vec{q},T) \;  = \;   \frac{1}{V}  \; \sum_{\vec{k},\sigma}  \; 
 \frac{f(\varepsilon^{(e)}_{\vec{k}}) \; - \; f(\varepsilon^{(e)}_{\vec{k} +\vec{q}})}{ \varepsilon_{\vec{k}}^{(e)} \, - \,  
  \varepsilon_{\vec{k}+\vec{q}}^{(e)}} \; ,
\end{equation}
where $ f(\varepsilon)$ is the Fermi-Dirac distribution function Eq. (\ref{4.31}). \\
For a one-dimensional electron gas it turns out that   $\chi(\vec{q})$ diverges at the Fermi surface
nesting wavevectors $| \vec{q} | = Q = 2 k_F$. The divergence of $\chi(Q)$ implies that at zero temperature the one-dimensional
electron gas is unstable with respect to the formation of a periodically varying electron charge density. At finite temperatures
 $\chi(\vec{q},T)$ attains its maximum value at  $| \vec{q} | = Q$. Consequently  the softening of the renormalized phonon 
 frequencies  $\tilde{\Omega}(\vec{q})$ will be most significant at these wavevectors. In fact, for the one-dimensional
 electron gas the renormalized phonon frequency at   $| \vec{q} | = Q$ goes to zero at a critical temperature:
\begin{equation}
\label{5.27} 
 \tilde{\Omega}^2(Q, T_{CDW}) \; =  \;  0 \; \; .  
\end{equation}
At the critical temperature $T_{CDW}$ there is a phase transition to a state with a periodic static lattice distortion and
an electron charge density modulation. This transition is generally referred to as the Peierls transition.
\subsection{Charge-ordering wavenumber vectors }
\label{s5.1}
To determine the critical temperature of the Peierls instability in the nodal quasielectron gas we need to establish the divergences
 of the zero-temperature static Lindhard response function at some wavevector $\vec{q} = \vec{Q}$.
%
\begin{figure}
\vspace{0.7cm}
\hspace{0.67cm}
\resizebox{0.45\textwidth}{!}{%
\includegraphics{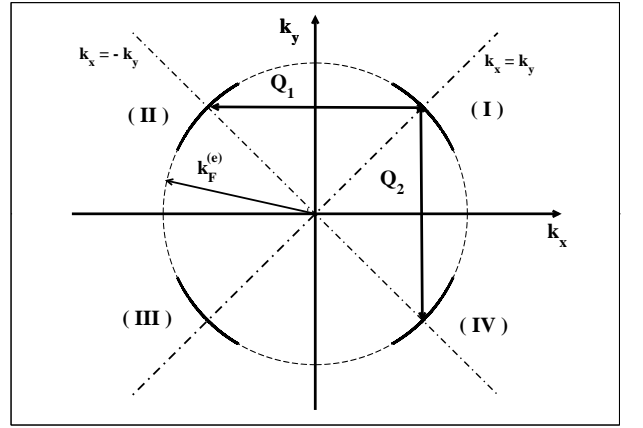} }
\caption{\label{Fig18} The quasielectron Fermi arcs  in the first Brillouin zone labelled as in Fig.~\ref{Fig12}.
$Q_1$ and $Q_2$ are the nesting wavenumbers leading to the charge density wave instability in the planar lattice displacements
$ u_1(\vec{r})$  and  $u_2(\vec{r})$ respectively.}
\end{figure}
%
In Appendix~\ref{AppendixD} we show that, when
\begin{eqnarray}
\label{5.36} 
\vec{q} \; = \; \pm \; \vec{Q}_1 \; , \;   \pm \; \vec{Q}_2 \;  \; \; ; \; \;   
\\ \nonumber
\vec{Q}_1 \; = \; (\sqrt{2} \, k_F^{(e)} ,  0)  \; \; , \; \;  
  \vec{Q}_2 \; = \; (0 , \sqrt{2} \, k_F^{(e)} ) \; ,
\end{eqnarray}
the static response function diverges logarithmically. \\
In Fig.~\ref{Fig18}  we display the nodal quasielectron Fermi arcs in the first Brillouin zone. It is worthwhile to recall that the $k_x, k_y$
axes are oriented along the copper-oxygen bonds.  $Q_1$ and $Q_2$ are the wavenumbers that are relevant
to trigger the density wave instabilities in the longitudinal bond-stretching planar modes.
\subsection{Critical temperature and gap}
\label{s5.2}
In the previous Section we argued that the nodal quasielectron density fluctuations at $q = Q_1, Q_2$ cause the  linear response
function to diverge at low temperatures. In the pseudogap region, where $\theta_{FA} \ll 1$, the physical motivations for these
instabilities reside on the fact that the nodal quasielectrons behave like the conduction electrons in quasi one-dimensional 
conductors. In fact, the wavevectors $\pm \vec{Q}_1, \pm \vec{Q}_2$ are the nesting vectors connecting the Fermi arc sectors
$(I) - (II)$, $(III) - (IV)$ and  $(I) - (IV)$, $(II) - (III)$ respectively (see Fig.~\ref{Fig18}). \\
As discussed early, with decreasing temperature the renormalized longitudinal phonon frequency decreases and eventually it
goes to zero. Indeed, using Eqs.~(\ref{5.52}) and (\ref{5.25}) one readily obtains:
\begin{equation}
\label{5.54} 
 \tilde{\Omega}^2(Q_1,T) \simeq  \Omega^2(Q_1)    \left [ 1 \; - \; 
   \frac{4 g^2(\vec{q})  {\cal{N}}(0)  }{\hslash \Omega(Q_1) V_u}  \;   
   \ln \left ( \frac{2 e^\gamma}{\pi} \; {\frac{\varepsilon_c}{ k_B T}} \right ) \right ]  \; ,
\end{equation}
where, according to Eq.~(\ref{5.16})
\begin{equation}
\label{5.55} 
g(Q_1) \; \simeq \;   \sqrt{ \frac{\hslash}{  2 M \Omega(Q_1)} }  \; \;  2  \pi  e^2 \;   .
\end{equation}
Therefore, the critical temperature given by Eq.~(\ref{5.27}) is:
\begin{equation}
\label{5.56} 
 k_B \; T_{CDW} \; \simeq \;  \frac{2 e^\gamma}{\pi}  \;  \varepsilon_c \;
e^{ - \;  
   \frac{ \hslash \Omega(Q_1) V_u   }{ 4 g^2(Q_1)  {\cal{N}}(0)    } }  \;  .
\end{equation}
Taking into account Eqs.~(\ref{2.15}) and (\ref{5.53}) we find:
\begin{equation}
\label{5.57} 
T_{CDW}(\delta)  \; \simeq \;  \frac{4 e^\gamma}{\pi}  \;  T^*(\delta) \;
e^{ - \;  
   \frac{ \hslash \Omega(Q_1) V_u   }{ 4 g^2(Q_1)  {\cal{N}}(0)    } }  \;  .
\end{equation}
Note that the doping dependence of the charge density wave critical temperature arises from the pseudogap temperature:
\begin{equation}
\label{5.58} 
T^*(\delta)  \; \simeq \;  T^*(0) \; \left [ 1 \; - \;  (\frac{\delta}{\delta_c})^{\frac{3}{2}}  \right ]     \;  ,
\end{equation}
and the density of states at the Fermi energy:
\begin{equation}
\label{5.59} 
 {\cal{N}}(0)  \; \simeq \;  \frac{1}{2 \pi}  \;  \frac{m^*_e}{\hslash} \; \frac{\delta}{1 - \delta}    \;  .
\end{equation}
In fact, we can rewrite Eq.~(\ref{5.57}) as:
\begin{eqnarray}
\label{5.60} 
T_{CDW}(\delta)  \; \simeq \;  \frac{4 e^\gamma}{\pi}  \;  T^*(0) \; \left [ 1 \; - \;  (\frac{\delta}{\delta_c})^{\frac{3}{2}}  \right ]    \;
\\ \nonumber
e^{ - \;  \frac{1}{4 \pi} \; 
   (\frac{ \hslash \Omega a_0 }{e^2})^2 \frac{M}{m^*_e}   \; \frac{1 - \delta}{\delta}   }  
\end{eqnarray}
where $ \Omega = \Omega(Q_1)$. The sudden vanishing of the renormalized longitudinal phonon frequency at the critical
temperature gives rise to a macroscopically occupied phonon modes with wavenumbers $Q_i$ which, physically, corresponds
to a static periodic distortion of the lattice displacements $u_i(\vec{r})$~\cite{Rice:1973}.  Accordingly, we may introduce the
order parameters:
\begin{equation}
\label{5.61} 
\Delta_{CDW}(Q_i)  \; = \; \frac{g(Q_i)}{V_u}  \; \left [  < \hat{b}_{i}(\vec{Q}_i ) > \; + \;  < \hat{b}^{\dagger}_{i}( - \vec{Q} )_i >
   \right ]    \; .
\end{equation}
Since  $< \hat{b}^{\dagger}_{i}( - \vec{Q} )_i >  =  < \hat{b}_{i}(\vec{Q}_i ) >$ we have:
\begin{equation}
\label{5.62} 
\Delta_{CDW}(Q_i)  \; = \; \frac{ 2 g(Q_i)}{V_u}  \;   < \hat{b}_{i}(\vec{Q}_i ) >      \; .
\end{equation}
Evidently we also have:
\begin{equation}
\label{5.63} 
\Delta_{CDW}(Q_1)  \; = \; \Delta_{CDW}(Q_2)  \; \equiv \;   \Delta_{CDW} \; .
\end{equation}
Moreover, both the phonons with wavenumbers $\pm \vec{Q}_i$ are involved, so that the order parameter may be assumed to be
a real number.  According to Eq.~(\ref{5.9}) the resulting lattice displacement distortions are:
\begin{eqnarray}
\label{5.64} 
 \Delta u_i(\vec{r})   \; =  \;  < u_i(\vec{r}) > \; = \; 
   \sqrt{ \frac{\hslash}{  2 M \Omega(Q_i) N_u} } \; 
 \\ \nonumber  
    \frac{2 \Delta_{CDW} V_u}{g(Q_i)}    \;  \cos (\vec{Q}_i  \cdot \vec{r})  \; .
\end{eqnarray}
To determine the order parameter we minimize the condensation energy per unit cell. To this end, in Appendix~\ref{AppendixE}
we show that, in the mean field approximation, the effective Hamiltonian is given by:
\begin{eqnarray}
\label{5.75} 
\hat{H}_{eff} \;  \simeq  \;   \frac{ V_u^2 \Delta_{CDW}^2 }{ g^2(Q_1)}  \; \frac{\hslash \Omega(Q_1)}{2}  \hspace{1.5cm}
\\ \nonumber
  +  \;  V_u  \;   {\cal{N}}(0) 
 \sum_{\sigma}   \int  d \varepsilon^{(e)}_{\vec{k}} \;  \tilde{\varepsilon}^{(e)}_{\vec{k}}  \hspace{1.0cm}
\\ \nonumber  
 \bigg [  \hat{\tilde{\psi}}^{\dagger}_{1}(\vec{k},\sigma) 
   \hat{\tilde{\psi}}_{1}(\vec{k},\sigma)  \;  +  \;  \hat{\tilde{\psi}}^{\dagger}_{2}(\vec{k},\sigma)    \hat{\tilde{\psi}}_{2}(\vec{k},\sigma)  \bigg ]
   \; ,
\end{eqnarray}
where:
\begin{equation}
\label{5.76} 
\tilde{\varepsilon}^{(e)}_{\vec{k}} \; = \; sign (k \; - \; k_F^{(e)} ) \; 
\sqrt{  \xi^2_{\vec{k}}+ \Delta^2_{CDW}}   
\end{equation}
is the energy measured with respect to the Fermi energy. \\
The Hamiltonian Eq.~(\ref{5.75}) shows that in the charge density wave state the spectrum of the nodal quasielectron excitations
develops a gap  $\Delta_{CDW}$.  The opening of the gap leads, in turns, to the lowering of the electron energy:
\begin{equation}
\label{5.77} 
\Delta E_c^{(e)} \simeq  2 \,  V_u  \,   {\cal{N}}(0)  \int_0^{\varepsilon_c}  
d \xi_{\vec{k}} \left [  \xi_{\vec{k}}  -   \sqrt{  \xi^2_{\vec{k}}+ \Delta^2_{CDW}}  \right ] \; .
\end{equation}
 A straightforward integration gives:
\begin{eqnarray}
\label{5.78} 
\Delta E_c^{(e)} \simeq  2 \,  V_u  \,   {\cal{N}}(0)  \Bigg \{ \frac{  \varepsilon_c}{2} \; - \;
\frac{1}{2}   \varepsilon_c \sqrt{ \varepsilon^2_c +  \Delta^2_{CDW}} \; 
\\ \nonumber
- \; \frac{\Delta^2_{CDW}}{2}
 \ln \bigg [ \frac{ \varepsilon_c +\sqrt{ \varepsilon^2_c +  \Delta^2_{CDW}}}{ \Delta_{CDW}} \bigg ]
  \Bigg \} \; .
\end{eqnarray}
In the weak coupling limit $\Delta_{CDW} \ll   \varepsilon_c$ we obtain:
\begin{eqnarray}
\label{5.79} 
\Delta E_c^{(e)} \; \simeq \;  2 \,  V_u  \,   {\cal{N}}(0)  \Bigg \{ - \;  \frac{\Delta^2_{CDW}}{4} \; 
\\ \nonumber
- \; \frac{\Delta^2_{CDW}}{2} \;  \ln \bigg ( \frac{ 2 \varepsilon_c }{ \Delta_{CDW}} \bigg )
  \Bigg \} \; .
\end{eqnarray}
The total  condensation energy is therefore:
\begin{eqnarray}
\label{5.80} 
\Delta E_c \;  \simeq  \;   \frac{ V_u^2 \Delta_{CDW}^2 }{ g^2(Q_1)}  \; \frac{\hslash \Omega(Q_1)}{2}  \; + \;  \hspace{3.0cm}
\\ \nonumber
2 \,  V_u  \,   {\cal{N}}(0)  \Bigg \{ - \;  \frac{\Delta^2_{CDW}}{4} \; - \;
 \frac{\Delta^2_{CDW}}{2}   \ln \bigg ( \frac{ 2 \varepsilon_c }{ \Delta_{CDW}} \bigg)
  \Bigg \} \; .
\end{eqnarray}
Minimizing the condensation energy with respect to  $\Delta_{CDW}$ we get:
\begin{equation}
\label{5.81} 
\Delta_{CDW} \; \simeq \;  2 \,  \varepsilon_c \;
e^{ - \;  
   \frac{ \hslash \Omega(Q_1) V_u   }{ 2 g^2(Q_1)  {\cal{N}}(0)    } }  \;  .
\end{equation}
The whole approach can be extended easily to finite temperatures. In fact, the thermodynamics has been worked out for the first time in
Ref.~\cite{Kuper:1955}. It can be seen that the thermodynamics of the phase transition and the temperature dependence
of the charge density wave gap are the same as those in the s-wave weak coupling BCS superconductors~\cite{Rice:1973,Nielsen:1980}.
Therefore, for the critical temperature one finds:
\begin{equation}
\label{5.82} 
k_B \; T_{CDW} \simeq  \frac{e^\gamma}{\pi} \; \Delta_{CDW}  \simeq   \frac{2 e^\gamma}{\pi}  \,  \varepsilon_c \;
e^{ - \;  
   \frac{ \hslash \Omega(Q_1) V_u   }{ 2 g^2(Q_1)  {\cal{N}}(0)    } }  \;  .
\end{equation}
The temperature dependence of the gap is given by:
\begin{eqnarray}
\label{5.83}
   \ln  \bigg (  \frac{T } {T_{CDW} } \bigg )  \; \simeq  \;  -  \;   
    \int_{0}^{\infty}  d x    \Bigg \{ \; 
 \frac{1}{ x }  \; \tanh (\frac{x}{2})   \hspace{1.0cm} 
 \\ \nonumber  
  - \; \frac{ 1} { \sqrt{ x^2 +  z^2 } }  \; \tanh (\frac{ \sqrt{ x^2 +  z^2 }}{2}) \; \Bigg \}  \; \; , \; \;   
  z = \frac{\Delta_{CDW}(T)}{k_B \, T} \;  . 
\end{eqnarray}
It is noteworthy to  emphasize   that the charge density wave critical temperature given by Eq.~(\ref{5.82}) does not agree with
Eq.~(\ref{5.56}). In fact, the argument of the exponential in  Eq.~(\ref{5.56}) is one half the one in Eq.~(\ref{5.82}). This means 
that  Eq.~(\ref{5.56}) overestimates by a considerable amount the charge density wave critical temperature. This difference
resides into the fact that in the estimate  Eq.~(\ref{5.56})  one does not include the quasielectron correlations built in the
ground state wavefunction. On the other hand,   Eq.~(\ref{5.82}) includes these correlation effects, albeit in the mean field
approximation. \\
It is interesting to display explicitly the doping dependence of the charge density wave critical temperature. Instead of  Eq.~(\ref{5.60}),
now we have:
\begin{eqnarray}
\label{5.84} 
T_{CDW}(\delta)  \; \simeq \;  \frac{4 e^\gamma}{\pi}  \;  T^*(0) \; \left [ 1 \; - \;  (\frac{\delta}{\delta_c})^{\frac{3}{2}}  \right ]    \;
\\ \nonumber
e^{ - \;  \frac{1}{2 \pi} \; 
   (\frac{ \hslash \Omega(Q_1) a_0 }{e^2})^2 \frac{M}{m^*_e}   \; \frac{1 - \delta}{\delta}   }   \; .
\end{eqnarray}
Eq.~(\ref{5.84}) can be contrasted to experimental observations. To this end, in Fig.~\ref{Fig19} we display the observed phase
diagram for YBCO in the pseudogap region. The dashed curve outlines  the doping dependence of the critical temperature
$T_c(\delta)$. The data for the critical superconductive temperature have been taken from Fig.~3 in Ref.~\cite{Liang:2006}. 
The pseudogap  temperatures $T^*$ have been determined by neutron diffraction measurement  (full circles) and resonant
ultrasound (full diamonds).  The data have been taken from Fig.~3 in Ref.~\cite{Shekhter:2013}.
%
\begin{figure}
\vspace{0.7cm}
\centering
\resizebox{0.5\textwidth}{!}{%
\includegraphics{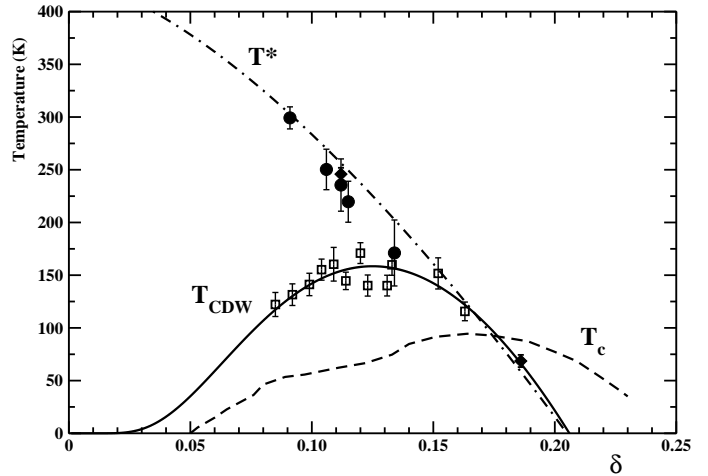} }
\caption{\label{Fig19} The phase diagram in the pseudogap region for YBCO.  The dashed curve is the superconductive critical
temperature $T_c$. Full circles and diamonds are the pseudogap  temperatures $T^*$ as determined by neutron diffraction 
measurement  and resonant ultrasound  respectively.   The dot-dashed line is  Eq.~(\ref{5.58}) with $T^*(0)$ given by
 Eq.~(\ref{5.85}). The charge density wave critical temperatures $T_{CDW}$ (open squares) have been estimated by
 means of X-ray diffraction experiments. The full curve is our  Eq.~(\ref{5.84}). }
\end{figure}
%
Note that the data are in satisfying agreement with  our Eq.~(\ref{5.58}),  displayed as  the dot-dashed line
in   Fig.~\ref{Fig19},  by assuming:
\begin{equation}
\label{5.85} 
T^*(0) \; \simeq \;  430 \;  K     \;  .
\end{equation}
As regard the charge density wave critical temperature, the onset of charge order are detected in X-ray diffraction below
the critical temperature $T_{CDW}$. The data (open squares) have been taken from Fig.~10 in Ref.~\cite{Blanco:2014b} and
Table~I in Ref.~\cite{Hucker:2014}.  To compare our theoretical estimate of the charge density critical temperature, Eq.~(\ref{5.84}),
we used the numerical values of the model parameters and $T^*(0)$ given in Eq.~(\ref{5.85}).  The reduced  atomic mass of the 
$Cu$ and $O$ ions is evaluated by  Eq.~(\ref{5.3}) using   $M_{Cu}  \simeq 63.5 \, M_P$ and  $M_{O} \simeq 16 \, M_P$, $M_P$
being the proton mass.  To match  the data we used:
\begin{equation}
\label{5.86} 
 \hslash \, \Omega(Q_1) \; \simeq \;  35.5 \;  mev     
\end{equation}
corresponding to a frequency $\sim \, 10 \; THz$,  which compares well with the observed bond-stretching frequencies in 
cuprates. Indeed, Fig.~\ref{Fig19} shows that Eq.~(\ref{5.84}) is in satisfying agreement with the charge density wave critical
temperature data. In particular we see that our theoretical calculations reproduce the shallow maximum at $\delta \simeq 0.12$
displayed by the data.  It is worth  to note that the charge density wave long range order in cuprates seems to set in in
a rather small  hole doping range, $0.08 \, \lesssim \, \delta \, \lesssim 0.17$,  around the charge density wave critical temperature
maximum at   $\delta \simeq 0.12$.  
In our opinion this could be easily explained if  the coherence length in charge density wave  
 cannot grow too much. Indeed,  the coherence length may be hindered by crystalline  imperfections and defects which are
 inevitably present in hole doped cuprates.  In general, the appearance of the energy gap  $\Delta_{CDW}$ also leads to a
 finite coherence length $\xi_{CDW}$. The  expression of the coherence length is the same as  the BCS coherence length 
 and, at zero temperature, it is given  by: 
\begin{equation}
\label{5.87} 
\xi_{CDW} \; \simeq \;     \frac{ \hslash \; v_F^{(e)}}{\pi \, \Delta_{CDW}}      \; , 
\end{equation}
 where $ v_F^{(e)}$ is the  quasielectron Fermi velocity.  From Eq. (\ref{5.87}) we may easily infer the doping dependence of
 the charge density wave coherence length:
\begin{equation}
\label{5.88} 
\xi_{CDW}(\delta) \; \simeq \;     \frac{ 4 \; a_0}{\sqrt{ 2 \, \pi}}  \; \sqrt{1 \; - \delta}    \;  \;  \frac{t}{  \Delta_{CDW}(\delta)}  \; ,
\end{equation}
 where, according to Eq.~(\ref{5.82}):
\begin{equation}
\label{5.89} 
 \Delta_{CDW}(\delta)  \; \simeq \;  \frac{\pi}{e^\gamma} \; k_B \; T_{CDW}(\delta) \;   \;  .
\end{equation}
%
\begin{figure}
\vspace{0.7cm}
\hspace{-0.3cm}
\resizebox{0.5\textwidth}{!}{%
\includegraphics{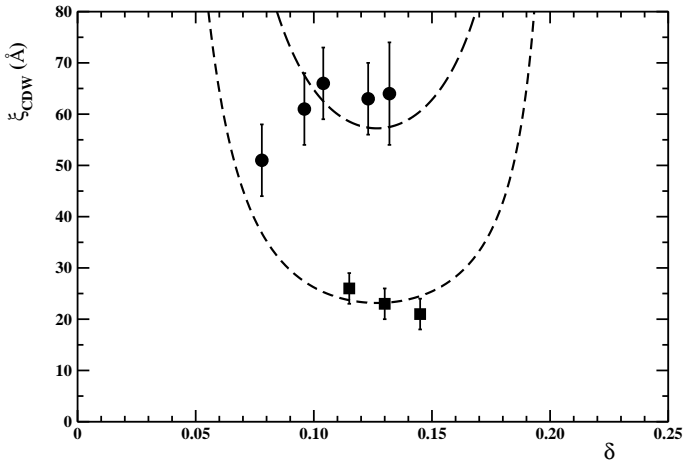} }
\caption{\label{Fig20} The charge density wave coherence length versus the hole doping for YBCO (full circles) and Bi-2201 (full squares).
The dashed curves correspond to Eq.~(\ref{5.88}) scaled by a factor $2.1$ and $0.85$ for YBCO and Bi-2201 respectively.}
\end{figure}
%
In Fig.~\ref{Fig20} we show the doping dependence of the coherence length and compare with available data for 
YBCO and $Bi_2Sr_{2-x}La_xCuO_{6+\delta}$ (Bi-2201). The data for YBCO have been taken from Table~I of Ref.~\cite{Hucker:2014}
where the doping dependence of the charge density wave order was analyzed  with a bulk-sensitive high-energy x-ray study.
The data for Bi-2201   have been taken from  Table~1 of Ref.~\cite{Comin:2014} where charge order was observed by combining
resonant x-ray scattering, scanning-tunneling microscopy, and angle resolved photoemission spectroscopy. To mach the experimental
data we scaled  $\xi_{CDW}(\delta)$ by a factor $2.1$ and $0.85$ for YBCO and Bi-2201 respectively. In any case, we see
that in the doping range $0.08 \, \lesssim \, \delta \, \lesssim 0.17$ the coherence length does not vary appreciably with the doping.
However, outside that doping range   $\xi_{CDW}$ increases suddenly.  Now, it is reasonable to assume that
the presence of lattice   imperfections and defects tend to  impede the  grow of the coherence length beyond a certain length.
Therefore, outside the  small  hole doping range  around  $\delta \simeq 0.12$ the charge order cannot  develop as a true 
long-range order, but it should manifest itself through  short-range  correlations. \\
Let us consider the small Peierls lattice deformations which, after using  Eq.~(\ref{5.64}), can be written as:
\begin{eqnarray}
\label{5.90} 
 \Delta \vec{u}(\vec{r})   \; =  \;  < \vec{u}(\vec{r}) > \; \simeq \; 
   \sqrt{ \frac{\hslash}{  2 M \Omega(Q) N_u} } \;  \hspace{1.0cm}
   \\ \nonumber
    \frac{2 \Delta_{CDW} V_u}{g(Q)}    \; 
  \bigg [  \cos (Q x ) \; \hat{x}   \; + \;   \cos (Q y) \; \hat{y} \bigg ] \; ,
\end{eqnarray}
where we used
\begin{equation}
\label{5.91} 
   \vec{Q}_1 \; = \; (Q ,  0)  \; \; , \; \;    \vec{Q}_2 \; = \; (0 , Q)  \; \; , \; \;  Q \; \simeq \;    \sqrt{2} \, k_F^{(e)} \; ,
\end{equation}
and
\begin{equation}
\label{5.92} 
  g(Q_1)  \simeq  g(Q_2)   \equiv g(Q)  \; \; , \; \;     \Omega(Q_1)   \simeq 
  \Omega(Q_2)  \equiv  \Omega(Q)  \; .
\end{equation}
The charge-order wavenumber  vectors are conventionally  expressed in terms of  the reciprocal lattice constant:
\begin{equation}
\label{5.93} 
a_0^* \;  =  \; \frac{2 \pi}{a_0}  
\end{equation}
as:
\begin{equation}
\label{5.94} 
   \vec{Q}_1 \; = \;  a_0^* \, (H ,  0)  \; \; , \; \;    \vec{Q}_2 \; = \;  a_0^* \, (0 ,  K)  \; \; .
\end{equation}
 After using Eq.~(\ref{4.11}), we find:
\begin{equation}
\label{5.95} 
H \; = \; K \;  \simeq \; \sqrt{\frac{1 - \delta}{\pi} } \; .
\end{equation}
From Eq.~(\ref{5.95}) it follows that $Q$ smoothly decreases upon increasing the hole doping $\delta$ in fair  agreement with observations.
However,  the observed charge-order wavenumbers  is not much more than  half the value as here determined. This is, mainly,
due to our over exemplified model which leads to an overestimate of the nesting wavevectors of the nodal quasielectron
Fermi circle. \\
In our approach the charge density wave  always exhibits  wavenumber vectors  parallel to the planar Cu-O bonds. It is interesting to note 
that in  Ref.~\cite{LeBoeuf:2013} it is  provided the first thermodynamic signature of the charge-order phase transition in underdoped
YBCO.  In particular, the comparison of different acoustic modes indicated that the char-ge modulation were biaxial, namely  directed
both along the $\hat{x}$  and  $\hat{y}$ axes. Remarkably, in Ref.\cite{Comin:2015b},  by using resonant X-ray scattering
to resolve the charge modulations in the two cuprate families Bi-2201 and YBCO in the underdoped region, it was  established that
the charge modulations run parallel to the copper-oxigen bond directions.  Moreover the pattern of the charge density wave turned out
to be compatible with Eq.~(\ref{5.90}). \\
Finally, it is worthwhile to estimate quantitatively the amplitude of the atomic planar displacement induced  by the charge density wave.
Let us consider the small Peierls lattice deformations which, after using  Eq.~(\ref{5.64}), can be written as:
\begin{equation}
\label{5.96} 
 \Delta \, u    \;   \simeq \;   \sqrt{ \frac{\hslash}{  2 M \Omega(Q)} } \;  \frac{2 \Delta_{CDW} V_u}{g(Q)}    \; .
\end{equation}
Ref.~\cite{Chang:2012} reported  the  X-ray diffraction study of a  detwinned   single crystal of YBCO  with 
hole concentration per planar Cu  $\delta \simeq 0.12$ ($T_c \simeq 67 \, K$).
The authors of   Ref.~\cite{Chang:2012}  by using high-energy X-ray diffraction showed  that a charge density wave  
develops  in the normal state of superconducting YBCO below the critical temperature
 $T_{CDW}  \simeq 135 \, K$.  In particular,  from the intensity ratio between the incommensurate satellite peaks and Bragg 
 reflection  peaks these authors were able to  estimate the amplitude of the lattice distortion in the charge density wave. 
 In fact, they reported the upper limit~\cite{Chang:2012}:
\begin{equation}
\label{5.97} 
\frac{\Delta u}{a_0}  \;  \lesssim \;  10^{-3}   \;   \; .
\end{equation}
To determine $\Delta u$, we need the charge density wave gap. Since for the YBCO crystal used in Ref.~\cite{Chang:2012} 
the charge density wave critical temperature was   $T_{CDW}  \simeq 135 \, K$, using  Eq.~({\ref{5.89}) we find:
\begin{equation}
\label{5.98} 
 \Delta_{CDW}   \;  \simeq  \;   20.5  \; mev   \; \; .
\end{equation}
Once we known the charge density wave gap, we may easily evaluate the lattice displacement by Eq.~({\ref{5.96}):
\begin{equation}
\label{5.99} 
\frac{\Delta u}{a_0}  \;  \simeq  \;  1.8   \; 10^{-3}   \; \;,
\end{equation}
which, indeed, is in reasonable agreement with Eq.~(\ref{5.97}).
\subsection{Competition between charge density wave and superconductivity }
\label{s5.3}
We said that there is growing evidence of a charge order existing in the pseudogap state of several cuprate families.
Actually, charge density wave order and superconductivity are competing phases.  It is widely believed that understanding
the interplay between superconductivity and charge order is essential to clarify the origin of the high temperature
superconductivity in cuprate materials. In fact, there are several experimental observations showing that superconductivity
weakens the charge density order and, conversely, the charge order tends to weaken superconductivity. Moreover,
application of a magnetic field restores the charge density wave amplitude below the superconductive critical temperature,
while it has no appreciable effect for temperatures above the critical temperature. The effects of applied magnetic fields
on the charge density wave order will be discussed in the next  Section. In the present Section we focus on the competition
between charge density wave and superconductivity. \\
As we have already discussed,  the charge-order state is driven by the nodal Fermi arc instability. Moreover, we know that
nodal quasielectron excitations are possible if the paired holes are phase disordered.  Therefore, by  using an argument similar 
to that employed in Sect.~\ref{s4.3}, we see that the number of  the nodal quasielectron excitations is reduced in the
superconductive region by:
\begin{equation}
\label{5.100} 
\frac{n_n(T)}{n_s(0)}  \;    \simeq    \; 
1 -  e^{  - \, b' \left [ \frac{1}{\sqrt{1 \, - \, T/T_c}}  -  1 \right ] }   \;  \; , \; \;  
T \; \le \; T_c  \; .
\end{equation}
Qualitatively, it is evident that below the superconductive critical temperature the number of available nodal quasielectron
excitations decreases thereby suppressing the char-ge density wave gap. Since the fully microscopic approach becomes
difficult, the simplest way to determine quantitatively the expected depletion of the charge density wave gap in the
superconductive phase is to deal with the free energy functional within the Ginzburg-Landau theory. The free energy functional
appropriate to the regime of small charge density wave order parameter $\Delta(\vec{r})$ can be written as:
\begin{equation}
\label{5.101} 
F_{CDW}[\Delta]     \simeq    F[0] \; +  \int d \vec{r}  \Big \{
a(T) \; | \Delta(\vec{r})|^2 +   b(T) \; | \Delta(\vec{r})|^4  \Big \}   \; \;  .
\end{equation}
In Eq.~(\ref{5.101}) we are neglecting the energy associated with the spatial variation of the order parameter. Indeed, we 
are mainly interested in the limit of homogeneous charge density order parameter where $\Delta(\vec{r})$ reduces
to $\Delta_{CDW}$.  The two coefficients in the Ginzburg-Landau free energy functional can be evaluated near the
transition temperature $T_{CDW}$ within the weak coupling s-wave BCS microscopic theory. One finds (see, eg, 
Ref.~\cite{Leggett:2012b}):
\begin{equation}
\label{5.102} 
a(T)    \;    \simeq  \;   {\cal{N}}(0)  \;  \Big ( \frac{T}{T_{CDW}} \; - \; 1 \Big ) 
\end{equation}
and
\begin{equation}
\label{5.103} 
b(T)    \;    \simeq   \;   {\cal{N}}(0)  \;  \frac{ 7 \;  \zeta(3)}{8 \pi^2 k^2_B T^2_{CDW} }    \; \;  ,
\end{equation}
where $\zeta(z)$ is the  Euler-Riemann zeta function~\cite{Gradshteyn:1980}.  Note that, within this approximation, the coefficient
$b$ is almost independent on the temperature while $a(T)$ is positive above the charge density critical  temperature, vanishes at $T_{CDW}$,
and it becomes negative below the transition temperature. \\
The temperature dependence of the order parameter is obtained from the equilibrium condition:
\begin{equation}
\label{5.104} 
\frac{ \delta \,  F_{CDW}[\Delta]}{\delta \; \Delta(\vec{r})}    \;   =  \;  0 \; \; . 
\end{equation}
For homogeneous order parameter, above the transition temperature the minimum of the free energy functional is at $\Delta = 0$.
For temperatures below $T_{CDW}$ we get instead:
\begin{equation}
\label{5.105} 
\Delta^2(T)    \;    \simeq   \;   \frac{8 \pi^2  }{ 7  \zeta(3)} \; k^2_B T^2_{CDW} \; 
\left [ 1 \; - \; \frac{T}{T_{CDW}}  \right  ] \; \; .
\end{equation}
It  can easily seen that, as expected,   Eq.~(\ref{5.105}) agrees with the solution of the BCS Eq.~(\ref{5.83}) at least for temperatures
not too far from the charge density wave critical temperature.  \\
In the superconductive region we may take care of the depletion of  the nodal quasielectron excitations by allowing the
Ginzubg-Landau parameter $a(T)$ to be reduced according to Eq.~(\ref{5.100}). Therefore, we are led to assume: 
\begin{eqnarray}
\label{5.106} 
\nonumber
  a(T)   \simeq   {\cal{N}}(0)   \Big ( \frac{T}{T_{CDW}}  -  1 \Big )  \hspace{0.5cm}   T_c  \le    T   \le  T_{CDW}   
 \\ 
 \\ \nonumber
a(T)  \simeq    {\cal{N}}(0)  \Bigg \{ 1 -  e^{  - \, b' \big [ \frac{1}{\sqrt{1 \, - \, T/T_c}}  -  1 \big] }  \Bigg \}   \hspace{1.3cm}
 \\ \nonumber
 \Big ( \frac{T}{T_{CDW}} \; - \; 1 \Big )  \;  \;   \; \; \;   T \; \le \; T_c  \; \; \; 
\end{eqnarray}
while the parameter $b$ is still  kept temperature independent.  Correspondingly,  for homogeneous order parameter
and for temperatures below $T_{CDW}$ we obtain:
\begin{eqnarray}
\label{5.107} 
\nonumber
   \Delta(T)    \;    \simeq   \; \Delta_{CDW}(T)   \hspace{1.5cm}  T_c \; \le  \;  T  \; \le \; T_{CDW}   \hspace{0.5cm}
 \\
\\ \nonumber 
    \Delta(T)    \simeq   \sqrt{1 -  e^{  - \, b' \left [ \frac{1}{\sqrt{1 \, - \, T/T_c}}  -  1 \right ] } }  
    \Delta_{CDW}(T)   \;  \;  \;   T \; \le \; T_c 
\end{eqnarray}
where  $\Delta_{CDW}(T)$ is the the solution of  Eq.~(\ref{5.83}).  In fact,  Eq.~(\ref{5.107}) implies that the charge density wave
gap is quickly reduced in the superconductive phase in qualitative agreements with observations.  For illustrative purposes,
we show here how our results  offer a consistent interpretation of the recent  angle resolved photoemission spectroscopy studies
presented in Ref.~\cite{Kondo:2015}.  These authors examined the momentum-resolved single particle spectra of hole
doped  cuprate Bi-2212 in the optimal doping region by utilizing a low-energy laser source which allowed to obtain  high-quality
spectra with very sharp line shapes.  The high quality of the data allowed to determine  the energy gap as the peak energy of 
the single particle spectra. In Fig.~\ref{Fig21} we report the temperature dependence of the spectral gap for optimal doped
Bi-2212 ($T_c \simeq 92 \, K$). The data have been extracted from Fig.~2, panel a), of  Ref.~\cite{Kondo:2015}.
%
\begin{figure}
\vspace{0.7cm}
\hspace{-0.4cm}
\resizebox{0.5\textwidth}{!}{%
\includegraphics{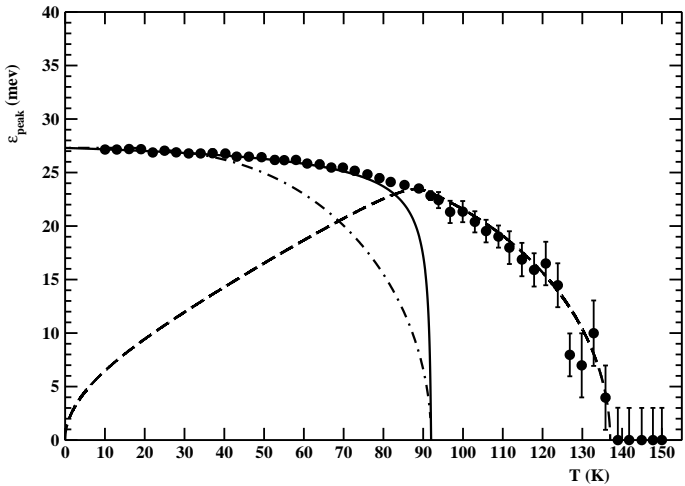} }
\caption{\label{Fig21} The spectral peak energy versus the temperature  for optimal doped Bi-2212, $T_c \simeq 92 \, K$.
The data have been extracted from Fig.~2, panel a), of  Ref.~\cite{Kondo:2015}. The solid line is the nodal gap Eq.~(\ref{4.21}) 
with $b \simeq 0.08$ and $\Delta_{nodal}(0) \simeq 27.3 \, mev$.  For comparison we also display the weak coupling d-wave BCS gap 
(dot-dashed line). The dashed curve is the charge density wave gap  Eq.~(\ref{5.107}) with $\Delta_{CDW}(0) \simeq 27.0 \, mev$,  
$T_{CDW}  \simeq 137 \, K$, and $b' \simeq 1.0$.}
\end{figure}
%
The data correspond to $\varepsilon_{peak}$ at $\phi \simeq 24 ^0$, where $\phi$ gives the direction of the Fermi wavevector as
defined in Fig.~3, panel d), of  Ref.~\cite{Kondo:2015}. As it is evident from  Fig.~\ref{Fig21}, the spectral energy gap does not close
at the superconductive critical temperature $T_c \simeq 92 \, K$, but  it survives beyond $T_c$ until the higher  temperature 
$T_{pair} \simeq 137 \, K$.  According to Ref.~\cite{Kondo:2015} the spectral gap $\varepsilon_{peak}$  is interpreted as the
energy pairing-gap in the nodal region persisting even above the superconductive critical temperature, and vanishing only at the critical
temperature $T_{pair}$ far above $T_c$.  Moreover, this pairing-gap seems to evolve with temperature according to the weak coupling
s-wave BCS gap function with onset temperature  $T_{pair}$. Actually, this interpretation is in contradiction with several observations which
pointed to a phase-incoherent superconductivity in the pseudogap region. In order to reach a sensible picture of these puzzling results
we must assume that the temperature  $T_{pair}$ signals the onset of charge order,  $T_{pair} = T_{CDW}$. In fact, we have already seen
in Sect.~\ref{s5.2} that long-range charge order in cuprates sets in a small hole doping range near the optimal doped region (see 
Fig.~\ref{Fig19}). Note that within this interpretation we may explain naturally both the absence of the gapless Fermi arcs above
the superconductive temperature and the presence of Bogoliubov quasiparticle low-lying excitations~\cite{Kondo:2015}
due to the charge density wave gapped state. In addition, the temperature evolution of the spectral gap as implied by
 Eq.~(\ref{5.107}) is in satisfying  agreement with experimental data  for temperatures above the superconductive critical temperature
 $T_c$ (see the dashed line in  Fig.~\ref{Fig21}).  In the superconductive region the charge density wave gap is rapidly suppressed. 
 In this region the observed single-particle gap is caused by the nodal gap as discussed at length in Sect.~\ref{s4.1}. Remarkably, 
 the thermal evolution of the nodal gap, Eq.~(\ref{4.21}), is quite consistent with the data at least for temperatures not too close to the  
  superconductive critical temperature (see the full curve in Fig.~\ref{Fig21}). Near the critical temperature $T_c$ there is competition 
  between the two gaps, so that the spectral gap evolves smoothly from the nodal gap to the charge density gap.
\subsection{Effects of the magnetic field}
\label{s5.4}
Several experimental observations indicated that the  charge density wave  is unaffected by applied magnetic fields in the normal state. 
In fact, for temperatures above the superconductive critical temperature $T_c$, a  magnetic field applied perpendicular to the $CuO_2$ plane
has no appreciable effects. However, in the superconductive region $T <  T_c$  an applied magnetic field  causes an increase of the 
intensity of  the charge density wave  signal. The magnetic field also seem to make  the charge density wave order  more coherent.
These experimental observations are usually interpreted as a clear  evidence for competition between  charge density wave  
and superconducting orders.  \\
Let us, now, discuss in our model the effects of the magnetic field on the charge density wave.  It will be shown that
the effects of magnetic fields on the charge density wave ground state are sizable in the superconductive region, while
they are  negligible small in the normal region. Indeed, in the normal non-superconductive phase the magnetic field acts
on the spins of the nodal quasielectrons only since, in general, the effect of spin-orbit interactions are negligible.
The Zeeman splitting of the quasielectron energy at the Fermi level will reduce the pairing interaction and eventually
it leads to a non-condensate metallic state where the charge density wave energy gap is driven to zero. It turns out
that the problem is quite similar to the influence of applied external fields on the Peierls instability in quasi  one-dimensional
conductors.  As we said, there is a formal resemblance between the energy gap in the Peierls state and the energy gap
in weak coupling s-wave BCS superconductors. In fact,  in analogy to the calculations in the theory of 
superconductivity~\cite{Maki:1964} one finds~\cite{Dieterich:1973}:
\begin{eqnarray}
\label{5.108} 
\ln \frac{T_{CDW}(H)}{T_{CDW}} \; + \; \Re \; \psi \bigg ( \frac{1}{2} + i \; \frac{\mu_B H}{2 \pi k_B T_{CDW}(H)} \bigg ) 
\hspace{1.0cm}
\\ \nonumber
\; - \; \psi (\frac{1}{2}) \; = \; 0  \; \; , \hspace{0.8cm}
\end{eqnarray}
where $\psi(z)$ is the  digamma function~\cite{Gradshteyn:1980}. In Eq.~(\ref{5.108}) $T_{CDW}(H)$ is the charge density wave
critical temperature in presence of a transverse  magnetic field $H$. By expanding to the first non-trivial order in $\mu_BH$
we find:
\begin{equation}
\label{5.109} 
\frac{T_{CDW}(H)}{T_{CDW}} \; \simeq  \;  1 \; + \; \frac{1}{8 \pi^2} \; \psi ''( \frac{1}{2}) \; \left ( \frac{\mu_B H}{k_B T_{CDW}} 
\right )^2  ,
\end{equation}
where the prime  denotes the derivative. Using~\cite{Gradshteyn:1980}:
\begin{equation}
\label{5.110} 
 \psi ''( \frac{1}{2}) \; = \; - \; 2 \; (2^3 - 1) \;  \zeta(3) \; \; \; , 
 \end{equation}
we obtain:
\begin{equation}
\label{5.111} 
\frac{T_{CDW}(H) - T_{CDW}}{T_{CDW}}  \simeq    -  \frac{7}{4 \pi^2}  \;  \zeta(3)  \left ( \frac{\mu_B H}{ k_B T_{CDW}} 
\right )^2  .
\end{equation}
Note that this last result can be obtained directly by repeating the calculations presented in Ref.~\cite{Rice:1973} for the case
in which the quasielectron spin interacts with an external magnetic field~\cite{Tiedje:1975}. 
From Eq.~(\ref{5.111}) it follows that even for magnetic field strength up to $H \sim 10^2 \; T$  the charge density wave critical temperature
decreases by a few percent only. In other words,  in the non-superconductive region  external magnetic fields do not affect appreciably
the charge density wave ground state. However, we will see in the next Section that the dependence of the charge density
wave gap on the magnetic field implied by Eq.~(\ref{5.111})  affects in a non-trivial  manner the frequencies in  the quantum oscillation
phenomena in underdoped cuprate superconductors. \\
In the superconductive region we have already remarked that the number of available nodal quasielectron excitations decreases 
 leading to a quickly suppression of  the charge density wave instability.  
However, there are several experimental observations  suggesting the restoration of charge order in presence of  applied
magnetic fields.  Intriguingly, in Ref.~\cite{Wu:2013} the charge order in underdoped YBCO was evidenced  by using the line splitting of 
the nuclear magnetic resonance of some of copper and oxygen sites in $CuO_2$ planes. Observations in the
superconductive region indicated a sharp set in of charge order starting above a threshold transverse magnetic field.
In particular the authors of Ref.~\cite{Wu:2013} found that charge order occurs for temperatures below an onset temperature
$T_{charge}$ and for magnetic fields above a threshold field $H_{charge}$. This threshold magnetic field turned out to be
weakly dependent on the hole doping level with values varying in the range $ H_{charge} = 9 - 15 \, T$. Evidently  the finite value
of the threshold magnetic field implies that there is no static long-range charge order in absence of external magnetic field.
This gives a clear indication for a field-dependent transition to the charge ordered state. As the authors of Ref.~\cite{Wu:2013}
argued, the finite threshold magnetic field can be attributed to the presence of Abrikosov vortices.  A magnetic field applied 
perpendicular to the $CuO_2$ planes generates vortices. Now,  the vortex cores represent 
normal region of radius $\xi_V$ within the superconductor. So that  it is expected that the charge order fluctuations detected in the normal
region above the superconductive critical temperature continue to develop at low temperatures $T < T_c$ within the 
cores~\cite{Wu:2013,Hoffman:2002}.
Therefore the halos of charge order are centered on the Abrikosov vortex cores and they extend over a typical distance
of order of the charge density wave coherence length $\xi_{CDW}$.  By increasing the strength of the magnetic field more 
Abrikosov vortices are added. Thus, the  long-range charge order may be expected to appear once these halos start to overlaps.
We said  in Sect.~\ref{s3.5} that in our theory the upper critical magnetic field is much smaller than the Landau-Ginzburg critical magnetic
field where the Abrikosov vortices become to overlap. So that we will employ  the dilute vortex approximation with  the density of Abrikosov 
vortices  given by Eq.~(\ref{4.74}). Since the average distance between vortices is  $d_H  \simeq  \sqrt{\frac{\phi_0}{H}}$, the
onset of the long-range static charge density order  happens when: 
\begin{equation}
\label{5.112} 
d_H \; \simeq \; \sqrt{\frac{\phi_0}{H}} \;  \lesssim \; 2 \; \xi_{CDW}  \; \; .
\end{equation}
From Eq.~(\ref{5.112}) we easily obtain:
\begin{equation}
\label{5.113} 
H \;  \gtrsim \;  \frac{\phi_0}{4 \; \xi^2_{CDW}}  \; \simeq \; H_{charge} \; \;  .
\end{equation}
It is useful to estimate quantitatively our determination of the threshold magnetic field. For YBCO from Fig.~\ref{Fig20} we
see that $ \xi_{CDW} \simeq 65 \, \text{\AA} $ almost independently on the hole doping  around $\delta \simeq 0.12$. 
So that we get:
\begin{equation}
\label{5.114} 
 H_{charge} \;  \simeq  \;  \frac{\phi_0}{4 \; \xi^2_{CDW}}  \; \simeq \;  12.3 \; T \; \; ,
\end{equation}
which compares rather well with observations. 
The long-range charge order, which sets in the superconductive region for magnetic fields above the threshold field, persists for
temperatures below a critical temperature $T_{CDW}(H)$  (denoted as $T_{charge}$ in  Ref.~\cite{Wu:2013}) that  is different from the 
charge density wave critical temperature in the normal region.  Indeed, it resulted that the maximum of  $T_{CDW}(H)$ occurs at
$\delta \simeq 0.11 - 0.12$ within the superconductive dome and the charge density wave and superconductive transition temperatures
were similar, $T_{CDW}(H) \sim T_c$.
In Fig.~\ref{Fig22} we display  the charge density wave  critical temperature for YBCO in an external magnetic field above the threshold
field.  The data have been taken from Fig.~2, panel b, of Ref.~\cite{Wu:2013}.  Note that, once the long-range order is established, on
increasing the magnetic field further the splitting of the nuclear magnetic resonance line saturates. This indicates that the charge density wave
gap and transition temperature become field independent for magnetic fields well above the threshold field. \\
 To determine the field-induced charge order transition temperature in our approach  we could work within the Gizburg-Landau
 theory as we did in Sect.~\ref{s5.3}. However, in the present case we need to include in the free energy functional the energy associated
 with the spatial variation of the order parameter and to take care of the interactions of nodal quasielectrons with the Abrikosov vortices.
 Actually, since we do not have at our disposal the relevant  microscopic calculations we shall follow a very simple description 
 which nevertheless should capture the essential aspects of the problem.   As repeatedly alluded before, we know that  the number of the 
 nodal  quasielectron excitations is  strongly reduced in the superconductive region at low temperatures.  The nodal quasielectron
 low-lying excitations relevant to the charge density wave instability are tied to the vortex cores which represent normal region within the superconductor.  
 The nodal quasielectron excitations affect the charge density wave gap through the density of state per spin at the
Fermi  level. We may roughly estimate the effective density of state as:
\begin{equation}
\label{5.115} 
< {\cal{N}}(0) >_{eff} \; \simeq \;  N_{charge} \; < {\cal{N}}(0) >_{vor}   \;\; ,  
 \end{equation}
where $ N_{charge}$ is the number of Abrikosov vortices in the  coherence region once the long-range charge order sets in.
Evidently we have:
\begin{equation}
\label{5.116} 
N_{charge} \;  \simeq  \;  \frac{H_{charge}}{\phi_0} \times   \; \pi \;   \xi^2_{CDW} \; \;  .
\end{equation}
After taking into account  Eq.~(\ref{4.53}) we find  that  Eq.~(\ref{5.57}) is modified as:
\begin{equation}
\label{5.117} 
 T_{CDW}(H)  \; \simeq \;   \frac{4 e^\gamma}{\pi}  \;  T^*(\delta) \;
e^{ - \;  
   \frac{ \hslash \Omega(Q_1) V_u   }{ 2 g^2(Q_1)  {\cal{N}}(0)  N_{charge}  } }  \;  .
\end{equation}
%
%
\begin{figure}
\vspace{0.7cm}
\hspace{-0.4cm}
\resizebox{0.5\textwidth}{!}{%
\includegraphics{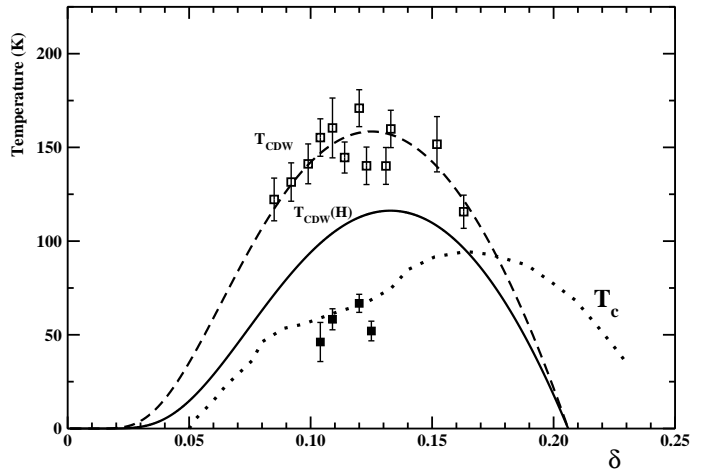} }
\caption{\label{Fig22} Charge density wave critical temperature versus the hole doping. Full squares correspond to the charge density wave
 critical temperature for YBCO in an external magnetic field. The data have been taken from Fig.~2, panel b, of Ref.~\cite{Wu:2013}. 
 The full line is Eq.~(\ref{5.117}) with $N_{charge} \simeq 0.785$.  For comparison we also display  the charge density wave critical temperatures 
 $T_{CDW}$ (open squares)  at zero magnetic field.  The dashed line is as in Fig.~\ref{Fig19}.  The dotted line represents the 
 superconducting transition temperature $T_c$. }
\end{figure}
%
To contrast Eq.~(\ref{5.117}) with available experimental informations, we  need to evaluate  $N_{charge}$.  Combining Eqs. (\ref{5.113}) 
and (\ref{5.116}) we reach the estimate:
\begin{equation}
\label{5.118} 
N_{charge} \;  \simeq  \; \frac{\pi}{4} \; \simeq \;  0.785  \; \;  .
\end{equation}
In Fig.~\ref{Fig22} we compare Eq.~(\ref{5.117}) with  $N_{charge}$ given by  Eq.~(\ref{5.118}) to the available
measurements. It is worthwhile to recall that  charge order is sensitive to disorder, which is a prominent feature of all cuprates. 
So that our qualitative estimate of the  charge density wave critical temperature Eq.~(\ref{5.117}) is intended to be valid only  in a small 
hole doping range  around $\delta \simeq 0.12$. We see that, indeed, the charge density critical temperature $T_{CDW}(H)$  is reduced
with respect to the charge order  transition temperature in the normal region. However, our estimate is about a factor two higher
with respect to the data. Nevertheless, the doping dependence turns out to be in fair agreement with observations.
Note that  increasing the magnetic field does not vary appreciably $N_{charge}$ since the addition of more Abrikosov vortices is
expected to do not further affect the long-range charge order. Therefore we expect that the charge density critical temperature  Eq.~(\ref{5.117}) 
should saturate at strong enough magnetic field in qualitative agreement with observations. Despite the obviously simplistic nature 
of the description, the qualitative agreement suggests that our picture could provide  a good starting point for explaining 
the field-induced charge order transition.
\subsection{Quantum Oscillations} 
\label{s5.5}
We conclude this Section by discussing the physics behind quantum oscillations in the pseudogap region 
of hole doped cuprates. Actually,  quantum oscillations are widely studied measurements to probe the Fermi 
surface~\cite{Shoenberg:1984}. In fact, according to the Onsager quantization condition~\cite{Onsager:1952,Lifshitz:1956}
the quantum oscillation frequency is directly proportional to  the cross-sectional area of the Fermi surface  normal to the applied magnetic
field direction. Due to this relationship, the observation of quantum oscillations is  attributed to the presence of  
closed  orbits on the Fermi surface.  \\
A breakthrough in the area of high temperature superconductivity was the observation of quantum oscillations in
cuprates~\footnote{An up-to-date discussion can be found in the already quoted
 Refs.~\cite{Sebastian:2011,Sebastian:2012,Vignolle:2013,Sebastian:2015}.}. In these experiments a strong magnetic field 
 was applied to suppress the superconductivity, which most likely revealed the normal ground state, leading
to the  unambiguous identification of quantum oscillations both in the underdoped and overdoped regions. 
Actually, in the underdoped region  the measured low oscillation frequencies 
revealed  a Fermi surface made of small pockets as inferred by the  Luttinger's  theorem and the Onsager relation.  
More interestingly, it turned out that the area of the pocket corresponded  to about a few percent of the first Brillouin zone area
in sharp contrast to that of overdoped cuprates where the frequency corresponded  to a large hole Fermi surface. 
Moreover, the clear evidence  of  negative Hall and Seebeck effects pointed to the conclusion that these pockets 
were electron-like rather than hole-like.  Finally, it resulted that these pockets were  associated with states near the nodal 
region of the Brillouin zone.  \\
From these experimental studies  we are led to suppose that the nodal quasielectron
low-lying excitations are responsible for quantum oscillations in  underdoped high temperature cuprate superconductors. 
However, we have seen  in Sect.~\ref{s4} that the  low-energy excitations of the quasielectron liquid are characterized
by disconnected Fermi arcs in the first Brillouin zone as depicted in Fig.~\ref{Fig12}.  
According to the widely used Onsager paradigm, which is based on the quasi-classical quantization, 
 the observation of quantum oscillations in thermodynamic quantities
is possible thank to the presence of closed orbits on the Fermi surface.  Of course, since the low-lying nodal quasielectron
excitations live on four disconnected Fermi arcs they cannot give rise to closed orbits.  Nevertheless, when a quasiparticle
Bragg diffracts at the Brillouin zone boundary it will have a momentum change given by a reciprocal lattice vector. The
quasiparticle, then, can jump to a different slice of the Fermi surface.  Looking at Fig.~\ref{Fig18} we see that the nodal
quasielectrons could be able to perform close orbits if they suffer Bragg diffractions with reciprocal lattice vectors given by
the nesting wavenumber vectors $\vec{Q}_1$ and $\vec{Q}_2$.  In general, these wavevectors are not commensurate with the 
reciprocal lattice of the planar $CuO_2$ lattice. On the other hand,   $\vec{Q}_1$ and $\vec{Q}_2$ are commensurate
with the lattice distorted by the charge density wave since they are the nesting vectors triggering the charge density
wave instability. We are led to conclude that Bragg reflections ensuring closed orbits are only possible in presence of the charge 
density wave. In the charge density wave ground state we known that the low energy nodal quasielectron excitations are gapped.
Naively one expects that quantum oscillations would be suppressed due to the presence of a gap at the Fermi surface.
Nevertheless, long time ago there have been measurements of Haas-van Alphen~\cite{Graebner:1976} and 
Haas-Shubnikov~\cite{Fleming:1976} quantum oscillation effects in transition-metal chalcogenide layer compounds which
were known to have charge density wave ground states~\footnote{A good account on transition-metal chalcogenide  compounds 
can be found in Refs.~\cite{Wilson:1975,Rossnagel:2011}.}. 
We see, then, that in our approach to observe quantum oscillations in underdoped cuprates it is necessary the presence
of long-range charge order. As discussed before at low temperatures this is assured for applied magnetic fields above the threshold
field $H_{charge} \sim 10 \, T$.  In fact, quantum oscillations in hole doped cuprates are observed for magnetic field strengths 
well above  $H_{charge}$. In the usual theory of quantum oscillations it is predicted that each extremal closed cross section
of the Fermi surface gives rise to a well defined frequency in the oscillatory part of the thermodynamic potential, and hence
in almost all thermodynamic  as well as transport  properties, so that each frequency appears together with its higher harmonic
without mixing of frequencies. However, it has been suggested~\cite{Schlottmann:1977} that in charge density wave or spin
density wave systems the interaction between Landau levels and many-body effects could produce nonlinear quantum oscillations.
Indeed, we will suggest later on that the nonlinear dependence of the charge density wave order parameter on the magnetic field
may be at the heart of the experimental observation of multiple quantum oscillation frequencies in underdoped high temperature
cuprate superconductors. \\
The quantum mechanical treatment of the motion of free electrons in an uniform magnetic field leads to quantized energy levels
labelled by an integer $n$~\footnote{See, e.g., Refs.~\cite{Peierls:1955,Landau:1991}.}. The set of all levels with a given $n$ is
referred to as the $n$-th Landau level. The energy of Landau levels with high quantum numbers can be evaluated within
the semiclassical approximation. Let $\varepsilon_n$ be the energy of the  $n$-th Landau level. For two-dimensional
quasielectrons in a transverse uniform magnetic field $H$, the difference in energy of two adjacent levels is given by
the Planck constant divided by the period of the semiclassical closed orbit:
\begin{equation}
\label{5.119} 
\Delta  \, \varepsilon_n \; \equiv \;  \varepsilon_{n+1} \; - \;   \varepsilon_n \;   \simeq  \; 
\frac{h}{T(\varepsilon_n )}   \; \;  .
\end{equation}
On the other hand, we have also:
\begin{equation}
\label{5.120} 
T(\varepsilon_n )   \;  \simeq \;   \frac{\hslash^2 c}{e H} \; \frac{\partial  {\cal{A}}_{\vec{k}}(\varepsilon_n)}{\partial  \varepsilon_n} \; \; ,
\end{equation}
where  ${\cal{A}}_{\vec{k}}(\varepsilon)$ is the k-space area enclosed by the closed orbit. Combining Eqs.~(\ref{5.119}) and
(\ref{5.120}) we get:
\begin{equation}
\label{5.121} 
 \varepsilon_{n+1} \; - \;   \varepsilon_n \;   \simeq  \;   \frac{2 \pi \, e H}  {\hslash \,  c} \;    
\frac{1}{\frac{\partial  {\cal{A}}_{\vec{k}}(\varepsilon_n)}{\partial  \varepsilon_n}} \; \; .
\end{equation}
Introducing the quasielectron cyclotron effective mass:
\begin{equation}
\label{5.122} 
m_e^c   \;  = \;   \frac{\hslash^2}{2 \, \pi} \; \frac{\partial  {\cal{A}}_{\vec{k}}(\varepsilon_n)}{\partial  \varepsilon_n}  \; \; ,
\end{equation}
we may rewrite  Eq.~(\ref{5.121}) as:
\begin{equation}
\label{5.123} 
 \varepsilon_{n+1} \; - \;   \varepsilon_n \;   \simeq  \;  \hslash \,  \omega_c  \;   \; \; , \; \; \;  
\omega_c \; = \; \frac{eH}{m_e^c \, c } \; \; ,
\end{equation}
where $\omega_c$ is the cyclotron frequency. It must be emphasized that the cyclotron effective mass is not, in general, the same
as the quasielectron effective mass $m^*_e$. From  Eq.~(\ref{5.123}) one gets:
\begin{equation}
\label{5.124} 
\varepsilon_n \;   \simeq  \;  \hslash \,  \omega_c  \; (n \; + \; \tilde{\gamma} )  
\end{equation}
where  $\tilde{\gamma}$ is a constant independent on $n$. In our simplified model the nodal quasielectrons satisfy the free electron
dispersion relation. In this case it turns out that  $\tilde{\gamma} \simeq \frac{1}{2}$~\cite{Shoenberg:1984}. Note that the Landau energy
levels are highly degenerate with degeneracy:
\begin{equation}
\label{5.125} 
g_L \;  =  \;  \frac{V}{2 \pi^2}  \; \frac{eH}{ \hslash  c } \; \; .
\end{equation}
We are interested in energies $\varepsilon_n$ of the order of the Fermi energy  $\varepsilon_F^{(e)}$. Observing that 
 $\hslash \, \omega_c \, \ll \, \varepsilon_F^{(e)}$, to a good approximation we can write:
\begin{equation}
\label{5.126} 
 \frac{\partial  {\cal{A}}_{\vec{k}}(\varepsilon_n)}{\partial  \varepsilon_n}  \;  \simeq \; 
\frac{ {\cal{A}}_{\vec{k}}(\varepsilon_{n+1}) \; - \;   {\cal{A}}_{\vec{k}}(\varepsilon_n)}{ \varepsilon_{n+1} \; - \;   \varepsilon_n } \; .
\end{equation}
Therefore, from Eq.~(\ref{5.121}) we are led to:
\begin{equation}
\label{5.127} 
{\cal{A}}_{\vec{k}}(\varepsilon_{n+1}) \; - \;   {\cal{A}}_{\vec{k}}(\varepsilon_n)  \; \simeq \;   \frac{2 \pi \, eH}{ \hslash  \, c } \;  ,
\end{equation}
which in turns gives the Onsager's relation:
\begin{equation}
\label{5.128} 
{\cal{A}}_{\vec{k}}(\varepsilon_{n})   \; \simeq \;   \frac{2 \pi \, eH}{ \hslash  \, c } \;   ( n \; + \; \frac{1}{2} ) \; .
\end{equation}
The quantization condition Eq.~(\ref{5.128}) results in an oscillatory structure in the thermodynamic potential. The reason for this
resides on the fact that,  whenever the value of the magnetic field causes an orbit on the Fermi surface to satisfy that quantization
condition with  $\varepsilon_{n} \simeq \varepsilon_F^{(e)}$, then the density of states at the Fermi level is enormously enhanced. 
It follows that the density of states at the Fermi level will be singular at regularly spaced intervals in $ \frac{1}{H} $ given by:
\begin{equation}
\label{5.129} 
\Delta (\frac{1}{H}) \; \simeq  \;  \frac{2 \pi \, e}{ \hslash  \, c } \;  
\frac{1}{{\cal{A}}_{\vec{k}}(\varepsilon_F^{(e)})}   \; \;  .
\end{equation}
Evidently, the oscillatory behavior as a function of  $\frac{1}{H}$ with period   Eq.~(\ref{5.129}) appears in any quantity that depends
on the density of states at the Fermi energy. To check this, let us consider the thermodynamic potential at zero temperature:
\begin{equation}
\label{5.130} 
\Omega(H)  \;  =  \;  g_L \; \sum_{r=0}^{n} \; ( \varepsilon_r \; - \;  \varepsilon_F^{(e)} ) 
\end{equation}
where the summation over $r$ is to be taken only on Landau levels such that $\varepsilon_r \; \lesssim  \;  \varepsilon_F^{(e)}$.
Actually, the summation in  Eq.~(\ref{5.130}) can be worked out by the Poisson summation formula or by the Euler-MacLaurin
formula. Using standard arguments one finds for the oscillating term in the thermodynamic potential~\cite{Shoenberg:1984}:
\begin{eqnarray}
\label{5.131} 
\Omega_{osc}(H)  \;  \simeq  \;   \frac{V}{4 \pi^2}  \; \frac{e^2 \, H^2}{ m_e^c \, c^2 }  \; 
\sum_{n=1}^{\infty } \; \frac{1}{\pi^2 \, n^2} \;
\\ \nonumber
 \cos \left ( 2 \pi \, n \left [ \frac{ \hslash c \;  {\cal{A}}_{\vec{k}}(\varepsilon_F^{(e)})}{2 \pi \, e H} \; 
- \; \frac{1}{2} \right ] \right ) \; \; .
\end{eqnarray}
 Eq.~(\ref{5.131}) shows that the thermodynamic potential is, in fact, a periodic function of $\frac{1}{H}$ with period given by
 Eq.~(\ref{5.129}).   At finite temperatures the probability of occupation of a state with energy $\varepsilon = \varepsilon_n$ is
 given by the Fermi-Dirac distribution. It can be seen that the effect of a finite temperature is to reduce the oscillation amplitude
 by the temperature-dependent factor:
\begin{equation}
\label{5.132} 
R_T(H,T) \;  \simeq  \;  \frac{2 \pi^2 \, n \; \frac{k_BT}{ \hslash \omega_c}}
{\sinh  \left ( 2 \pi^2 \, n  \;  \frac{k_BT}{ \hslash \omega_c}  \right )} \; \; .
\end{equation}
Moreover, if the quasielectrons have a finite relaxation time $\tau$ due to scattering, then the Landau energy levels
become broadened. This, in turns, leads to a further reduction of the oscillation amplitude by the so-called Dingle 
factor~\cite{Dingle:1952}:
\begin{equation}
\label{5.133} 
R_D  \;  \simeq  \;  e^{ - \pi \, n \; \frac{m_e^c \, c}{ e H \, \tau} }  \; \; .
\end{equation}
To determine the oscillation frequencies according to Eq. (\ref{5.129}) we need to evaluate the k-space area enclosed by the
semiclassical orbit at the Fermi energy. In the charge density wave ground state the low-energy quasielectron excitations are
gapped as in Eq.~(\ref{5.75}).  Therefore we find:
\begin{eqnarray}
\label{5.134} 
{\cal{A}}_{\vec{k}}(\varepsilon_F^{(e)})   \simeq    4 \; \frac{m^*_e}{ \hslash^2}  \theta_{FA}  \int_{\Delta_{CDW}}^{\varepsilon_F^{(e)} }   
d \xi_{\vec{k}} \; \frac{ \xi_{\vec{k}}}{  \sqrt{\xi_{\vec{k}}^2  -  \Delta_{CDW}^2 } }  \hspace{1.0cm}
\\ \nonumber
\simeq 
  4 \; \frac{m^*_e}{ \hslash^2} \;  \theta_{FA} \; \sqrt{ (\varepsilon_F^{(e)})^2   -  \Delta_{CDW}^2 } \, , \hspace{1.0cm}
\end{eqnarray}
while the fundamental oscillation frequency is:
\begin{equation}
\label{5.135} 
F \; \simeq  \;  \frac{ \hslash  \, c } {2 \pi \, e} \;  {\cal{A}}_{\vec{k}}(\varepsilon_F^{(e)})   \; \;  .
\end{equation}
Usually it is enough to retain the fundamental component of oscillation in the thermodynamic potential:
\begin{equation}
\label{5.136} 
\Omega_{osc}(H)  \;  \simeq  \;   A  \;  R_T \; R_D \;
\cos \left ( 2 \pi  \, \frac{F}{ H}  \right ) 
\end{equation}
where $R_T$ and $R_D$ are given by Eqs.~(\ref{5.132}) and (\ref{5.133}) with $n=1$ respectively, and $A$ is an overall amplitude.
From Eq.~(\ref{5.135}) we obtain our estimate for the fundamental frequency:
\begin{equation}
\label{5.137} 
F \; \simeq  \;  \frac{ 2 \, m^*_e \, c } { \pi \, e \hslash} \;   \theta_{FA}  \; \varepsilon_F^{(e)}   \; 
\sqrt{1 \; - \; \left ( \frac{\Delta_{CDW}}{\varepsilon_F^{(e)}} \right )^2 } \;  .
\end{equation}
Since $\Delta_{CDW} \ll  \varepsilon_F^{(e)}$ to a good approximation we can write:
\begin{equation}
\label{5.138} 
F \; \simeq  \;  F_0 \; \simeq \;  \frac{ 2 \, m^*_e \, c } { \pi \, e \hslash} \;   \theta_{FA}  \; \varepsilon_F^{(e)}   \;  \;  .
\end{equation}
In terms of the microscopic parameters of our model we end with the quite simple result:
\begin{equation}
\label{5.139} 
F \; \simeq  \;  F_0 \; \simeq \; \phi_0 \;  \frac{ \delta}{a_0^2}  
\end{equation}
where $\phi_0$ is the magnetic flux quantum Eq.~(\ref{3.98}). We see, then, that the main frequency in quantum oscillation phenomena
grows almost linearly with the hole doping in qualitative agreement with observations (see, eg, Fig.~2, panel B in Ref.~\cite{Ramshaw:2015}).
Moreover, in the underdoped region $\delta \sim 0.1$, using Eq.~(\ref{5.139}) we obtain $F_0 \simeq 1300 \, T$ that differs by a factor of few
from the quantum oscillation fundamental frequency measured in YBCO, $F_0 \simeq 530 \, T$.
Our previous discussion suggests that apparently the influence of the charge density wave gap on the oscillation frequencies is negligible
small. However, we showed in Sect.~\ref{s5.4} that the charge density wave gap is slightly affected by the applied magnetic field. In fact,
combing Eqs.~(\ref{5.82}) and (\ref{5.111}) we have:
\begin{equation}
\label{5.140} 
\Delta_{CDW}(H)  \; \simeq  \;  \Delta_{CDW} \;  -  \; \frac{7}{4}  \; \frac{ \zeta(3)}{e^{2 \gamma}} \; \frac{(\mu_B H)^2}{ k_B T_{CDW}}  \; \; .
\end{equation}
This last equation implies an explicit dependence of the fundamental frequency on the magnetic field:
\begin{equation}
\label{5.141} 
F(H)  \; \simeq  \; F_0 \; \left [  1 \;  +  \; \frac{7}{4}  \; \frac{ \zeta(3)}{e^{2 \gamma}} \; \left ( \frac{\mu_B H)}{ \varepsilon_F^{(e)}} \right )^2  \;
\right ] \; .
\end{equation}
Although the effect of the magnetic field is tiny, Eq.~(\ref{5.141}) introduces a non-linear dependence of the fundamental frequency
on the "time" $t = \frac{1}{H}$. As a consequence, the thermodynamic potential is no more  a strictly periodic function, but it becomes
a quasi-periodic function of the time $t$. In nonlinear quantum oscillations there is a mixing of frequencies which could account for
multiple oscillation frequencies observed in underdoped  high temperature cuprate  superconductors. To illustrate this point,
let us consider the following thermodynamic potential:
\begin{equation}
\label{5.142} 
\Omega(t)  \;  =  \;   \cos \left [ 2 \pi  \, F(t) \, t   \right ] \; \; ,
\end{equation}
where   $t = \frac{1}{H}$ is measured in $\text{Tesla}^{-1}$  and the frequency $F$ in $\text{Tesla}$. According to Eq.~(\ref{5.141}) we
can write:
\begin{equation}
\label{5.143} 
F(t)  \; =  \; F_0 \; \left [  1 \;  +  \;  \frac{\alpha}{t^2}  \right ] \;  \; .
\end{equation}
The parameter $\alpha$ can be inferred by comparing Eq.~(\ref{5.143}) with Eq.~(\ref{5.141}). If we rewrite the frequency as:
\begin{equation}
\label{5.144} 
F(t)  \; = \; F_0 \;   +  \;  \delta F(t)    \;  \; \; , \; \; \;   \delta F(t)  \; = \; \alpha \; \frac{F_0}{t^2} \; \; ,
\end{equation}
then in the Fourier transform of $\Omega(t)$ one expects to find  the superposition of the dominant $F_0$ oscillation with
weak-er amplitude oscillations of frequencies $F_0 \,  \pm  \, \delta F(t)$ and $\delta F(t)$. However the small frequency 
$\delta F(t)$  depends on $t$, so that to check the above expectations we need to evaluate numerically the Fourier transform
of the thermodynamic potential:
\begin{equation}
\label{5.145} 
{\tilde \Omega}(\nu)  \;  =  \;   \int_{- \infty}^{+\infty}  \; dt \; e^{ -  2 \pi \, i \,  \nu t}  \;  \Omega(t)   \; .
\end{equation}
It is convenient to write this last equation in the following form:
\begin{equation}
\label{5.146} 
{\tilde \Omega}(\nu)    =   \frac{1}{\pi}   \int_{0}^{+\infty}  dt \; \cos \nu t \; \cos \left ( F_0  \, t  +  
4 \pi^2 \, \alpha \, \frac{F_0}{t} \right  ) \;  \; .
\end{equation}
For $\alpha = 0$ it is easy to check that:
\begin{equation}
\label{5.147} 
{\tilde \Omega}(\nu)  \;  =  \;  \delta (  \nu \; - \; F_0 )  \; \; .
\end{equation}
This shows that  ${\tilde \Omega}(\nu)$ is a tempered distribution. Therefore, to evaluate numerically the Fourier transform of the thermodynamic
potential we must smear the distribution with a suitable test function. The most natural choice is to consider test functions with compact
support. Accordingly, we consider:
\begin{eqnarray}
\label{5.148} 
{\tilde \Omega}(\nu)  \;  =  \;  \frac{1}{\pi} \; \int_{\nu - \frac{\Delta \nu}{2}}^{\nu + \frac{\Delta \nu}{2}} d \nu' 
 \int_{0}^{+\infty}  dt \; \cos \nu' t \; 
 \\ \nonumber
 \cos \left ( F_0  \, t \; + \; 
4 \pi^2 \, \alpha \, \frac{F_0}{t} \right  ) \; ,
\end{eqnarray}
which corresponds to a test function $f_t = 1$ within the bin $\Delta \nu$ centered at the given frequency $\nu$, while $f_t = 0$ otherwise.
We further introduce the smeared spectral power function:
\begin{equation}
\label{5.149} 
P(\nu) \; = \; | {\tilde \Omega}(\nu) |^2  \; \; .
\end{equation}
%
%
\begin{figure}
\vspace{0.7cm}
\hspace{0.5cm}
\resizebox{0.45\textwidth}{!}{%
\includegraphics{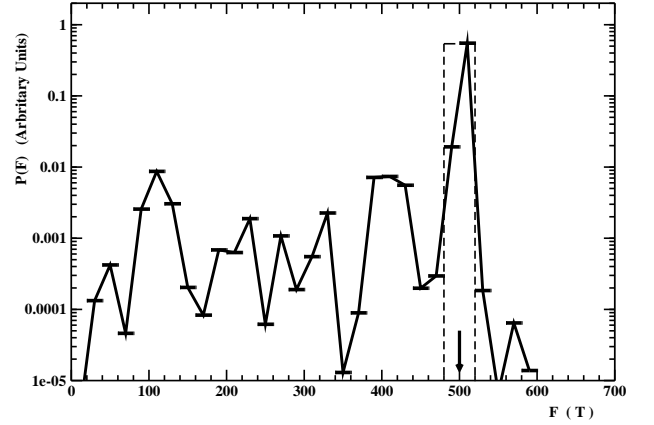} }
\caption{\label{Fig23}  The spectral power function Eq.~(\ref{5.149}) versus the quantum oscillation frequency (measured in Tesla).
The dotted line is the reference smeared spectral power corresponding to Eq.~(\ref{5.147}) with frequency $F_0 = 500 \, T$ 
(indicated by the arrow) binned with bin size $\Delta \nu = 20 \, T$.  The continuum line is the spectral power function corresponding to Eq.~(\ref{5.148})
adopting  the same normalization as  the reference spectral power.}
\end{figure}
%
In Fig.~\ref{Fig23} we present the spectral power function  Eq.~(\ref{5.149}) by assuming:
\begin{equation}
\label{5.150} 
F_0  \; = \;  500 \; T \; \; \; , \; \; \; \alpha \; = \; 10^{-3} \; T^{-2}   \; \; .
\end{equation}
Note that the main frequency $F_0$ is close to the observed oscillation frequency in underdoped YBCO. As concern the parameter $\alpha$, to
magnify the effect of the mixing of frequencies we assumed a value much greater than the one obtained by  Eq.~(\ref{5.141}). For comparison
we also display in  Fig.~\ref{Fig23} the spectral power for $\alpha=0$. In this case the spectral power should correspond to a smeared
delta-function. In fact, we see (dashed line in  Fig.~\ref{Fig23}) that all the spectral power is concentrated within the two bins around
the main frequency $F_0$. On the other hand, for a non-zero value of the parameter $\alpha$ the spectral power turns out to be considerably
modified (see the continuum curve  in  Fig.~\ref{Fig23}). Indeed, the main peak in the Fourier transform is still at the frequency $F_0$.
This peak accounts for about one half of the total spectral power. The other half of the spectral power is spread over multiple peaks.
Interestingly enough, there are two peaks near the main frequency $F_0 = 500 \, T$ corresponding to frequencies 
 $F_1 \simeq 440 \, T$   and $F_2 = 550 \, T$. It should be noted that the typical Fourier-transformed quantum oscillation
 amplitude in underdoped cuprates does show this characteristic almost symmetrically split three peaks structure. 
Usually, one explanation for the close frequencies in the Fourier transform is achieved  by assuming a cylindrical Fermi surface that
extends along the inter-layer direction with a small warping~\cite{Yamaji:1989}. Indeed, warping manifests itself as a splitting
corresponding to the so-called neck and belly frequencies. In this case the thermodynamic potential is customarily  written as:
\begin{equation}
\label{5.151} 
\Omega(t)  \;  =  \;   \cos ( 2 \pi  \, F_0 \, t )  \times  J_0(2 \pi  \, \Delta F \, t)   \; \;  ,
\end{equation}
where $J_0$ is the Bessel function. In Eq.~(\ref{5.151}) $\Delta F$ is the first harmonic of warping in the momentum direction
perpendicular to the copper-oxide planes.  Our results, however, are suggesting that the origin of multiple
frequencies near the spectrally dominant frequency can be accounted for by the non-linearities in quantum oscillation
phenomena due to the magnetic field dependence of the charge density wave gap. Another interesting feature in  Fig.~\ref{Fig23}
is due to the presence of sizable spectral power in the low frequency region. In particular, there is a peak in the Fourier transform
at $F_3 \simeq 100 \, T$. In the conventional interpretation this peak would be attributed to an additional small Fermi pocket
of quasielectrons or quasiholes, which would  imply a drastic Fermi surface reconstruction. We feel that the qualitative agreement of
Fig.~\ref{Fig23} with the Fourier transform of the quantum oscillation amplitude in underdoped cuprates is suggesting that the origin
of multiple frequencies resides in the intrinsic non-linearity caused by the charge density wave gap. \\
To conclude this Section, we briefly  discuss the cyclotron effective mass. After taking into account Eq.~(\ref{5.126}) we rewrite  Eq.~(\ref{5.122})
as:
\begin{equation}
\label{5.152} 
m_e^c   \;  \simeq \;   \frac{\hslash^2}{2 \, \pi}  \; 
\frac{ {\cal{A}}_{\vec{k}}(\varepsilon_{n+1}) \; - \;   {\cal{A}}_{\vec{k}}(\varepsilon_n)}{ \varepsilon_{n+1} \; - \;   \varepsilon_n }  \; \; .  
\end{equation}
In quantum oscillations the involved Landau levels have energies very close to the Fermi energy. So that we may employ  Eq.~(\ref{5.134}) to get:
\begin{equation}
\label{5.153} 
m_e^c   \;  \simeq \;   \frac{2}{\pi}  \;  m^*_e  \;   \theta_{FA}  \; 
\sqrt{1 \; - \; \left ( \frac{\Delta_{CDW}}{\varepsilon_F^{(e)}} \right )^2 } \;  .
\end{equation}
So that, to a good approximation, we are left with the quite simple result:
\begin{equation}
\label{5.154} 
m_e^c   \;  \simeq \;   \frac{\delta}{1 - \delta}  \;  m^*_e   \;  \;  .
\end{equation}
Evidently the cyclotron effective mass is smaller than the quasielectron effective mass and it increases almost linearly with the hole doping fraction $\delta$.
However, the cyclotron effective mass measured by quantum oscillations, generally, turns out to be greater than the effective mass and strongly dependent
on the hole doping. Indeed, the cyclotron mass is seen to extrapolate to a mass divergence at two hole doping points, $\delta_1$ and $\delta_2$ (compare,
for instance, with Fig.~11, panel d, in Ref.~\cite{Sebastian:2015}).  The precise values of these hole doping depend on the cuprate family. Nevertheless,
several recent studies~\cite{Suchitra:2010,LeBoeuf:2011,Baek:2012,Vishik:2012,Peng:2013,Sacuto:2013,Benhabib:2015,Badoux:2016}  indicated that:
\begin{equation}
\label{5.155} 
\delta_1  \;  \simeq \; 0.08 \; - \; 0.10 \; \; \; , \; \; \;  \delta_2 \; \simeq \; 0.18 \; - \; 0.20    \;  \;  .
\end{equation}
It is widely believed that the point $\delta = \delta_1$ is a Lifshitz critical point due to  a variation of the  topology of the Fermi
surface~\cite{Lifshitz:1959}, while  the point $\delta = \delta_2$ is identified with a quantum critical point, i.e. a phase transition
at zero temperature induced by tuning some external parameter of the system~\cite{Sachdev:2011}.  \\
To extract the cyclotron mass from quantum oscillations the relevant oscillating amplitude is assumed to be of the form:
\begin{equation}
\label{5.156} 
A_{osc}  \;  =  \; A \; R_T \; R_D   \cos ( 2 \pi  \, \frac{F}{H} ) \;  J_0(2 \pi  \, \frac{\Delta F}{H} )   \; \;  ,
\end{equation}
with
\begin{equation}
\label{5.157} 
R_D  \;  \simeq  \;  e^{ - \pi \, \frac{m_e^c \, c}{ e H}  \, \frac{1}{\tau} }  \; \; ,
\end{equation}
\begin{equation}
\label{5.158} 
R_T  \;  \simeq  \;  \frac{2 \pi^2 \; k_B T \;   \frac{m_e^c c}{ \hslash \, eH}}
{\sinh  \left ( 2 \pi^2  \; k_B T \;  \frac{m_e^c c }{ \hslash \, eH}  \right )} \; \; .
\end{equation}
The cyclotron mass is, then, obtained by fitting for a particular hole doping the temperature dependence of the oscillating amplitude to
Eq.~(\ref{5.156}), keeping $\tau$ and $F$ fixed. We feel, however, that the assumption of a temperature independent relaxation time in the
pseudogap region of hole doped cuprates is highly questionable. In normal metals at low temperatures the relaxation time is dominated
by the scattering of low-lying excitations off impurities. In that case one can safely neglect the temperature dependence of the relaxation
time at low enough temperatures. However, it must be remarked that in our approach the nodal quasielectrons are the effective low-lying
excitations which are able to retain the needed quantum coherence only on the Fermi arcs. Therefore the quasielectron relaxation time
is determined by scattering processes that preserve the quasielectron coherence. At low temperatures it results that the nodal quasielectron
relaxation time is entirely due to the Coulomb interaction. In fact, Coulomb umklapp scattering between quasielectrons on opposite nodal Fermi arcs 
allows the quasielectrons to retain coherence and hence to contribute to the conduction processes.  Notably  the calculation of the relaxation time 
has been recently discussed in details in Ref.~\cite{Gorkov:2013}. As a result one finds:
\begin{equation}
\label{5.159} 
\frac{1}{\tau_{ee}}   \;  \sim   \;   T^2  \; \; .
\end{equation}
Note that this temperature dependence is in agreement with the spectroscopic evidence for the Fermi liquid-like behavior of the relaxation rate
 in the pseudogap phase of hole doped cuprates~\cite{Mirzali:2013,Barisic:2013}. It is amusing to mention that the Coulomb umklapp processes
 are allowed in the rather narrow interval:
\begin{equation}
\label{5.160} 
\tilde{\delta}_1  \; \lesssim  \; \delta  \; \lesssim \;  \tilde{\delta}_2    \;   \;  \;  .
\end{equation}
The meaning of Eq.~(\ref{5.160}) is that outside this interval the nodal quasielectrons cannot contribute to the conduction processes.
We found $\tilde{\delta}_1   \simeq  0.15$,      $\tilde{\delta}_2  \simeq  0.21$, which are remarkably close to the putative critical points Eq.~(\ref{5.155}).
Actually, the slight difference can be presumably ascribed to the extreme simplifications of our model. So that, we are led to suspect that the origin
of the puzzling critical points could reside in the peculiar doping dependence of the nodal quasielectron relaxation time. \\
For the purposes of determining the cyclotron mass from quantum oscillations we need to keep track of the temperature dependence of the
relaxation time, Eq.~(\ref{5.159}). Then, a remarkably thing that occurs is that the Dingle factor Eq.~(\ref{5.157}) becomes temperature dependent.
Therefore, before drawing any conclusions regarding  the cyclotron mass, it is necessary to reanalyze the data with a fitting procedure which allows
 the Dingle factor to depend on the temperature.
\section{Summary  and Conclusions}
\label{s6}
Understanding  high temperature superconductivity in copper oxides remains one of the most challenging problem in condensed matter physics.
Since the discovery of superconductivity in cuprates~\cite{Bednorz:1986}  there has been a tremendous advance in understanding their physical properties
and  many experimental and theoretical points have been clarified. It has emerged that these compounds possess a number of unusual normal state
and superconducting properties due to a complicated interplay of electronic, spin, and lattice degrees of freedom. In view of the fact that cuprates are 
very complex materials, no consensus on a common accepted interpretation of all the physical phenomena and the mechanism for formation of the 
superconducting state has yet been achieved. Notwithstanding,  since long time P. W. Anderson convincingly stated~\cite{Anderson:1987,Anderson:1997}  
that the essential physics, including  superconductivity, was contained in an effective two dimensional square lattice Hubbard model. The electronic structure 
is such that at low energies there is a single spin degenerate band of correlated electrons with a two-dimensional dispersion in the copper-oxygen plane, 
while the dispersion  in the third direction may in first approximation be neglected. Indeed, numerical approaches have given strong reason to believe that
 this basic picture  was correct, namely  the two-dimensional Hubbard model captures the basic physics of the superconductivity and the pseudogap 
(for a recent comprehensive overview, see Ref.~\cite{Gull:2015} and references therein). \\
The cuprate challenge could best met by the construction of a phenomenological simplified model inspired by experiment which, 
nevertheless, is able to capture at least qualitatively the physics of cuprate superconductors. In fact, in our previous paper~\cite{Cea:2013}
we attempted  to find a picture that was as simple as possible and that represented phenomena as accurately as possible. 
The driving principle of the approach presented in I  has been that the high temperature superconductivity could be understood 
within some framework along the line of the microscopic theory of Bardeen, Cooper, and Schrieffer.  
To construct a phenomenological  model able to recover  the most salient  aspects of the unusual behavior seen in the various regions of the phase diagram
 of  hole doped cuprates,  we relied  heavily on some assumptions. First, we assumed that the physics of the high temperature cuprates was deeply rooted 
 in the copper-oxide planes. This allowed us to  completely neglect the motion along the direction perpendicular to the $CuO_2$ planes. 
 In addition, we assumed that the single-band effective Hubbard  model is sufficient to account for all the essential physics of the copper-oxide planes.
 Accordingly,  in I we proposed an effective Hamiltonian aimed to describe the dynamics of the holes injected into the
undoped copper-oxide planes. We arrived at our effective Hamiltonian by using  known arguments  on the motion of charge carriers in an
antiferromagnetic background. Notwithstanding, we were unable to offer a truly microscopic derivation of the effective Hamiltonian. 
Thereby  our arguments, albeit suggestive, cannot be considered as a first principle derivation. In spite of that, we showed that the effective
 Hamiltonian offered  a  consistent picture of the high transition temperature cuprate superconductors. Firstly, due to the reduced dimensionality the two-body 
 attractive potential turned out to admit real-space  d-wave bound states. The binding energy of these bound states, which plays the role of the pseudogap, 
 decreases with increasing doping until it vanishes at a certain  critical  doping   $\delta = \delta^*$. This allowed us to reach the conclusion 
 that the key features of the underdoped side of the phase diagram were controlled by very  strong pairing that is phase-disordered by 
 thermal fluctuations.  In our model the overdoped region is realized for hole doping in excess of the critical doping $\delta^*$ where the pseudogap 
 vanishes.  It is well established that the long range antiferromagnetic order in the underdoped region is rapidly lost with increasing $\delta$, but
 nevertheless  two-dimensional  short-range order  persists  up to the overdoped region. This led us to conclude
 that in this region the conventional  d-wave BCS framework account for many of the low-energy and low-temperature properties of the 
 copper oxides, in nice  agreement with several observations.  Finally, we pointed out that in the optimal doped region the 
 competition between the pseudogap and the d-wave BCS gap together  with the enhanced role of the phase fluctuations makes  
 the mean field approximation of doubtful validity.  \\
The aim of the present paper was to discuss in greater details the physics of the underdoped region in hole doped high temperature cuprate
superconductors. Indeed, recently the enigmatic cuprate superconductors have attracted resurgent interest with several reports and discussions 
of competing orders in the underdoped side.  Even though in I we presented  a partial account of the  pseudogap region,  we did not attempt a complete 
discussion of the strange behavior of the cuprates in this region.  In this paper we attempted  a serious effort to bring together 
 several  observational features of the hole doped high temperature cuprate superconductors in the pseudogap region.
We discussed in greater details with respect to I the structure of the hole pair condensate ground state. Moreover, we argued that the
low-lying excitations of the condensate were the analog of rotons in $^4He$. One crucial point to enlighten was that these elementary
excitations relied heavily on the phase coherence of the hole pair condensate. Taking into account  that in the pseudogap region the superconductive
transition is described by the Berezinskii-Kosterlitz-Thouless order-disorder transition and using the Kosterlitz's recursion relation for
the screening length, we were able to establish the temperature dependence of both the superfluid velocity and the hole pair condensate fraction.
We also considered the microscopic dynamics of the hole pair condensate in presence of applied magnetic fields transverse to the copper-oxide
planes. We worked out the thermodynamics of the roton gas near the superconductive critical temperature. Remarkably, we found that even in the
ideal gas approximation the specific heat anomaly in the critical region could be accounted for qualitatively and
quantitatively by the roton gas. We discussed the electrodynamics of the charged hole pair condensate. In particular, we determined the dependence
of the London penetration length on the hole doping fraction and on the temperature. We critically compared  our peculiar temperature dependence
of the penetration length with experimental observations for different class of cuprate superconductors in the underdoped and optimal doped
regions. We carefully investigated the structure of the Abrikosov vortex and obtained the lower and upper critical magnetic fields. We found that
the doping and temperature dependence of the critical magnetic fields compared in satisfying agreement with several experimental studies.
We also determined and compared with selected observations the temperature dependence of the critical current. \\
Inspired by the cuprate phenomenology  that points to a clear experimental evidence for nodal quasielectron quantum liquid, we developed some
arguments to justify how the presence of the pseudogap were responsible for the formation of the quasielectron nodal Fermi liquid.
In our approach the nodal quasielectrons are effective low energy excitations  in the phase disordered hole pair condensate
 which retain their quantum coherence only  in the nodal directions. We showed that the interplay between the condensate roton excitations
and the nodal quasielectrons laid at  the heart of the so-called nodal gap. We further determined the doping and temperature dependence
of the nodal gap and contrasted successfully with available experimental data in  literature.  We also discussed the nodal Fermi velocity
detected by angle resolved photoemission spectroscopy. We critically examined the contribution of the nodal quasielectrons
to the low temperature specific heat with and without external transverse magnetic fields. We proposed that the charge density wave instabilities in
hole doped high temperature superconductors were triggered by the interactions of nodal quasielectrons with longitudinal bond-stretching phonon
modes of the planar copper-oxide lattice. Our proposal allowed us to determine and to compare with experimental data both the doping dependence
of the charge density wave critical temperature and the temperature dependence of the charge density wave gap. We also were able to clarify
the puzzling competition between charge order and superconductivity. Finally, we discussed the physics behind quantum oscillations in the
pseudogap region. In particular, we suggested that the origin of multiple frequencies in quantum oscillations resided in the intrinsic non-linearity
implied by the dependence of the charge density wave gap on applied magnetic fields. \\

The  inherent complexity of the hole doped cuprates has hidden key features of the pairing mechanism in most experiments, preventing a satisfactory 
understanding of high temperature superconductivity. In our previous and in the present paper  we have shown that a relatively simple model for the 
effective Hamiltonian based on plausible physical assumptions allowed to reach a coherent and fairly complete picture of hole doped high
temperature cuprate superconductors. 
We would like to stress once more that the results presented in this paper have been obtained in the mean field approximation. This means that
we are neglecting systematically the fluctuation effects. However, it is known since long time that in high temperature cuprate superconductors 
 phase fluctuations could play a  significant role. In any case, in our opinion, the results presented in our papers should be useful to reach a truly microscopic
  explanation of the high temperature superconductivity in cuprates.  To conclude, we would like to stress that we did not discussed in the present paper 
the transport phenomena in cuprate superconductors. Actually, even though we already reached some partial results on this subject, 
 we refrained to include even a partial account of these problematics  in order to avoid the increase of the length of the present paper 
 beyond reasonable limits. 
\appendix
\section{D-Wave BCS Gap and Penetration Depth}
\label{AppendixA}
In the overdoped region the superconductive instability is driven by the short-range attractive interaction between the quasiholes.
 This means that  the pairing is  in momentum space, so that the relevant superconducting ground state is the BCS variational ground state.
 At zero temperature the relevant gap equation has been discussed since long time~\cite{Anderson:1961}:
\begin{equation}
\label{A.1}  
\Delta(\vec{k}) \; = \; - \frac{1}{2} \; \sum_{\vec{k}'} \; \; \frac{ V(\vec{k} - \vec{k}') \; \Delta(\vec{k}')} { \sqrt{ \xi^2_{\vec{k}'} + |  \Delta(\vec{k}')|^2 } } 
\; ,
\end{equation}
where:
\begin{equation}
\label{A.2}  
\xi_{\vec{k}} \;  =  \; \frac{\hbar^2 \vec{k}^2}{2 m^*_h} \; -  \; \varepsilon_F  \; ,
\end{equation}
$ \varepsilon_F$ being the Fermi energy, and 
\begin{equation}
\label{A.3}  
V(\vec{k} - \vec{k}')  \;  =  \;  \int d \vec{r} \; e^{ - i ( \vec{k} - \vec{k}') \cdot \vec{r} } \; V(r) \; .
\end{equation}
Using the expansion:
\begin{equation}
\label{A.4}  
e^{ i \vec{k} \cdot \vec{r} } \; = \;  e^{  i k r \cos{\theta_{kr} }  }  \; = \; \sum_{m=-\infty}^{+ \infty}  \; 
e^{ i m (\theta_{kr} + \frac{\pi}{2} ) }   \; J_m( kr)   \; ,
\end{equation}
we may rewrite  Eq.~(\ref{A.1}) as:
\begin{equation}
\label{A.5}  
\Delta(\vec{k}) \;  = \; - \frac{1}{2} \sum_{\vec{k}'} \; \sum_{m=-\infty}^{+\infty}  e^{ i m \theta_{kk'} } \; 
 \; \frac{ V_m(k, k')  \Delta(\vec{k}')} { \sqrt{ \xi^2_{\vec{k}'} + |  \Delta(\vec{k}')|^2 } }  \; ,
\end{equation}
with:
\begin{equation}
\label{A.6}  
V_m(k, k')  \;  =  \;  2 \pi \;  \int_0^{\infty}  dr \; r \; V(r) \; J_m(kr) \; J_m(k'r)  \; .
\end{equation}
Equation~(\ref{A.5}) suggests that:
\begin{equation}
\label{A.7}  
\Delta(\vec{k}) \;  = \; \Delta(k,\theta_k) \;  = \; \sum_{\ell = - \infty}^{+\infty}  \; e^{(i \ell \theta_{k})} \;  \Delta_{\ell}(k)   \; ,
\end{equation}
In the weak coupling limit  Eq.~(\ref{A.5}) reduces to~\cite{Anderson:1961}:
\begin{equation}
\label{A.8}  
\Delta_m(k,\theta_k) \; \simeq \; - \frac{1}{2} \sum_{\vec{k}'} \;   e^{ i m \theta_{kk'} } \; 
 \; \frac{ V_m(k, k')  \Delta_m(k',\theta_{k'})} { \sqrt{ \xi^2_{\vec{k}'} + |  \Delta_m(k',\theta_{k'})|^2 } }  \; .
\end{equation}
As discussed in I,  Eq.~(\ref{A.8}) admits non trivial solutions for the d-wave gap ($m= \pm 2$). Moreover, we have~\cite{Cea:2013}:
\begin{equation}
\label{A.9}  
\Delta_2(k,\theta_k)  \; = \;  \Delta_{BCS}(k) \; \cos{(2 \theta_k)} \; .
\end{equation}
Thus, we end with the following gap equation:
\begin{eqnarray}
\label{A.10}  
 \Delta_{BCS}(k) \;  \simeq \; -  \sum_{\vec{k}'} \;  [\cos{(2 \theta_{k'})]^2} \; 
 \\ \nonumber
 \; \frac{ V_2(k, k')  \Delta_{BCS}(k')} { \sqrt{ \xi^2_{\vec{k}'} + [  \Delta_{BCS}(k') \cos{(2 \theta_{k'})}]^2 } }  \; .
\end{eqnarray}
Since the gap is sizable on the Fermi surface, we may further simplify Eq.~(\ref{A.10}) as:
\begin{equation}
\label{A.11}  
1 \;  \simeq \; -   V_2  \; \int \frac{d\vec{k}'}{(2 \pi)^2}  \; 
 \; \frac{  [\cos{(2 \theta_{k'})}]^2 } { \sqrt{ \xi^2_{\vec{k}'} + [  \Delta_{BCS} \cos{(2 \theta_{k'})}]^2 } }  
\end{equation}
where $\Delta_{BCS} =  \Delta_{BCS}(k_F)$ and $V_2= V_2(k_F,k_F)$. After some standard manipulations we obtain: 
\begin{eqnarray}
\label{A.12}  
1 \;  \simeq \; -   \frac{V_2 m^*_h}{\hbar^2}  \; \int_{-\varepsilon_c}^{+\varepsilon_c}  \frac{d \xi}{(2 \pi)^2}  \;  \int_{0}^{2 \pi} d\theta
\\ \nonumber
 \; \frac{  [\cos{(2 \theta)}]^2 } { \sqrt{ \xi^2 + [  \Delta_{BCS} \cos{(2 \theta)}]^2 } }  
\end{eqnarray}
where $\varepsilon_c$ is a high-energy cut-off which, usually  is much smaller than the Fermi energy. 
Performing the integrals and using the approximation:
\begin{equation}
\label{A.13}  
archsinh \left [   \frac{\varepsilon_c}{ \Delta_{BCS} |\cos{(2 \theta)}| }   \right ]     \simeq     
 \ln{  \left  [  \frac{2 \varepsilon_c}{  \Delta_{BCS} |\cos{(2 \theta)| } }\right ] } \; ,
\end{equation}
we get:
\begin{equation}
\label{A.14}  
\Delta_{BCS} \;  \simeq \;   \frac{4 \varepsilon_c}{\sqrt{e}}   \;  e^{ -  \frac{1}{\lambda_2} }  \; ,
\end{equation}
where:
\begin{equation}
\label{A.15}  
\lambda_2  \; = \; \frac{m^*_h V_0}{\hbar^2} \; \int_{0}^{r_0(\delta)} dr \; r \left [ J_2(k_F r) \right ]^2 \; .
\end{equation}
To obtain the critical temperature, we note that~\cite{Won:1994} (see below):
\begin{equation}
\label{A.16}  
\frac{\Delta_{BCS}} {k_B T_c} \; \simeq \; \pi e^{\ln{2} - \frac{1}{2} - \gamma}  \simeq 2.140  \; .
\end{equation}
 We obtain, thus,  the BCS critical temperature $T_c \equiv T_{BCS}$:
\begin{equation}
\label{A.17}  
k_B \; T_{BCS} \;  \simeq \;   \frac{2  e^{\gamma}}{\pi}   \; \varepsilon_c \;  e^{  -  \frac{1}{\lambda_2} }  \; .
\end{equation}
In I we argued that  it is natural to identify the cut-off energy $\varepsilon_c$ with $\Delta_2(\delta=0)$ which,
indeed, is much smaller than the Fermi energy in the range of hole doping fraction of interest. \\
To obtain the thermal corrections to the BCS gap  we use the well-known gap equation at finite
 temperatures~\cite{Anderson:1961,Won:1994}. With the same approximations as before, we find that
 at finite temperature Eq.~(\ref{A.11}) is replaced by:

\begin{eqnarray}
\label{A.18}  
1  \simeq  -   V_2   \int \frac{d\vec{k}'}{(2 \pi)^2}  
 \; \frac{  [\cos{(2 \theta_{k'})}]^2 } { \sqrt{ \xi^2_{\vec{k}'} + [  \Delta_{BCS} \cos{(2 \theta_{k'})}]^2 } }  
 \hspace{1.0cm}
 \\ \nonumber
  \tanh \left (  \frac{ \sqrt{ \xi^2_{\vec{k}'} + [   \cos{(2 \theta_{k'})}]^2 } }{2  k_B \; T } \right ) \; .
\end{eqnarray}
In principle, solving Eq.~(\ref{A.18}) one obtains the BCS energy gap at finite temperature,  $\Delta_{BCS}(T)$.
Note that at zero temperature   Eq.~(\ref{A.18}) reduces to Eq.~(\ref{A.11}), so that  $\Delta_{BCS}(T=0)$ is given
by   Eq.~(\ref{A.14}).
To proceed further, we rewrite  Eq.~(\ref{A.18}) in the following form:
\begin{eqnarray}
\label{A.19}
  \int \frac{d\vec{k}'}{(2 \pi)^2}  
 \; \frac{  [\cos{(2 \theta_{k'})}]^2 } { \sqrt{ \xi^2_{\vec{k}'} + [  \Delta_{BCS} \cos{(2 \theta_{k'})}]^2 } }  \; = \hspace{1.0cm}   
 \\ \nonumber
 \int \frac{d\vec{k}'}{(2 \pi)^2}  
 \; \frac{  [\cos{(2 \theta_{k'})}]^2 } { \sqrt{ \xi^2_{\vec{k}'} + [  \Delta_{BCS} \cos{(2 \theta_{k'})}]^2 } }  
\\ \nonumber  
\; +  \; \int \frac{d\vec{k}'}{(2 \pi)^2}  
 \frac{  [\cos{(2 \theta_{k'})}]^2 } { \sqrt{ \xi^2_{\vec{k}'} + [  \Delta_{BCS} \cos{(2 \theta_{k'})}]^2 } } 
 \\ \nonumber
\left [  \tanh (  \frac{ \sqrt{ \xi^2_{\vec{k}'} + [  \Delta_{BCS} \cos{(2 \theta_{k'})}]^2 } }{ 2 k_B \; T }  ) - 1 \right ] .
\end{eqnarray}
Combining Eqs.~(\ref{A.11})   and (\ref{A.18}), and in the same approximations as before, after some manipulations
we obtain:
\begin{eqnarray}
\label{A.20}
 \int_{0}^{\varepsilon_c}  d \xi     \int_{0}^{2 \pi} d\theta
  \frac{  [\cos{(2 \theta)}]^2 } { \sqrt{ \xi^2 + [  \Delta_{BCS}(0) \cos{(2 \theta)}]^2 } }  \; =  \hspace{1.0cm}
  \\ \nonumber
  \int_{0}^{\varepsilon_c}  d \xi   \;  \int_{0}^{2 \pi} d\theta
 \; \frac{  [\cos{(2 \theta)}]^2 } { \sqrt{ \xi^2 + [  \Delta_{BCS}(T) \cos{(2 \theta)}]^2 } }   \; \; \; 
\\ \nonumber  
   +  \;   \int_{0}^{\varepsilon_c}  d \xi   \;  \int_{0}^{2 \pi} d\theta
  \frac{  [\cos{(2 \theta)}]^2 } { \sqrt{ \xi^2 + [  \Delta_{BCS}(T) \cos{(2 \theta)}]^2 } } 
  \\ \nonumber
\left [  \tanh (  \frac{ \sqrt{ \xi^2 + [  \Delta_{BCS}(T) \cos{(2 \theta)}]^2 } }{ 2 k_B \; T }  ) - 1 \right ] . 
\end{eqnarray}
Now, we observe that:
\begin{eqnarray}
\label{A.21}
 \int_{0}^{\varepsilon_c}  d \xi   \int_{0}^{2 \pi} d\theta
  \frac{  \cos^2{(2 \theta)}} { \sqrt{ \xi^2 + [  \Delta  \cos{(2 \theta)}]^2 } }  
 \simeq    \hspace{1.0cm}
 \\ \nonumber
  \int_{0}^{2 \pi} d\theta
  \cos^2(2 \theta) \; \log \left ( \frac{ 2 \varepsilon_c}{ \Delta |\cos{(2 \theta)}| } \right ) 
\\ \nonumber  
  = \;  \pi \; \left [  \log  ( \frac{ 2 \varepsilon_c}{ \Delta } ) \; + \;  \log 2 \; - \; \frac{1}{2} \right ] \; . 
\end{eqnarray}
This allows us to put Eq.~(\ref{A.20}) into:
\begin{eqnarray}
\label{A.22}
 \pi  \ln \left [ \frac{ \Delta_{BCS}(T)}{ \Delta_{BCS}(0)} \right ] = - 2    
 \int_{0}^{\varepsilon_c}  d \xi   \;  \int_{0}^{2 \pi} d\theta \hspace{1.0cm}
 \\ \nonumber
  \frac{  \cos^2{(2 \theta)}} { \sqrt{ \xi^2 + [  \Delta_{BCS}(T) \cos{(2 \theta)}]^2 } }  
  \\ \nonumber
\frac{1}{ e^{(  \frac{ \sqrt{ \xi^2 + [  \Delta_{BCS}(T) \cos{(2 \theta)}]^2 } }{  k_B \; T }  ) } + 1 }  \; \; . 
\end{eqnarray}
Putting $x=\frac{\xi}{k_BT}$ and using $ \varepsilon_c \gg k_B T$, we obtain:
\begin{eqnarray}
\label{A.23}
 \pi  \ln \left [ \frac{ \Delta_{BCS}(T)}{ \Delta_{BCS}(0)} \right ] \; \simeq  \hspace{3.0cm}
\\ \nonumber 
  - 2  \;   \int_{0}^{\infty}  d x   \;  \int_{0}^{2 \pi} d\theta
  \frac{  \cos^2{(2 \theta)}} { \sqrt{ x^2 +  z^2 } }  \; \; 
\frac{1}{ e^{( \sqrt{ x^2 +  z ^2 })} + 1 }  \; \; ,
\end{eqnarray}
where:
\begin{equation}
\label{A.24}
z \; = \; \frac{\Delta_{BCS}(T)}{k_B \, T} \; |  \cos{(2 \theta)}| \;  .
\end{equation}
Let us consider:
\begin{eqnarray}
\label{A.25}
F(z) \; = \; 2 \;  \int_{0}^{\infty}  d x   \;
  \frac{ 1} { \sqrt{ x^2 +  z^2 } }  \; \; 
\frac{1}{ e^{( \sqrt{ x^2 +  z ^2 })} + 1 }  \;  \hspace{1.0cm}
\\ \nonumber
 = \; F_1(z) \; + F_2(z) \; \; ,
\end{eqnarray}
where:
\begin{equation}
\label{A.26}
F_1(z) \; =    \;   \int_{0}^{\infty}  d x   \;  \left [
  \frac{ 1} { \sqrt{ x^2 +  z^2 } }  \; \; - 
\frac{1}{ x }  \; \tanh (\frac{x}{2}) \; \right ] \;  , 
\end{equation}
and
\begin{eqnarray}
\label{A.27}
F_2(z) \; =    \;   \int_{0}^{\infty}  d x   \;  \Bigg [
 \frac{1}{ x }  \; \tanh (\frac{x}{2})  \;  \hspace{1.5cm}
 \\ \nonumber
 - \;  \frac{ 1} { \sqrt{ x^2 +  z^2 } }  \; \tanh (\frac{ \sqrt{ x^2 +  z^2 }}{2}) \; \Bigg ] \;  .
\end{eqnarray}
Since:
\begin{equation}
\label{A.28}
F_1(z) \; =    \;  \ln  \left [  \frac{ \pi } { e^{\gamma}  z }  \right ]\;  \;  , 
\end{equation}
we can rewrite  Eq.~(\ref{A.23}) as: 
\begin{eqnarray}
\label{A.29}
 \pi  \ln \left [ \frac{ \Delta_{BCS}(T)}{ \Delta_{BCS}(0)} \right ]  \simeq  - 
   \int_{0}^{2 \pi} d\theta \; \cos^2{(2 \theta)} \;  \ln  \left [  \frac{ \pi } { e^{\gamma}  z }  \right ]  \hspace{1.0cm} 
 \\ \nonumber
 -  \;  \int_{0}^{2 \pi} d\theta \; \cos^2{(2 \theta)} \; F_2(z)   \; .
\end{eqnarray}
Finally, observing that:
\begin{eqnarray}
\label{A.30}
  \int_{0}^{2 \pi} \; d\theta \; \cos^2{(2 \theta)}  \;  \ln  \bigg [  \frac{ \pi } { e^{\gamma}  z }  \bigg ] \; = \; \hspace{1.5cm}
 \\ \nonumber    
 \pi \; \Bigg (   \ln  \bigg [  \frac{ \pi \; k_B T } { e^{\gamma}  \Delta_{BCS}(T) } \bigg] \; 
 + \; \ln 2 \; - \; \frac{1}{2}  \Bigg  ) \; ,
\end{eqnarray}
we obtain:
\begin{eqnarray}
\label{A.31}
 \pi   \;  \ln  \left [  \frac{ 2 \pi \; k_B T } { e^{\gamma} \sqrt{e} \;  \Delta_{BCS}(0) } \right ] \; = \;   \hspace{1.5cm}
 \\ \nonumber
 - \;   \int_{0}^{2 \pi} d\theta \; \cos^2{(2 \theta)} \; 
 F_2\left (\frac{ \Delta_{BCS}(T) }{ k_B T }  |  \cos{(2 \theta)}|  \right ) \; .
\end{eqnarray}
The superconductive ordering sets in at a critical temperature  where  the energy gap vanishes. Since $F_2(0)=0$, from
Eq.~(\ref{A.31}) we obtain at once: 
\begin{equation}
\label{A.32}
 \frac{  k_B T_{BCS} } { \Delta_{BCS}(0) }  \; =  \;      \frac{ e^{\gamma} \;  \sqrt{e} }{ 2 \; \pi  } \; \; , 
\end{equation}
which agrees with  Eq.~(\ref{A.16}). This last equation allows us to write  Eq.~(\ref{A.31}) in the more  useful form:
\begin{eqnarray}
\label{A.33}
 \pi \;    \ln  \left (  \frac{T } {T_{BCS} } \right ) \; = \;  \hspace{3.0cm}
 \\ \nonumber
  - \;    \int_{0}^{2 \pi} d\theta \; \cos^2{(2 \theta)} \; 
 F_2\left (\frac{ \Delta_{BCS}(T) }{ k_B T }  |  \cos{(2 \theta)}|  \right ) \; .
\end{eqnarray}
The remaining integral in   Eq.~(\ref{A.33}) must be handled  numerically. As a result one obtains the BCS energy gap as
a function of the temperature. In Fig.~\ref{FigA}, top panel,  we display the   BCS energy gap $\Delta_{BCS}(T)$ versus the 
reduced temperature. \\
%
\begin{figure}
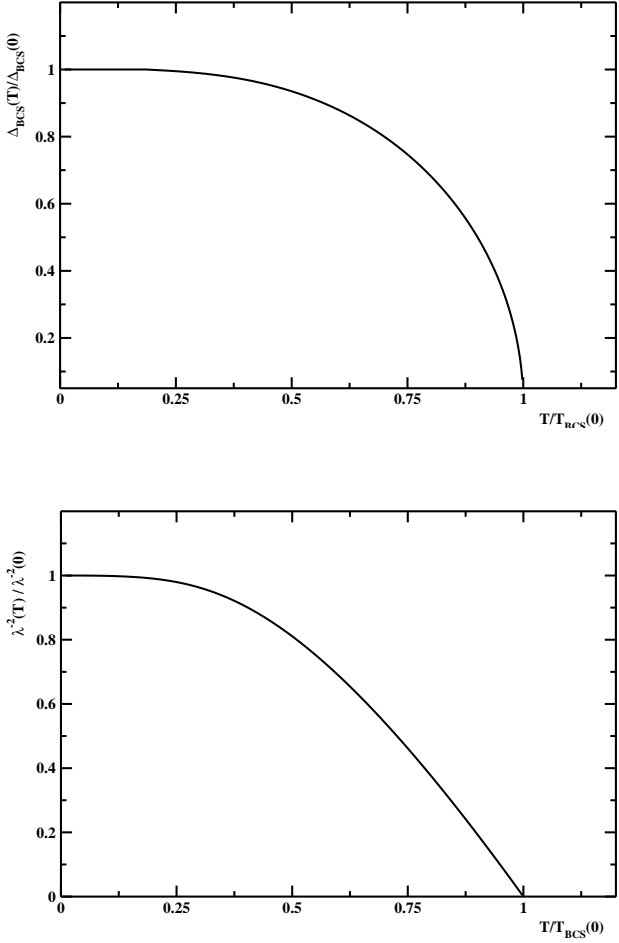

\vspace{0.7cm}
\centering
\resizebox{0.45\textwidth}{!}{%
\includegraphics{FigA1.eps} }

\vspace{1.0cm}

\resizebox{0.45\textwidth}{!}{%
\includegraphics{FigA2.eps}  }
\caption{\label{FigA} (Top) Temperature dependence of the BCS energy gap $\Delta_{BCS}(T)$. 
  (Bottom) Temperature dependence of the  BCS weak-coupling  penetration depth  $\lambda^{-2}(T)$ in the London limit.}
\end{figure}
%
To conclude this Section, we address ourselves  to the problem of evaluating the thermal  behavior of the
penetration length. In the London limit the BCS weak coupling d-wave theory gives for the penetration length the
following expression (see, for instance, Ref.~\cite{Tinkham:1996}):
\begin{eqnarray}
\label{A.34}
 \frac{\lambda^{-2}(T)} {\lambda^{-2}(0)} \; \simeq  \; 1 \; - \frac{2}{\pi \; k_B T} \;    
 \int_{0}^{\infty}  d x   \;  \int_{0}^{2 \pi} d\theta \;  \hspace{1.0cm}
 \\ \nonumber
 \cos^2 (2 \theta) \; 
 \frac{e^{\frac{E}{k_B T}  } }{ \left [ 1 +  e^{\frac{E}{k_BT}} \right ]^2 } \; \; ,
\end{eqnarray}
where 
\begin{equation}
\label{A.35}
   E \; = \; \sqrt{ x^2 +  \Delta_{BCS}^2(T) \cos^2{(2 \theta)}}  \; .
\end{equation}
For vanishing temperatures the integral in  Eq.~(\ref{A.34}) vanishes as $e^{- \Delta_{BCS}/k_B T}$, so that the penetration
length tends exponentially to the the zero-temperature value. On the other hand, at the critical temperature $T_{BCS}$,
the BCS energy gap vanishes. Then, from  Eq.~(\ref{A.34}) we infer:
\begin{eqnarray}
\label{A.36}
 \frac{\lambda^{-2}(T_{BCS})} {\lambda^{-2}(0)}  \simeq  1 \; - \frac{2}{\pi \; k_B T_{BCS}} \;    
 \int_{0}^{\infty}  d x   \;  \int_{0}^{2 \pi}  d\theta \; \hspace{1.0cm}
 \\ \nonumber
 \cos^2{(2 \theta)} \; 
   \frac{e^{\frac{x}{k_BT_{BCS}} }} { \left [ 1 +  e^{\frac{x}{k_B \, T_{BCS}}} \right ]^2 }   \; \; .
\end{eqnarray}
Observing that:
\begin{eqnarray}
\label{A.37}
  \int_{0}^{2 \pi} d\theta \; \cos^2{(2 \theta)} \; = \; \pi \; \; \; \; , 
  \\ \nonumber  
 \frac{1}{k_B T_{BCS}}    \int_{0}^{\infty}  d x   \;    
   \frac{e^{\frac{x}{k_BT_{BCS}} }} { \left [ 1 +  e^{\frac{x}{k_B \, T_{BCS}}} \right ]^2 }   
  \; = \; \frac{1}{2} \;  \; ,
\end{eqnarray}
one  can easily check that, indeed, $ \lambda^{-2}(T_{BCS}) = 0$. The temperature dependence of the London penetration
length can be obtained by integrating  numerically the double integral in  Eq.~(\ref{A.34}), after taking into account the
temperature dependence of the BCS gap. In  Fig.~\ref{FigA}, bottom panel,  we display the  BCS London
penetration length  versus the  reduced temperature. 
\section{Electrodynamics of the superconductive condensate}
\label{AppendixB}
In this Appendix we discuss the electrodynamics of the charge superconductive condensate within our theory.
In particular, we discuss explicitly  the lower critical magnetic field and the structure of the Abrikosov vortices.
From   Eq.~(\ref{3.88}) we get 
\begin{equation}
\label{3.96}  
 \nabla \,  \times \,  \vec{h}(\vec{R}) \; =   \;   \frac{4 \pi e \hbar}{m^*_h c}  \,  n_s(T)  \;  \nabla  \; \Theta(\vec{R}) \;  - \; 
 \frac{8 \pi e^2}{ m^*_h c^2} \, n_s(T)  \; \vec{A}( \vec{R})   \; .
\end{equation}
Using   Eq.~(\ref{3.91}) we rewrite this last equation as:
\begin{equation}
\label{3.97}  
 \nabla \,  \times \,  \vec{h}(\vec{R}) \; =  \;  \frac{1}{\lambda^2(T)}  \;  \left [ \frac{\phi_0}{2 \pi}   \;  \nabla  \; \Theta(\vec{R}) \;  -  \; \vec{A}( \vec{R}) \right ]  \; ,
\end{equation}
where 
\begin{equation}
\label{3.98} 
\phi_0  \; = \;   \frac{ 2 \pi \hbar c}{2 e} \; \simeq \; 2.07 \; 10^{-7}  \; G \; cm^2 
 \end{equation}
is the elementary quantum flux. Note that our result is considerably different from conventional BCS superconductors where the vortex core energy is 
negligible in extreme type-II superconductors. In our model the lower critical magnetic field can be considerable
larger than in BCS superconductors. In fact, as we will see in Sect.~\ref{s3.6},  $ H_{c1}^{core}$ turns out to be comparable
to  $H_{c1}^{em}$. \\  
Let us consider the simplest case of an isolated singularity at the origin $ \vec{R} = 0$. 
From   Eq.~(\ref{3.97}) we obtain: 
\begin{equation}
\label{3.99}  
 \vec{h}(\vec{R}) \; + \;  \lambda^2(T) \;   \nabla \, \times  \, \nabla  \, \times \,  \vec{h}(\vec{R}) \; =  \; 
\frac{\phi_0}{2 \pi}  \; \,    \nabla \, \times  \,   \nabla  \; \Theta(\vec{R})  \; .
\end{equation}
Now, we consider  the following integral extended over the surface of a small circle centered at the origin:
\begin{equation}
\label{3.100}  
\int \left [   \nabla \, \times  \,  \nabla  \; \Theta(\vec{R})  \right ] \cdot d \vec{S} \; \; .
\end{equation}
Using the Stokes's theorem we have:
\begin{equation}
\label{3.101}  
\int \left [   \nabla \, \times  \, \nabla  \; \Theta(\vec{R})  \right ] \cdot d \vec{S} \; = \;  
\oint \;   \nabla  \; \Theta(\vec{R})  \cdot d \vec{l} \; = \; 2 \, \pi
\end{equation}
since the phase $\Theta$ changes by $2 \pi$ after a full circle around the singularity. Therefore, we are led to:
\begin{equation}
\label{3.102}  
 \nabla \, \times  \,  \Theta(\vec{R})    \; = \;   \; 2 \, \pi \; \delta(\vec{R}) \; ,
\end{equation}
so that the singularity in    $\nabla  \; \Theta(\vec{R})$ carries one flux quantum showing, thereby, that it
coincides with an isolated Abrikosov vortex. In fact,   Eq.~(\ref{3.99}) gives:
\begin{equation}
\label{3.103}  
 \vec{h}(\vec{R}) \; + \;  \lambda^2(T) \;   \nabla \, \times  \, \nabla  \, \times \,  \vec{h}(\vec{R}) \; =  \; 
 \phi_0  \;      \delta(\vec{R}) \; , 
\end{equation}
or, better:
\begin{equation}
\label{3.104}  
 h(\vec{R}) \; - \;  \lambda^2(T) \;   \nabla^2  \;  h(\vec{R}) \; =  \; 
 \phi_0  \;      \delta(\vec{R}) \; .
\end{equation}
The solution of   Eq.~(\ref{3.104}) is the well-known Abrikosov vortex field distribution:
\begin{equation}
\label{3.105}  
 h(R) \;  =  \; 
 \frac{\phi_0}{2 \pi   \lambda^2(T)}  \;   K_0(\frac{R}{\lambda(T)})  \; ,
\end{equation}
where $K_n(x)$ is the modified Bessel function of order $n$. Since
\begin{equation}
\label{3.106}  
K_0(x)  \;   \stackrel{ x  \to  0}{\sim}  \;  \ln (\frac{1}{x}) \; \; \; , \; \; \; 
K_0(x)  \;   \stackrel{ x  \to  \infty}{\sim}  \;  \frac{1}{\sqrt{x}}  \; e^{-x} \; , 
\end{equation}
we see that the magnetic field decreases exponentially at large distance, but it diverges logarithmically at the vortex center.
However,   Eq.~(\ref{3.105})  is no longer valid in the vortex core, namely in a region of linear size $\xi_V$ around the vortex center,
where the condensate loses the surperfluidity. According to our model, since we are considering temperatures $ T \lesssim T^*$,
we still have bound holes which, however, have lost the phase coherence. To determine   $\xi_V$  we use    Eq.~(\ref{3.88})  to 
determine the superconductivity current density outside the vortex core:
\begin{eqnarray}
\label{3.107}  
j(R) \; \equiv  \; | \vec{j}_{em}(\vec{R})| \; = \; \frac{c}{4 \pi} \;   |\nabla \,  \times \,  \vec{h}(\vec{R})| 
\\ \nonumber
\; \simeq \; 
 \frac{c \; \phi_0}{8 \pi^2 \lambda^2}  \;  \frac{1}{R} \; \; , \; \;  r \; \gtrsim \; \xi_V \; \; . 
\end{eqnarray}
Since outside the vortex core we have $j \simeq 2 e \, n_s \, v_s$, we may estimate the vortex core radius as:
\begin{equation}
\label{3.108}  
j(\xi_V) \; \simeq   \;  2 e \, n_s \, v_c  \; ,
\end{equation}
where $v_c$ is the critical velocity given by~\footnote{See Eq.~(4.28) in I.} 
\begin{equation}
\label{3.109}  
m^*_h  v_c^2  \; \simeq  \; \Delta_2(\delta) \; . 
\end{equation}
As discussed in I, the meaning of the critical velocity is that for velocities greater than $v_c$ the condensate phase coherence
is lost. Combining Eqs.~(\ref{3.107}) and  (\ref{3.108}) we obtain:
\begin{equation}
\label{3.110}  
 \frac{1}{\xi_V} \; \simeq   \;  \frac{2 m_h^*}{\hbar}  \, v_c  \;  \simeq \;  \sqrt{ \frac{4 m^*_h}{\hbar^2} \;  \Delta_2(\delta)} \; .
\end{equation}
Within our approximations, it is easy to check that   Eq. (\ref{3.110})  implies that $\xi_V \lesssim d_0$. Now, we known that
$d_0$ is the average distance between hole pairs. so that, obviously, the vortex core size cannot be smaller than $d_0$, thereby
we must argue that:
\begin{equation}
\label{3.111}  
\xi_V \; \simeq   \;  d_0 \; \;  .
\end{equation}
Having determined the vortex core size, presently we estimate the magnetic field at the center of the vortex by
using  Eq.~(\ref{3.105}) by setting a cut-off at $R \simeq \xi_V$:
\begin{equation}
\label{3.112}  
 h(0) \;  \simeq  \; 
 \frac{\phi_0}{2 \pi   \lambda^2(T)}  \;   \ln ( \frac{\lambda}{\xi_V})  \; \simeq \;  
  \frac{\phi_0}{2 \pi   \lambda^2(T)}  \;   \ln ( \frac{\lambda(T)}{d_0})  \; .
\end{equation}
For temperatures not too close to the critical temperature $T_c$, where the isolated Abrikosov vortex approximation is 
usually not tenable, we may neglect in the logarithm the temperature dependence of the London penetration length obtaining :
\begin{equation}
\label{3.113}  
 h(0) \;  \simeq  \;   \frac{\phi_0}{2 \pi   \lambda^2(T)}  \;   \ln ( \kappa)   \; ,
\end{equation}
where we introduced the analogous to the Ginzburg-Landau $\kappa$-parameter:
\begin{equation}
\label{3.114}  
\kappa \;  \simeq  \;  \frac{\lambda(0)}{d_0}   \; .
\end{equation}
It is  interesting to estimate the value of the parameter $\kappa$. With our approximations, one obtains:
\begin{equation}
\label{3.115}  
\kappa \;  \simeq  \;  \sqrt{ \frac{m^*_h c^2}{8 \pi} \, \frac{c_0}{e^2} }   \; .
\end{equation}
Using the numerical values of the microscopic parameters we see that $\kappa \sim 10^2$  confirming that the hole doped
cuprates are extreme type-II superconductors~\cite{Blatter:1994}. \\
We turn, now, on  the determination of the lower critical magnetic field $H_{c1}$, namely the lowest magnetic field strength 
at which formation of Abrikosov vortices are thermodynamically favorable. First, we need to evaluate the free energy density in presence
of an isolated Abrikosov vortex. Evidently we have:
\begin{equation}
\label{3.116}  
{\cal F}_{sH} \; =  \; {\cal F}_{s0} \; + \; \frac{1}{8 \pi} \;  \vec{h}^2(\vec{R}) \; +\; \varepsilon_{kin} \; ,
\end{equation}
where   $\varepsilon_{kin}$ is the kinetic energy density of the supercurrent. We  have seen that: 
\begin{eqnarray}
\label{3.117}  
 \varepsilon_{kin} \; \simeq \;    n_s \, m^*_h \,    \vec{v}_s^{\,2} (\vec{R}, \vec{A}) \; \; \; , \; \; \; 
 \\ \nonumber
   \vec{v}_s (\vec{R}, \vec{A}) \; = \; \frac{\hbar}{  2   m^*_h}  \;   \nabla  \; \Theta(\vec{R}) \; - \;  \frac{e}{ m^*_h c} \; \vec{A}( \vec{R})  \; .
\end{eqnarray}
Using  Eq.~(\ref{3.85}) we rewrite the kinetic energy as:
\begin{equation}
\label{3.118}  
 \varepsilon_{kin} \; \simeq \;   \frac{m^*_h}{4 \, e^2 n_s} \;    \vec{j}_{em}^{\,2}   \; .
\end{equation}
With the aid of the Maxwell equation   Eq.~(\ref{3.88}) we obtain:
\begin{equation}
\label{3.119}  
 \varepsilon_{kin} \; \simeq \;   \frac{1}{8 \, \pi} \; \lambda^2  \;  [\nabla \,  \times \,  \vec{h}(\vec{R})]^{\,2}   \; .
\end{equation}
Therefore we end with:
\begin{equation}
\label{3.120}  
{\cal F}_{sH} \; =  \; {\cal F}_{s0} \; + \; \frac{1}{8 \pi} \bigg \{ \vec{h}^2(\vec{R}) \; + \; 
\lambda^2  \;  [\nabla \,  \times \,  \vec{h}(\vec{R})]^{\,2} \bigg \}  \; .
\end{equation}
Whereupon, the variation of the free energy due to the presence of an isolated Abrikosov vortex is given by:
\begin{eqnarray}
\label{3.121}  
\Delta F_V \; = \; \int \, d\vec{R} \, \bigg ( {\cal F}_{sH} \; -  \; {\cal F}_{s0} \bigg ) \; \simeq  \;
\\ \nonumber
+ \,  \frac{1}{8 \pi}  \int \,  d\vec{R} \,  \bigg \{ \vec{h}^2(\vec{R}) \; + \; 
\lambda^2  \;  [\nabla \,  \times \,  \vec{h}(\vec{R})]^{\,2} \bigg \}  \; .
\end{eqnarray}
After some standard manipulations~\footnote{See, for instance, Ref.~\cite{Schmidt:1997}.}, we get:
\begin{equation}
\label{3.122}  
\Delta F_V   \simeq 
 \frac{1}{8 \pi}  \int  d\vec{R} \;   \vec{h}(\vec{R}) \cdot  \bigg \{ \vec{h}(\vec{R}) + 
\lambda^2  \;   \nabla \,  \times \,    \nabla \,  \times \,  \vec{h}(\vec{R})  \bigg \}  \; ,
\end{equation}
or, using   Eq.~(\ref{3.103}) 
\begin{equation}
\label{3.123}  
\Delta F_V  \; \simeq  \;  \frac{\phi_0}{8 \pi}  \; h(0)  \; \simeq   \;    
  \frac{\phi_0^2}{16 \pi^2   \lambda^2(T)}  \;   \ln (\kappa) \; \; . 
\end{equation}
However, we need to take care of the energy due to the vortex core. Usually, the core energy can be neglected due to
the fact that the vortex core being  in the normal non-superconductive phase does not contribute to $ \Delta F_V$.
In our theory the normal phase in the core region corresponds to phase-disordered hole pairs. On the other hand, we have 
estimate that $\xi_V \simeq d_0$. Then, the vortex core energy corresponds to the energy of a region of disordered hole pair
condensate of linear size $d_0$, implying  that the vortex core energy coincides with the excitation energy of rotons.
Therefore, we can write for the free energy of an isolated Abrikosov vortex:
\begin{equation}
\label{3.124}  
 \varepsilon_{vor}(T)  \; \simeq \;   \varepsilon_{em}(T) \; + \;  \varepsilon_{core}(T) \; \; ,
\end{equation}
where
\begin{equation}
\label{3.125}  
 \varepsilon_{em}(T)  \; \simeq \;   \frac{\phi_0^2}{16 \pi^2   \lambda^2(T)}  \;   \ln (\kappa)  \;  \; \; ,
\end{equation}
is the energy of the magnetic field and the supercurrent associated to the vortex, and 
\begin{equation}
\label{3.126}  
 \varepsilon_{core}(T)   \; \simeq \;  \frac{1}{c_0} \;  \varepsilon_{rot}(T) 
\end{equation}
is the vortex core energy per unit length.   Eq.~(\ref{3.124}) shows that the free energy of an isolated Abrikosov vortex is positive, i.e.
the vortex cannot exist in the superconductor in absence of an external magnetic field. To determine the minimum value
of the external magnetic field strength at which Abrikosov vortex formation becomes favorable, we need to minimize the
Gibbs free energy (see, for instance, Ref.~\cite{Landau:1984}):
\begin{equation}
\label{3.127}  
G \; =  \;   \varepsilon_{vor}(T)  \; - \;  \frac{1}{4 \pi}  \int  d\vec{R} \; \;   \vec{H}_0 \cdot   \vec{h}(\vec{R})  \; ,
\end{equation}
where $H_0$ is the applied  magnetic field with constant field strength. Since the Abrikosov vortex carries one magnetic flux quantum,
from   Eq.~(\ref{3.127}) one obtains at once:
\begin{equation}
\label{3.128}  
G \; =  \;   \varepsilon_{vor}(T)  \; - \;  \frac{1}{4 \pi}   \; \phi_0  H_0   \;  \; .
\end{equation}
It is evident from  Eq.~(\ref{3.128}) that for weak external field $G > 0$ and there is no vortex formation. For external fields
exceeding the lower critical field:
\begin{equation}
\label{3.129}  
H_{c1} \; =  \;   \frac{4 \pi}{\phi_0} \;   \varepsilon_{vor}(T)   
\end{equation}
it is energetically favorable to produce Abrikosov vortices. Using   Eqs.~(\ref{3.124}) - (\ref{3.126}) we find:
\begin{equation}
\label{3.130}  
H_{c1}(T) \; =  \;  H_{c1}^{em} (T) \; + \;   H_{c1}^{core}(T)  \; \; ,
\end{equation}
with:
\begin{equation}
\label{3.131}  
H_{c1}^{em}(T)  \; \simeq \;   \frac{\phi_0}{4 \pi  \lambda^2(T)}  \;   \ln (\kappa)  \; , 
\end{equation}
and
\begin{equation}
\label{3.132}  
H_{c1}^{core}(T) \; \simeq  \;   \frac{4 \pi}{c_0 \phi_0} \;   \varepsilon_{rot}(T)   \; .
\end{equation}
\section{Density of states in the vortex region}
\label{AppendixC}
 To determine the effects of the Abrikosov vortices on the density of states of  the low-lying excitations, we observe that
  at low energy only the nodal quasielectrons can be excited in the core of the vortex where
 the pair condensate is phase disordered.  Now, following  Ref.~\cite{Volovik:1993} we may employ the semiclassical approach 
 where the momentum and position of quasiparticles are  commuting variables. The effects of the supercurrent circulating around a vortex 
 is accounted for by a Doppler shift of  the quasiparticle energy according to  Eq.~(\ref{4.16}). Thus, for the nodal
 quasielectrons we have:
\begin{equation}
\label{4.46} 
  \varepsilon_{\vec{k}}( \vec{R})   \; \simeq  \;    \frac{\hbar^2 \vec{k}^2}{2 \; m^*_e}  \; + \;  \hbar \; \vec{k} \cdot \vec{v}_s(\vec{R})         \; ,
 \end{equation}
where $\vec{v}_s(\vec{R}) $ is the superfluid velocity around an Abrikosov vortex. Adopting polar coordinates, we have:
\begin{equation}
\label{4.47} 
 \vec{v}_s(\vec{R})  \; = \;  \vec{v}_s(R, \theta_R)  \; \simeq \frac{\hbar}{4 \; m^*_h} \; \frac{\hat{\theta}_R}{R}    \;   \; .
 \end{equation}
The Doppler shift  Eq.~(\ref{4.47}) could modify  the density of state. To see this, we need to evaluate the density of state per
spin  at the Fermi surface averaged over the vortex:
\begin{equation}
\label{4.48} 
 < {\cal{N}}(0) >_{vor}   \equiv   \int  \frac{d \vec{k}}{(2 \pi)^2}  \int  d\vec{R} \;
  \delta \left [  \varepsilon_{\vec{k}}( \vec{R})  -   \varepsilon_F^{(e)} \right ]   \frac{1}{\pi d_V^2}  \; ,
 \end{equation}
where  $d_V$ is the average distance between vortices.  Eq. (\ref{4.48}) refers to an isolated Abrikosov vortex. However,
since the high temperature superconductors are extreme type-II superconductors and  $d_V \gg \xi_V \simeq d_0$, 
the dilute vortex approximation should be reliable.  
Now, we observe that:
\begin{eqnarray}
\label{4.49} 
\varepsilon_{\vec{k}}( \vec{R})  -   \varepsilon_F^{(e)} \; = \;  \hspace{1.5cm}
\\ \nonumber
 \frac{\hbar^2}{2  m^*_e} \left [  k^2 \; + \; \frac{ m^*_e}{2 m^*_h} \,
\frac{k}{R} \, \sin (\theta_R + \theta_k)  - (k_F^{(e)})^2 
\right ] \; .
\end{eqnarray}
So that we obtain:
\begin{eqnarray}
\label{4.50} 
 < {\cal{N}}(0) >_{vor}   \simeq  \frac{ 2  m^*_e}{\hbar^2} 
\int_{FA}  \frac{d\theta_k}{4 \pi^2}  \int_{0}^{\infty} dk \, k   \int d\theta_R  \hspace{1.0cm}
\\ \nonumber
 \int_{\xi_V}^{d_V}  dR  \frac{R}{\pi d_V^2} \;
  \delta  \bigg [  k^2 \; + \; \frac{ m^*_e}{2 m^*_h} \, \frac{k}{R} \, \sin (\theta_R + \theta_k)  - (k_F^{(e)})^2 
\bigg ]  \; .
 \end{eqnarray}
The integration over $k$ can be done easily once one realizes that:
\begin{eqnarray}
\label{4.51}  
\nonumber
  k^2  +  \frac{ m^*_e}{2 m^*_h} \, \frac{k}{R} \, \sin (\theta_R + \theta_k)  - (k_F^{(e)})^2  \; = \; 
(k - k_+) (k-k_-) \; \; ,
\\  \nonumber 
k_{\pm} \;  =  \;   \frac{ m^*_e}{4 m^*_h R} \sin (\theta_R + \theta_k)  \hspace{3.5cm}
\\ 
 \pm \; \sqrt{ \left [ \frac{ m^*_e}{4 m^*_h R}  \sin (\theta_R + \theta_k) \right ]^2
+ \left ( k_F^{(e)} \right )^2 } \; . \hspace{1.0cm}
\end{eqnarray}
We get:
\begin{eqnarray}
\label{4.52} 
 < {\cal{N}}(0) >_{vor}   \simeq {\cal{N}}(0)  \; + \;  \frac{m^*_e}{\hbar^2}  
\int_{FA}  \frac{d\theta_k}{4 \pi^2}    \int d\theta_R  \hspace{1.5cm}
\\ \nonumber
 \int_{\xi_V}^{d_V}  dR \;
  \frac{R}{\pi d_V^2}  \frac{ \frac{ m^*_e}{4 m^*_h R}  \sin (\theta_R + \theta_k)}{ \sqrt{ \left [ \frac{ m^*_e}{4 m^*_h R}  \sin (\theta_R + \theta_k) \right ]^2 + \left ( k_F^{(e)} \right )^2 }} 
  \; .
 \end{eqnarray}
Now, it is easy to check that the angular integration over $\theta_R$ in the second term on the right hand side of 
 Eq.~(\ref{4.52}) vanishes. Therefore we end with the following remarkable result:
\begin{equation}
\label{4.53} 
 < {\cal{N}}(0) >_{vor}   \; \simeq \; {\cal{N}}(0) \; .  
 \end{equation}
Note that this last equation replaces  Eq.~(4.67) in I. Likewise, one can check that   Eq.~(\ref{4.53}) extends also to
thermal activated rotons which, in any case, are not relevant at low temperatures. 
\section{The static Lindhard function}
\label{AppendixD}
Let us consider:
\begin{equation}
\label{5.28} 
 \chi(\vec{q}) \;  = \;   2  \; \int  \frac{d \vec{k}}{(2 \pi)^2} \; 
 \frac{n(\vec{k}) \; - \; n(\vec{k}+\vec{q})}{ \varepsilon_{\vec{k}}^{(e)} \, - \,   \varepsilon_{\vec{k}+\vec{q}}^{(e)}} \; .
\end{equation}
Now, we note that:
\begin{equation}
\label{5.29} 
 \int  \frac{d \vec{k}}{(2 \pi)^2} \; 
 \frac{n(\vec{k}+\vec{q})}{ \varepsilon_{\vec{k}}^{(e)} \, - \,   \varepsilon_{\vec{k}+\vec{q}}^{(e)}} \; 
 = \;  \int  \frac{d \vec{k}}{(2 \pi)^2} \;  \frac{n(\vec{k})}{ \varepsilon_{\vec{k}-\vec{q}}^{(e)} \, - \,   \varepsilon_{\vec{k}}^{(e)}} 
 \; .
\end{equation}
Putting $\vec{k}  \rightarrow - \vec{k}$ and noting that   $  \varepsilon_{- \vec{k}}^{(e)} =   \varepsilon_{\vec{k}}^{(e)} $, we get:
\begin{equation}
\label{5.30} 
 \int  \frac{d \vec{k}}{(2 \pi)^2} \; 
 \frac{n(\vec{k}+\vec{q})}{ \varepsilon_{\vec{k}}^{(e)} \, - \,   \varepsilon_{\vec{k}+\vec{q}}^{(e)}} \; 
 = \;  \int  \frac{d \vec{k}}{(2 \pi)^2} \;  \frac{n(\vec{k})}{ \varepsilon_{\vec{k}+\vec{q}}^{(e)} \, - \,   \varepsilon_{\vec{k}}^{(e)}} 
 \; .
\end{equation}
So that we can write:
\begin{equation}
\label{5.31} 
 \chi(\vec{q}) \;  = \;   4  \; \int  \frac{d \vec{k}}{(2 \pi)^2} \; 
 \frac{n(\vec{k})}{ \varepsilon_{\vec{k}}^{(e)} \, - \,   \varepsilon_{\vec{k}+\vec{q}}^{(e)}} \; .
\end{equation}
Since
\begin{equation}
\label{5.32} 
 \varepsilon_{\vec{k}}^{(e)} \, - \,   \varepsilon_{\vec{k}+\vec{q}}^{(e)} \; = \; - \; \frac{\hslash^2 q}{m^*_e} \; 
 \big [ \frac{q}{2} \; + \; k \, \cos \theta_{kq} \big ] \; ,
\end{equation}
we readily  obtain:
\begin{equation}
\label{5.33} 
 \chi(\vec{q})  =  -  \frac{m^*_e}{\pi^2 \hslash^2 q}     \int_{FA} \, d \theta_k  \int_{0}^{k_F^{(e)}}  d k \,
 \frac{k}{\frac{q}{2} + k \cos \theta_{kq}} \; .
\end{equation}
It is convenient to rewrite Eq.~(\ref{5.33}) as:
\begin{eqnarray}
\label{5.34} 
 \chi(\vec{q}) \;  = \;  - \; \frac{m^*_e}{\pi^2 \hslash^2 q}    \; \int_{FA} \, d \theta_k  \int_{0}^{k_F^{(e)}}  d k \; 
 \\ \nonumber
\left [ \frac{1}{\cos \theta_{kq}} \; - \; \frac{q}{2 \cos \theta_{kq}} \,  \frac{1}{\frac{q}{2} + k \cos \theta_{kq}} 
\right ] \; .
\end{eqnarray}
We see that the singularity in  $ \chi(\vec{q})$ arises from the second term in the square brackets on the right hand side
of   Eq.~(\ref{5.34}) whenever:
\begin{equation}
\label{5.35} 
 q \; + \; 2 \,  k_F^{(e)}  \, \cos \theta_{kq}  \;  = \; 0 \; \; .
\end{equation}
 Let us consider, for instance,  $\vec{q}  = -  \vec{Q}_1$. The singular part of the Lindhard function is:
\begin{eqnarray}
\label{5.37} 
 \chi(Q_1) \;  \simeq  \;  + \; \frac{m^*_e}{2 \pi^2 \hslash^2 }    \; \int_{FA} \, d \theta_k  \int_{0}^{k_F^{(e)}}  d k \; 
 \\ \nonumber
 \frac{1}{\cos \theta_{kq}} \; \;  \frac{1}{\frac{q}{2} + k \cos \theta_{kq}}  \; \; .
\end{eqnarray}
We find:
\begin{equation}
\label{5.38} 
 \chi(Q_1) \;  \simeq  \;  + \; 2 \, \theta_{FA} \, \frac{m^*_e}{\pi^2 \hslash^2 }    \; \ln \left |
1 \; - \;  \frac{  \sqrt{2} \, k_F^{(e)} }{Q_1} 
\right | \; ,
\end{equation}
or, using Eq.~(\ref{4.33})
\begin{equation}
\label{5.39} 
 \chi(Q_1) \;  \simeq  \;  + \; 2 \,  {\cal{N}}(0)   \; \ln \left |
1 \; - \;  \frac{  \sqrt{2} \, k_F^{(e)} }{Q_1} 
\right | \; .
\end{equation}
Obviously,  Eq.~(\ref{5.39}) applies also to the wavevectors   $+ \vec{Q}_1$,  $\pm \vec{Q}_2$ since $ |\vec{Q}_1| =  |\vec{Q}_2|$.
The logarithmic divergence of the static response function points to the instability of the nodal quasielectron gas towards the
formation of a charge density wave.  To determine the  onset temperature for this  instability we need to evaluate the Lindhard response
function at finite temperatures:
\begin{equation}
\label{5.40} 
 \chi(\vec{q},T) \;  = \;   2  \; \int  \frac{d \vec{k}}{(2 \pi)^2} \; 
 \frac{f(\varepsilon^{(e)}_{\vec{k}}) \; - \; f(\varepsilon^{(e)}_{\vec{k} +\vec{q}})}{ \varepsilon_{\vec{k}}^{(e)} \, - \,  
  \varepsilon_{\vec{k}+\vec{q}}^{(e)}} \; .
\end{equation}
The main contributions to the integral in Eq.~(\ref{5.40}) come around the Fermi energy. So that we can write:
\begin{equation}
\label{5.41} 
 \varepsilon_{\vec{k}}^{(e)} \, - \,   \varepsilon_F^{(e)} \; \simeq \; + \; \frac{\hslash^2 k_F^{(e)}}{m^*_e} \; 
 ( k \; - \;  k_F^{(e)} ) \; ,
\end{equation}
\begin{equation}
\label{5.42} 
 \varepsilon_{\vec{k}+\vec{q}}^{(e)} \, - \,   \varepsilon_F^{(e)} \; \simeq \; + \; \frac{\hslash^2}{2 m^*_e} \; 
 ( q^2 \; + \; 2  k q \cos \theta_{kq} ) \; ,
\end{equation}
\begin{equation}
\label{5.43} 
\varepsilon^{(e)}_{\vec{k}}  \;  -  \;   \varepsilon_{\vec{k}+\vec{q}}^{(e)}
 \; \simeq \; - \; \frac{\hslash^2}{2 m^*_e} \;  ( q^2 \; + \; 2  k q \cos \theta_{kq} ) \; .
\end{equation}
Therefore, for   $\vec{q}  =  -  \vec{Q}_1$ we get:
\begin{equation}
\label{5.44} 
 \varepsilon_{\vec{k}+\vec{q}}^{(e)} \, - \,   \varepsilon_F^{(e)}  \simeq  - \; \frac{\hslash^2 k_F^{(e)} }{ m^*_e} \; 
 ( k \; - \;  k_F^{(e)} )   \simeq  - \; ( \varepsilon_{\vec{k}}^{(e)} \, - \,   \varepsilon_F^{(e)} ) \; ,
\end{equation}
\begin{equation}
\label{5.45} 
\varepsilon^{(e)}_{\vec{k}}  \;  -  \;   \varepsilon_{\vec{k}+\vec{q}}^{(e)}
  \simeq  + \; \frac{\hslash^2 k_F^{(e)} }{ m^*_e} \; 
 ( k \; - \;  k_F^{(e)} )   \simeq  + \; ( \varepsilon_{\vec{k}}^{(e)} \, - \,   \varepsilon_F^{(e)} ) \; .
\end{equation}
So that one finds:
\begin{eqnarray}
\label{5.46} 
 \chi(Q_1,T) \;  \simeq \;   2  \; \int  \frac{d \vec{k}}{(2 \pi)^2} \; 
 \frac{ f(\xi_{\vec{k}}) \; - \; f( - \xi_{\vec{k}}) }{ \xi_{\vec{k}} } \; \; , \; \; 
 \\ \nonumber
\xi_{\vec{k}} \; = \; \varepsilon_{\vec{k}}^{(e)} \, - \,   \varepsilon_F^{(e)}  \; . \; \;
\end{eqnarray}
Now, observing that
\begin{equation}
\label{5.47} 
 \int  \frac{d \vec{k}}{(2 \pi)^2} \;  \simeq \;    {\cal{N}}(0)   \int d   \varepsilon_{\vec{k}}^{(e)}   \; ,
\end{equation}
we obtain:
\begin{eqnarray}
\label{5.48} 
 \chi(Q_1,T) \;  \simeq \;   2  \; \ {\cal{N}}(0)   \int d   \varepsilon_{\vec{k}}^{(e)}  \; 
 \frac{ f(\xi_{\vec{k}}) \; - \; f( - \xi_{\vec{k}}) }{ \xi_{\vec{k}} } \;   \; \; 
 \\ \nonumber
 \simeq \; 
  2  \; \ {\cal{N}}(0)   \int_{0}^{\varepsilon_c}  d   \xi  \; 
 \frac{ f(\xi) \; - \; f( - \xi) }{ \xi} \;  , \; \;
\end{eqnarray}
where  $\varepsilon_c$ is a high-energy cutoff. Finally, using
\begin{equation}
\label{5.49} 
 f(\xi) \; - \; f( - \xi) \; = \; - \; \tanh \left ( \frac{\xi}{2 k_B T} \right ) \;   ,
\end{equation}
we recast Eq.~(\ref{5.48}) into:
\begin{equation}
\label{5.50} 
 \chi(Q_1,T) \;  \simeq \;  
- \;   2  \; \ {\cal{N}}(0)   \int_{0}^{\frac{\varepsilon_c}{2 k_B T}}  \frac{d  x}{x}   \; \tanh x \;  .
\end{equation}
After taking into account that:
\begin{equation}
\label{5.51} 
  \int_{0}^{\Lambda}  \frac{d  x}{x}   \; \tanh x \;  \simeq  \; \ln \Lambda \; - \; \ln \pi \; + 2 \, \ln 2 \; + \; \gamma \;  ,
\end{equation}
where $ \gamma  = 0.577216...$ is the Euler's constant~\footnote{We must mention that in the literature sometimes our Euler's constant
is denoted by $C$, while $\gamma = e^C$.}, we obtain finally:
\begin{equation}
\label{5.52} 
 \chi(Q_1,T) \;  \simeq \;  
- \;   2  \; \ {\cal{N}}(0)  \;  \ln \left ( \frac{2 e^\gamma}{\pi} \; {\frac{\varepsilon_c}{ k_B T}} \right ) \;  .
\end{equation}
Regarding the cutoff energy $\varepsilon_c$, usually it is assumed to be of the order of the Fermi energy $\varepsilon_F^{(e)}$.
However,  as discussed in Sect.~\ref{s4}, the nodal quasielectrons are intimately related to the hole  pairs. Therefore the natural
high-energy cutoff should be the binding energy of the hole pairs:
\begin{equation}
\label{5.53} 
\varepsilon_c  \; \simeq \; \Delta_2(\delta) \;  
\end{equation}
which, indeed, turns out to be much smaller than the Fermi energy,   $  \Delta_2(\delta) \ll  \varepsilon_F^{(e)}$.
\section{The CDW effective Hamiltonian}
\label{AppendixE}
 In the mean field approximation
the effective Hamiltonian Eq.~(\ref{5.14}) reads:
\begin{eqnarray}
\label{5.65} 
\hat{H}_{eff} \;  =  \;  \frac{ V_u^2 \Delta_{CDW}^2 \hslash \Omega(Q_1)}{2 g^2(Q_1)}  \;  \; \hspace{2.0cm}     
\\ \nonumber
+ \;  \sum_{\vec{k},\sigma} \; \bigg ( \varepsilon_{\vec{k}}^{(e)} \, - \,    \varepsilon_{F}^{(e)}  \bigg )
 \;   \hat{\psi}^{\dagger}_{e}(\vec{k},\sigma)  \;   \hat{\psi}_{e}(\vec{k},\sigma) \; \hspace{0.5cm}
 \\ \nonumber
   + \;  
  \frac{1}{V_u}  \; \sum_{\vec{k},\sigma}  
2 \; g(Q_1)  \; 
  \bigg \{ < \hat{b}_{1}(-\vec{Q}_1 ) >    \hat{\psi}^{\dagger}_{e}(\vec{k} - \vec{Q}_1,\sigma)   \hat{\psi}_{e}(\vec{k},\sigma)
 \\ \nonumber
  +   < \hat{b}_{1}(+\vec{Q}_1 ) >    \hat{\psi}^{\dagger}_{e}(\vec{k} +\vec{Q}_1,\sigma)   \hat{\psi}_{e}(\vec{k},\sigma) 
 \bigg \}  ,
 \end{eqnarray}
where, for definiteness, we are considering the charge density wave instability at $q=Q_1$. Using Eq.~(\ref{5.62}) we get:
\begin{eqnarray}
\label{5.66} 
\hat{H}_{eff} \;  =  \;  \frac{ V_u^2 \Delta_{CDW}^2 \hslash \Omega(Q_1)}{2 g^2(Q_1)}    \;  \; \hspace{1.0cm}     
\\ \nonumber
 +   \;   \sum_{\vec{k},\sigma} \; \bigg ( \varepsilon_{\vec{k}}^{(e)} \, - \,    \varepsilon_{F}^{(e)}  \bigg )
 \;   \hat{\psi}^{\dagger}_{e}(\vec{k},\sigma)  \;   \hat{\psi}_{e}(\vec{k},\sigma) \;   \; \;
\\ \nonumber 
 +  \;  \Delta_{CDW}  \;   \sum_{\vec{k},\sigma}  \;  \bigg \{  \hat{\psi}^{\dagger}_{e}(\vec{k} - \vec{Q}_1,\sigma)   \hat{\psi}_{e}(\vec{k},\sigma)
 \\ \nonumber
  \; + \;    \hat{\psi}^{\dagger}_{e}(\vec{k} +\vec{Q}_1,\sigma)   \hat{\psi}_{e}(\vec{k},\sigma) 
 \bigg \}  \;  .
 \end{eqnarray}
Let us consider the nodal quasielectrons in sectors $(I)$ and $(II)$ that are connected by the nesting wavevectors
$\pm \vec{Q}_1$. We label by subscripts $1$ and $2$ the quasielectrons in the Fermi arc sectors  $(I)$ and $(II)$
respectively. We consider only states near the Fermi level. After making use of Eq.~(\ref{5.44}) we get:
\begin{eqnarray}
\label{5.67} 
\hat{H}_{eff} \;  \simeq  \;   \frac{ V_u^2 \Delta_{CDW}^2 \hslash \Omega(Q_1)}{2 g^2(Q_1)}     \;  \; \hspace{1.0cm}   
\\ \nonumber
 +    \; V_u \;  \frac{{\cal{N}}(0)}{2} 
 \sum_{\sigma}   \int  d \varepsilon^{(e)}_{\vec{k}} \; \xi_{\vec{k}}  \hspace{0.8cm}
 \\ \nonumber
  \bigg [ \hat{\psi}^{\dagger}_{1}(\vec{k},\sigma) 
   \hat{\psi}_{1}(\vec{k},\sigma)  -  \hat{\psi}^{\dagger}_{2}(\vec{k},\sigma)    \hat{\psi}_{2}(\vec{k},\sigma)  \bigg ]
\\ \nonumber 
  +  \;  V_u \; \frac{{\cal{N}}(0)}{2} \;  \Delta_{CDW} \;  \sum_{\sigma}   \int  d \varepsilon^{(e)}_{\vec{k}} 
\\ \nonumber
 \bigg [  \hat{\psi}^{\dagger}_{2}(\vec{k},\sigma) 
   \hat{\psi}_{1}(\vec{k},\sigma)  +   \hat{\psi}^{\dagger}_{1}(\vec{k},\sigma)    \hat{\psi}_{2}(\vec{k},\sigma)  \bigg ] , 
\end{eqnarray}
where $\xi_{\vec{k}} =  \varepsilon_{\vec{k}}^{(e)}  -   \varepsilon_{F}^{(e)}$.  Of course, we must add also the contribution due to the
quasielectrons in the Fermi arc sectors   $(III)$ and $(IV)$. The final result can be written as:
\begin{eqnarray}
\label{5.68} 
\hat{H}_{eff} \;  \simeq  \;   \frac{ V_u^2 \Delta_{CDW}^2 \hslash \Omega(Q_1)}{2 g^2(Q_1)}     \;  \; \hspace{1.0cm}   
\\ \nonumber
 +  \; V_u  \;   {\cal{N}}(0) 
 \sum_{\sigma}   \int  d \varepsilon^{(e)}_{\vec{k}} \; \xi_{\vec{k}}  \hspace{0.8cm}
 \\ \nonumber
   \bigg  [ \hat{\psi}^{\dagger}_{1}(\vec{k},\sigma) 
   \hat{\psi}_{1}(\vec{k},\sigma)  -  \hat{\psi}^{\dagger}_{2}(\vec{k},\sigma)    \hat{\psi}_{2}(\vec{k},\sigma)  \bigg ]
\\ \nonumber
  +   \;  V_u   \;  {\cal{N}}(0) \;  \Delta_{CDW} \;  \sum_{\sigma}   \int  d \varepsilon^{(e)}_{\vec{k}} 
  \\ \nonumber
  \bigg [  \hat{\psi}^{\dagger}_{2}(\vec{k},\sigma) 
   \hat{\psi}_{1}(\vec{k},\sigma)  +   \hat{\psi}^{\dagger}_{1}(\vec{k},\sigma)    \hat{\psi}_{2}(\vec{k},\sigma)  \bigg ] \;  . 
  \end{eqnarray}
The Hamiltonian Eq.~(\ref{5.68}) can be diagonalized by means of the Bogoliubov-Valatin canonical 
transformations~\cite{Bogoliubov:1958b,Valatin:1958}:
\begin{eqnarray}
\label{5.69} 
\hat{\tilde{\psi}}_{1}(\vec{k},\sigma) =  \cos (\frac{\theta_{\vec{k}}}{2}) \;  \hat{\psi}_{1}(\vec{k},\sigma)   -  
 \sin (\frac{\theta_{\vec{k}}}{2})  \; \hat{\psi}_{2}(\vec{k},\sigma)  \; \; \; \; \;
\\ \nonumber
\hat{\tilde{\psi}}_{2}(\vec{k},\sigma)  =  \sin (\frac{\theta_{\vec{k}}}{2}) \;  \hat{\psi}_{1}(\vec{k},\sigma)  \; + \; 
 \cos (\frac{\theta_{\vec{k}}}{2})  \; \hat{\psi}_{2}(\vec{k},\sigma)  \; \; .
 \end{eqnarray}
After some algebra we find:
\begin{eqnarray}
\label{5.70} 
\hat{H}_{eff} \;  \simeq  \;   \frac{ V_u^2 \Delta_{CDW}^2 \hslash \Omega(Q_1)}{2 g^2(Q_1)}   \;  \; \hspace{1.0cm}   
\\ \nonumber
 +  \;   V_u  \;   {\cal{N}}(0)  \sum_{\sigma}   \int  d \varepsilon^{(e)}_{\vec{k}} \;  \hspace{1.0cm}
 \\ \nonumber 
 \Bigg \{  
 \bigg [ \xi_{\vec{k}} \cos \theta_{\vec{k}}  -   \Delta_{CDW} \sin \theta_{\vec{k}} \bigg ] 
 \\ \nonumber
   \bigg [ \hat{\tilde{\psi}}^{\dagger}_{1}(\vec{k},\sigma) 
   \hat{\tilde{\psi}}_{1}(\vec{k},\sigma)  -  \hat{\tilde{\psi}}^{\dagger}_{2}(\vec{k},\sigma)    \hat{\tilde{\psi}}_{2}(\vec{k},\sigma)  \bigg ]
\\ \nonumber 
  \; + \;  \bigg [ \xi_{\vec{k}} \sin \theta_{\vec{k}}   +  \Delta_{CDW} \cos \theta_{\vec{k}}    \bigg ] 
  \\ \nonumber
  \bigg [  \hat{\tilde{\psi}}^{\dagger}_{2}(\vec{k},\sigma) 
   \hat{\tilde{\psi}}_{1}(\vec{k},\sigma)  +   \hat{\tilde{\psi}}^{\dagger}_{1}(\vec{k},\sigma)    \hat{\tilde{\psi}}_{2}(\vec{k},\sigma)  \bigg ]
  \; \Bigg \}   \; .  
  \end{eqnarray}
To diagonalize the effective Hamiltonian we choose the angles $\theta_{\vec{k}}$ such that:
\begin{equation}
\label{5.71} 
  \xi_{\vec{k}} \;  \sin \theta_{\vec{k}} \;   + \;  \Delta_{CDW} \; \cos \theta_{\vec{k}} \; = \;  0 \; \; ,
\end{equation}
or equivalently:
\begin{equation}
\label{5.72} 
 \tan \theta_{\vec{k}} \; = \; - \; \frac{\Delta_{CDW}}{  \xi_{\vec{k}} } \; \; . 
\end{equation}
From Eq.~(\ref{5.72}) we get:
\begin{equation}
\label{5.73} 
 \cos \theta_{\vec{k}}  =   \frac{ \xi_{\vec{k}}}{\sqrt{  \xi^2_{\vec{k}}+ \Delta^2_{CDW}}} \;  , \; 
 \sin \theta_{\vec{k}}   =  - \;  \frac{    \Delta_{CDW} }{\sqrt{  \xi^2_{\vec{k}}+ \Delta^2_{CDW}}} 
 \; \; . 
\end{equation}
Therefore we end with:
\begin{eqnarray}
\label{5.74} 
\hat{H}_{eff} \;  \simeq  \;   \frac{ V_u^2 \Delta_{CDW}^2 }{ g^2(Q_1)}  \; \frac{\hslash \Omega(Q_1)}{2}  \hspace{1.5cm}
\\ \nonumber
 +  \;   V_u  \;   {\cal{N}}(0) 
 \sum_{\sigma}   \int  d \varepsilon^{(e)}_{\vec{k}} \;  \sqrt{  \xi^2_{\vec{k}}+ \Delta^2_{CDW}} \; \; \; 
\\ \nonumber  
 \bigg [  \hat{\tilde{\psi}}^{\dagger}_{1}(\vec{k},\sigma) 
   \hat{\tilde{\psi}}_{1}(\vec{k},\sigma)  \;  -  \;  \hat{\tilde{\psi}}^{\dagger}_{2}(\vec{k},\sigma)    \hat{\tilde{\psi}}_{2}(\vec{k},\sigma)  \bigg ]
   \; ,
\end{eqnarray}
which justifies  Eq.~(\ref{5.75}).
\end{document}